\documentclass[%
 reprint,
 superscriptaddress,
bibnotes,
 amsmath,amssymb,
 aps,
pra,
]{revtex4-1}

\usepackage{graphicx}
\usepackage{dcolumn}
\usepackage{bm}
\usepackage{booktabs}
\usepackage{makecell}
\usepackage[flushleft]{threeparttable}
\usepackage{natbib}
\usepackage{subcaption}
\usepackage{amsmath}
\usepackage[dvipsnames]{xcolor}

\usepackage{xr}

\providecommand{\e}[1]{\ensuremath{\times 10^{#1}}}

\begin{document}


\title{Thermal conductivity for III-V and II-VI semiconductor wurtzite and zinc-blende polytypes: the role of anharmonicity and phase space}

\author{Mart\'i Raya-Moreno}
 \affiliation{Departament d’Enginyeria Electr\`onica, Universitat Aut\`onoma de Barcelona, 08193 Bellaterra, Barcelona,Spain}
\author{Riccardo Rurali}%
\affiliation{Institut de Ci\`encia de Materials de Barcelona (ICMAB–CSIC), Campus de Bellaterra, 08193 Bellaterra,
	Barcelona, Spain}%
\author{Xavier Cartoix\`a}
\email{Xavier.Cartoixa@uab.cat}
\affiliation{Departament d’Enginyeria Electr\`onica, Universitat Aut\`onoma de Barcelona, 08193 Bellaterra, Barcelona,Spain}

\date{\today}

\begin{abstract}
We calculate the lattice thermal conductivity ($\kappa$) for cubic (zinc-blende) and hexagonal (wurtzite) phases for 8 semiconductors using {\it ab initio} calculations and solving the Phonon Boltzmann Transport Equation, explaining the different behavior of the ratio $\kappa_{\rm hex}/\kappa_{\rm cub}$ between the two phases.
We show that this behavior depends on the relative importance of two antagonistic factors: anharmonicity, which we find to be always higher in the cubic phase; and the accessible phase space, which is higher for the less symmetric hexagonal phase. Based on that, we develop a method that predicts the most conducting phase---cubic or hexagonal---where other more heuristic approaches fail. We also present results for nanowires made of the same materials, showing the possibility to tune $\kappa_{\rm hex}/\kappa_{\rm cub}$ over a wide range by modifying their diameter, thus making them attractive materials for complex phononic and thermoelectric applications/systems.
\end{abstract}

\maketitle


\section{\label{sec:Intro}Introduction}

Heat management stands out as one of the key problems for several technologies. The reduction in size~\cite{Moores_law} and the required increment in operating speeds of devices in electronics, or the necessity to enhance energy conversion ratio for thermoelectrics have increased the importance of phonon engineering or phononics~\cite{CahillAPL2014}.
For instance, for thermoelectric applications a material with low thermal conductivity and high electrical conductivity, the so called ``phonon-glass and electron-crystal'', is desired in order to have a good efficiency~\cite{Slack_Thermoelectrics_book}. Historically, one of the preferred approaches to thermoelectric materials was alloying, as it introduces point-mass defects that strongly scatter phonons, significantly reducing lattice thermal conductivity~\cite{KanatzidisChemMatter2010}. A more recent approach to solve this problem is the nanostructuration as an effective way of reducing the lattice thermal conductivity~\cite{MaldovanNature2013}---hereafter just referred to as thermal conductivity, $\kappa$---or a combination of both~\cite{KanatzidisChemMatter2010}.

At the same time, recent advances in semiconductor synthesis at the nanoscale have granted access to different phases that, in bulk, are only observed under extreme conditions. Namely, the wurtzite phase has theoretically been proved to be the most stable for several bulk-cubic~\cite{YehPRB1992} semiconductors when grown as nanowires (NWs)~\cite{AkiyamaJJAPL2006} provided their diameter is smaller than a given critical value, but it can also be obtained at larger sizes, though thermodynamically it is only metastable. Indeed, the wurtzite phase in NWs has been reported for a great range of bulk-cubic semiconductors: InP~\cite{MattilaAPL2006,PaimanNano2009}, Si~\cite{FranLopezACSNano2011,HaugeNanoLet2015,TangNanoscale2017}, InAs~\cite{HimuraJAP1995,CaroffNatNano2009}, GaAs~\cite{TchernychevaNANO2006,ZardoPRB2009} or GaP~\cite{AssaliNanoLett2013,BergJCG2014} to name but a few.

It is in that context where these new semiconductor phases can become a cornerstone for thermoelectricity and phononics in general, as they normally show a lower thermal conductivity than their more stable counterparts while maintaining similar electronic properties. For example, Togo~\textit{et al.}\ calculated thermal conductivity of wurtzite and zinc-blende phases for 33 different materials~\cite{phono3py}, Lindsay~\textit{et al.}~\cite{LindsayPRB2013} studied the zinc-blende/wurtzite thermal conductivity of GaN, and Li~\textit{et al.}\ addressed the thermal conductivity of bulk and nanowire InAs, AlN, and BeO polymorphs~\cite{LiJAP2013}. However, none of these works explained in detail the origin of the reduction in the thermal conductivity. In a more recent example, we have reported and discussed a reduction of $40\%$ in the thermal conductivity between the more stable 3C cubic-Si and 2H hexagonal-Si~\cite{MartiAPL2017}.

With regard to the qualitative behavior of the thermal conductivity of materials, Mukhopadhyay {\it et al.}~\cite{MukhopadhyayPRB2016}, building on earlier work by Slack~\cite{SlackJPCS1973} and Lindsay~\textit{et al.}~\cite{LindsayPRL2013}, have provided seven criteria to interpret the relative magnitude of the thermal conductivity between different materials.

In this work we study the thermal conductivity ($\kappa$) of several bulk and nanowire semiconductors for their cubic (zinc-blende, ZB) and the hexagonal (wurtzite, WZ) phases from first-principles. After seeing that the seven criteria provided in Ref.~\onlinecite{MukhopadhyayPRB2016} are unsuitable for rationalizing the $\kappa_{\rm hex}/\kappa_{\rm cub}$ value for the different materials, we propose an approach that successfully addresses this issue, providing insight into the factors determining the thermal conductivity of the materials we have studied.

The paper is structured as follows: after discussing the applied methodology for the thermal conductivity calculation in Sec.~\ref{sec:Method}, we present the results in Sec.~\ref{sec:Results}. Results are provided for bulk systems (Sec.~\ref{ssec:bulk}) and nanowires (Sec.~\ref{ssec:NWs}). Summary and conclusions are given in Sec.~\ref{sec:Conclusions}.

\section{\label{sec:Method} Methodology}

Harmonic and anharmonic interatomic force constants (IFCs), needed to calculate the thermal conductivity, were obtained using the supercell method, as implemented in Phonopy~\cite{Phonopy} for the harmonic IFCs and thirdorder.py~\cite{ShengBTE} for anharmonic IFCs. They were calculated in cubic polytypes in a 5x5x5 supercell for harmonic IFCs and 4x4x4 for the anharmonic IFCs. For hexagonal polytypes, the used supercell was 4x4x3 for both types of IFC. In order to minimize the computational burden, anharmonic IFCs were computed from interactions up to fourth nearest neighbors, while it has been previously reported that including up to third nearest neighbors was sufficient to give a satisfactorily converged value of $\kappa$~\cite{ShengBTE}.
The unit cells used to span the supercells were optimized until strict limits for stress ($3\cdot10^{-3}$GPa) and forces ($5 \cdot 10^{-4}$ eV/\AA) were attained. These optimizations were conducted within density-functional theory (DFT), using the VASP~\cite{VASP_1,VASP_2,VASP_3,VASP_4} code with projector augmented-wave (PAW) potentials~\cite{PAW_blochl,PAW_Kresse}. Local density approximation (LDA) for the exchange-correlation as parametrized by Perdew and Zunger~\cite{Perdew_PZ} to Ceperley-Alder~\cite{Ceperley_CA} data was used. For each system, the $k$-point mesh size had been previously optimized, taking into account that the supercell used to calculate 3rd-order IFCs should be commensurate with the mesh in order not to introduce spurious forces. Therefore, the selected primitive cell $k$-mesh for cubic materials was a 16$\times$16$\times$16 shifted mesh, and a 16$\times$16$\times$12 $\Gamma$-centered mesh for hexagonal polytypes, which are converged meshes for all systems. After optimization, a density functional perturbation theory (DFPT) run using VASP, with a doubled $k$-mesh was performed to obtain the Born charges ($Z^{*}$) and dielectric constant at high frequency ($\varepsilon^{\infty}$), needed to calculate the non-analytic term correction for the dynamical matrix near $\Gamma$.

After obtaining the IFCs, we solved the full linearized Boltzmann Transport Equation (LBTE) iteratively, as implemented in the almaBTE~\cite{AlmaBTE} code for bulk and the ShengBTE code~\cite{ShengBTE} for nanowires. Consequently, the thermal conductivity tensor ($\kappa^{\alpha\beta}$) is obtained as:
\begin{equation}
\kappa^{\alpha\beta}=\frac{1}{N k_B \Omega T^2} \sum_{\lambda} n_{0}(n_{0}+1)(\hbar\omega_{\lambda})^{2} \upsilon_{\lambda}^{\alpha} F_{\lambda}^{\beta},
\label{kappa_eq}
\end{equation}
where $\alpha$ and $\beta$ label the Cartesian axes $x$, $y$ and $z$; ${N,k_B,\Omega,T}$ are the number of q-points, the Boltzmann constant, the cell volume and the temperature, respectively. The summation is done over all phonon modes $\lambda$, which are characterized by the band label $n$ and their $q$-vector. $n_{0}$ is the equilibrium Bose-Einstein distribution function, $\hbar$ is the reduced Planck constant, $\omega_{\lambda}$ is the phonon mode frequency and $\upsilon_{\lambda}^{\alpha}$ is the group velocity of phonon mode $\lambda$ along the $\alpha$ direction. $F_{\lambda}^{\beta}$ is the generalized mean free path of the phonon mode along the $\beta$ direction, and it is calculated as $\tau_{\lambda}(\upsilon_{\lambda}^{\beta}+\Delta_{\lambda}^{\beta})$. $\tau_{\lambda}$ is the lifetime of the phonon in the relaxation time approximation (RTA) and $\Delta_{\lambda}^{\beta}$ is a measure of how much the population of a specific phonon mode and its associated heat current deviates from the RTA prediction~\cite{ShengBTE}. This correction is obtained iteratively starting from the RTA $(\Delta_{\lambda}^{\beta}=0)$.
In addition to anharmonic three-phonon scattering processes, mass-disorder scattering coming from isotope presence is treated using the Tamura model~\cite{Tamura_model}. Boundary scattering, whose effect is negligible in the temperature range considered, is not explicitly accounted for in the bulk systems (see Sec.~\ref{ssec:NWs} for a short description on the implementation of boundary scattering in NWs).

Regarding convergence with $q$-mesh, we found (see Figs.~\ref{conv-WZ} and \ref{conv-ZB} in the Supplemental Material~\cite{SI}) that for all materials but GaN and BN a $q$-mesh of 30$\times$30$\times$30 (ZB) and 30$\times$30$\times$19 (WZ) is enough to obtain converged values---less than 5\% change with respect to a higher accuracy 34$\times$34$\times$34/21 (ZB/WZ) mesh---of $\kappa$ and $\kappa_{pure}$ (i.e.\ without isotopes) for the 50--1000 K range. For GaN the converged $q$-mesh is found to be 34$\times$34$\times$34 (ZB) and 34x34x21 (WZ) for $\kappa$ and $\kappa_{pure}$, with respect to a 38$\times$38$\times$38/24 (ZB/WZ) mesh. Finally, the harder phonon modes of BN translate into a more demanding convergence behavior. On one hand, $q$-mesh convergence was only achieved for a 38$\times$38$\times$38 and 38$\times$38$\times$24 mesh, but only for the non-isotopically pure material, while the iterative calculation did not reach convergence for the isotopically pure material at low temperatures. Note that all these $q$-mesh values are significantly denser than the 24$\times$24$\times$24 mesh needed to achieve convergence in Si~\cite{ShengBTE}.

\section{\label{sec:Results} Results}

\subsection{\label{ssec:IFC_test} Interatomic Force Constants test}

Owing to the computational cost required to obtain the anharmonic IFCs (168 and 208 DFT runs for cubic and hexagonal polytypes, respectively), prior to their calculation an accuracy test to harmonic IFCs was done using Phonopy~\cite{Phonopy,Phonopy_NACinterpolation,phono3py}, by checking the optical phonon frequencies at the $\Gamma$ point. These results, comparing calculated and experimental values, are shown in the phonon dispersion relations (see Figs.~\ref{disp1}-\ref{disp3}~\cite{SI}), exhibiting, despite small differences, a good agreement between our calculations and the experimental values for the TO modes.
The non-analytical correction (NAC), needed in polar materials to get the LO-TO splitting, is underestimated because of it being inversely proportional to $\epsilon^{\infty}$~\cite{NAC_Pick}, which is overestimated due to LDA inability to take into account the polarization dependence on the exchange-correlation functional under a field~\cite{OrtizPRL1998}. This leads to an underestimation of the LO frequency with respect to the experiments.
It should be noted, nonetheless, that the effect of LDA and GGA shortcomings have been proven to have a smaller effect in the anharmonic properties than in the harmonic ones~\cite{LDAvsGGA_harmvsanharm,ArrigoniCMS2019}.

\subsection{\label{ssec:bulk} Bulk}
In this section we study the thermal conductivities of cubic (ZB) and hexagonal (WZ) phases for different compound semiconductors, using the methodology discussed in Sec.~\ref{sec:Method}. These two phases have the same coordination (4-fold, tetrahedral) and very similar first neighbor distances, the main difference being the stacking of atomic layers.

We further substantiate our results by comparing our calculated thermal conductivity to available experimental values (see Table~\ref{table-kappas}). It is reasonable to expect that, for some materials, our results overestimate the experimental values, because samples used in such experiments might contain defects (impurities, vacancies, dislocations, etc), which can strongly suppress thermal conductivity, and they have not been considered in our simulations. Moreover, there might be a dependence of the measured value on the experimental technique, and it is quite challenging to obtain experimental thermal conductivity values with less than 5\% error~\cite{ZhaoQianGu2016}. We note that our values are in excellent agreement with experimental results for GaAs and AlAs, while keeping a good agreement for InP, InAs, ZnSe and GaP. In the nitrides, BN and GaN, we obtain values within the dispersion of the experimentally reported magnitudes. Regarding comparison with other first-principles calculations, our results are in reasonable agreement with those of Lindsay et al.~\cite{LindsayPRB2013}, while there is a stronger disagreement with the values of Togo et al.~\cite{phono3py}, which are obtained by a different approach to the LBTE. These two approaches are known to provide different values for the thermal conductivity in transition metal dichalcogenides as well~\cite{Torres2DMAT2019}.

\begin{figure}
	\centering
	\includegraphics[width=\linewidth]{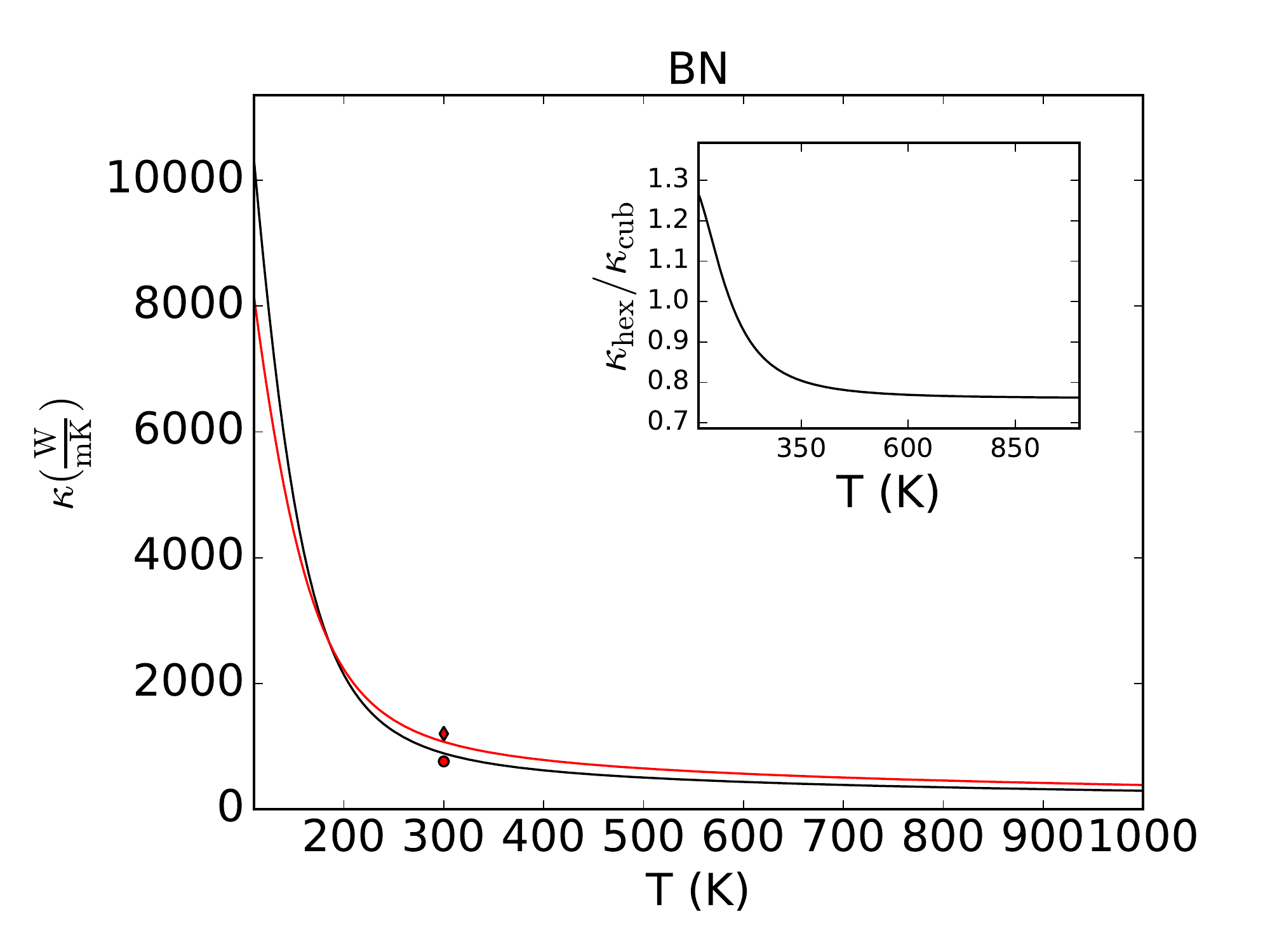}
	\caption{Hexagonal(black) and cubic(red) BN thermal conductivity $\kappa$ trace mean as a function of temperature. Experimental results for cubic phase are from Ref.~\citenum{HTCM_shinde} (circle) and Ref.~\citenum{KARIM1993502,SamantarayBN} (diamond). Inset: ratio between hexagonal and cubic thermal conductivity as a function of the temperature.}
	\label{bn-ktr}
\end{figure}

\begin{figure}
	\centering
	\includegraphics[width=\linewidth]{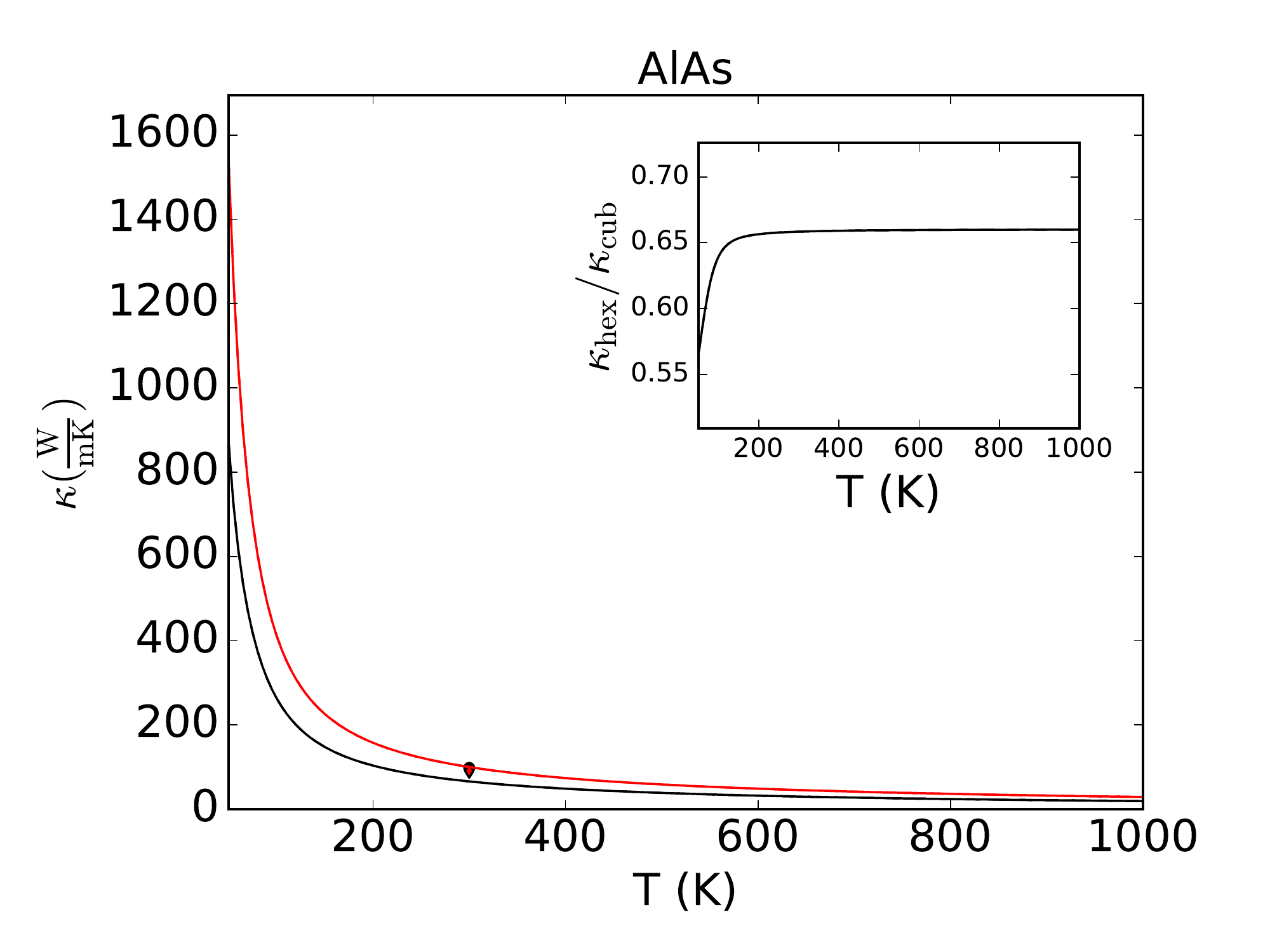}
	\caption{Hexagonal(black) and cubic(red) AlAs thermal conductivity $\kappa$ trace mean as a function of temperature. Experimental results for cubic phase are from Ref.~\citenum{HTCM_shinde}(circle) and Ref.~\citenum{AfromowitzJAP1973}.  Since both constituting elements are isotopically pure, AlAs presents no isotopic scattering. Inset: ratio between hexagonal and cubic thermal conductivity as a function of the temperature.}
	\label{alas-ktr}
\end{figure}

\begin{figure}
	\centering
	\includegraphics[width=\linewidth]{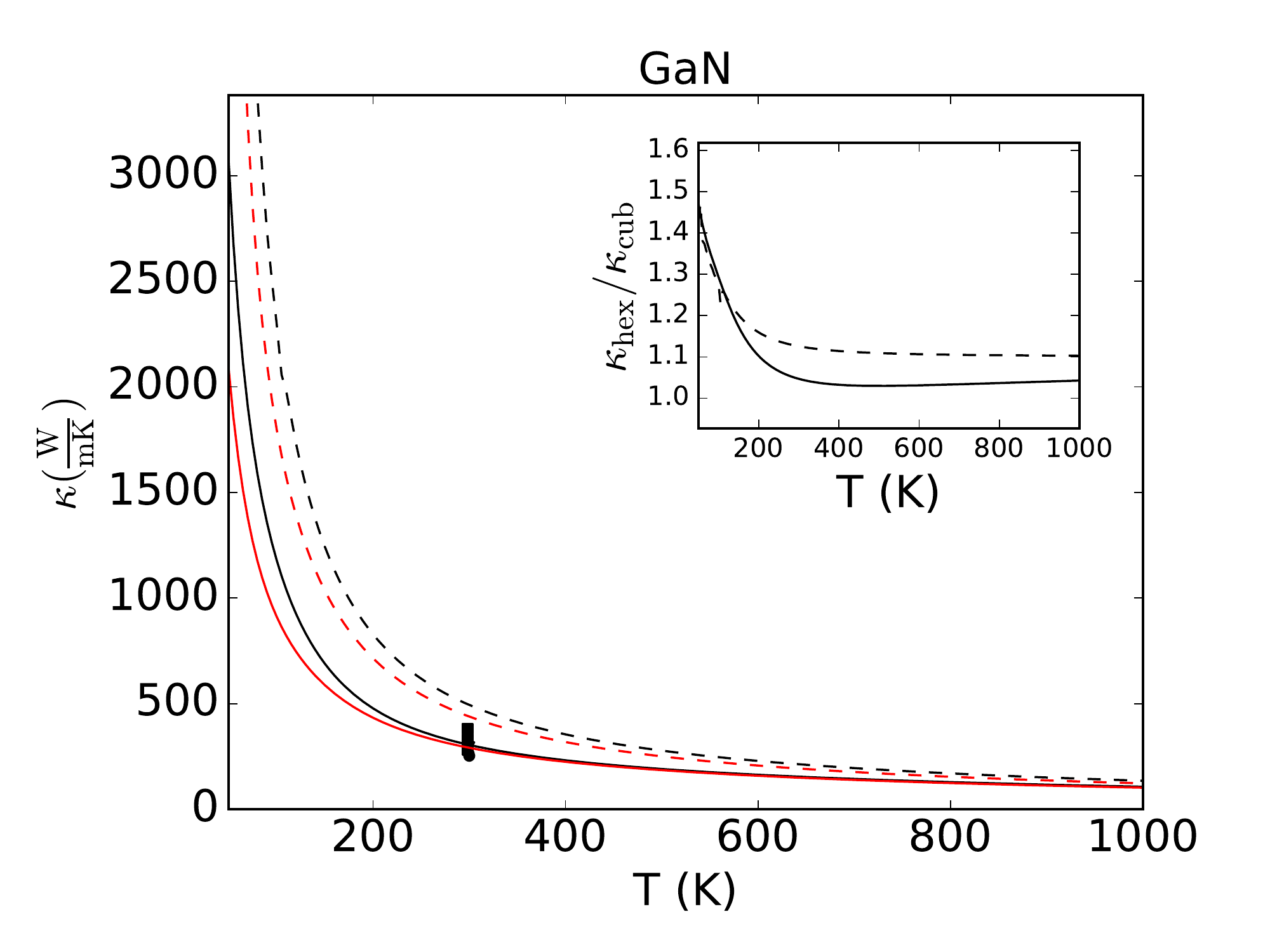}
	\caption{Hexagonal(black) and cubic(red) GaN thermal conductivity $\kappa$ (solid) and $\kappa_{pure}$ (dashed) trace mean as a function of temperature. Experimental results for hexagonal phase are from: Ref.~\citenum{ShibataMRP2007} (circle), Ref.~\citenum{JezowskiMRE2015} (diamond), Ref.~\citenum{RichterPSSC2011} (triangle) and Ref.~\citenum{ShibataUS20090081110A1} (square). Inset: ratio between hexagonal and cubic thermal conductivity as a function of the temperature with (solid) and without (dashed) isotopic scattering. We notice that some numerical instabilities/noise were found in isotope free simulations at low temperature for wurtzite phase.}
	\label{gan-ktr}
\end{figure}

\begin{figure}
	\centering
	\includegraphics[width=\linewidth]{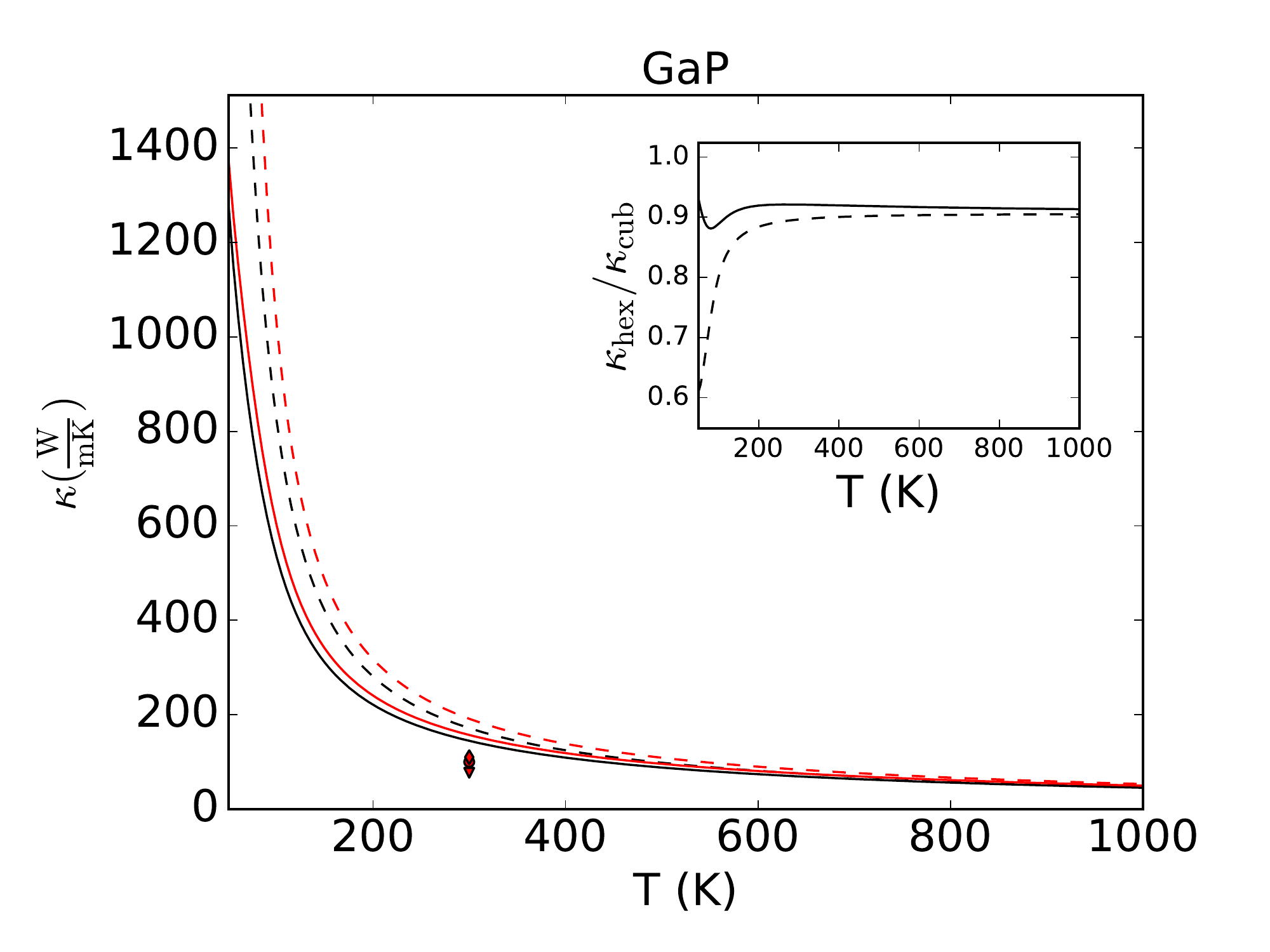}
	\caption{Hexagonal(black) and cubic(red) GaP thermal conductivity  $\kappa$ (solid) and $\kappa_{pure}$ (dashed) trace mean as a function of temperature. Experimental results for cubic phase are from Ref.~\citenum{SteigmeierPR1966} (triangle), Ref.~\citenum{HTCM_shinde}(circle) and Ref.~\citenum{HandbookElecMat}(diamond). Inset: ratio between hexagonal and cubic thermal conductivity as a function of the temperature with (solid) and without (dashed) isotopic scattering.}
	\label{gap-ktr}
\end{figure}

\begin{figure}
	\centering
	\includegraphics[width=\linewidth]{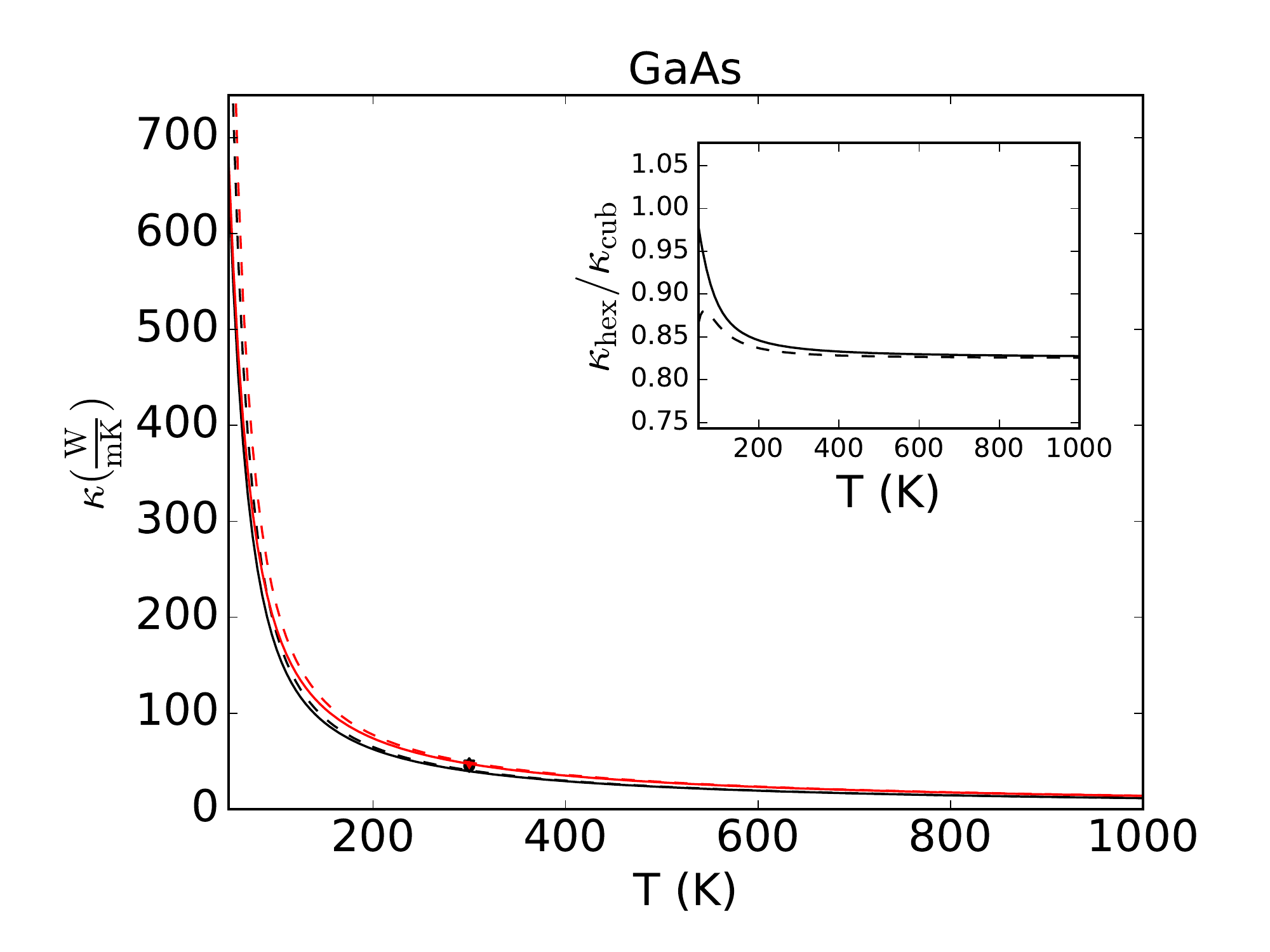}
	\caption{Hexagonal(black) and cubic(red) GaAs thermal conductivity $\kappa$ (solid) and $\kappa_{pure}$ (dashed) trace mean as a function of temperature. Experimental results for cubic phase are from Ref.~\citenum{HTCM_shinde}(circle) Ref.~\citenum{POSSzeWiley}(triangle) and Ref.~\citenum{SteigmeierPRev1963}(diamond). Inset: ratio between hexagonal and cubic thermal conductivity as a function of the temperature with (solid) and without (dashed) isotopic scattering.}
	\label{gaas-ktr}
\end{figure}

\begin{figure}
	\centering
	\includegraphics[width=\linewidth]{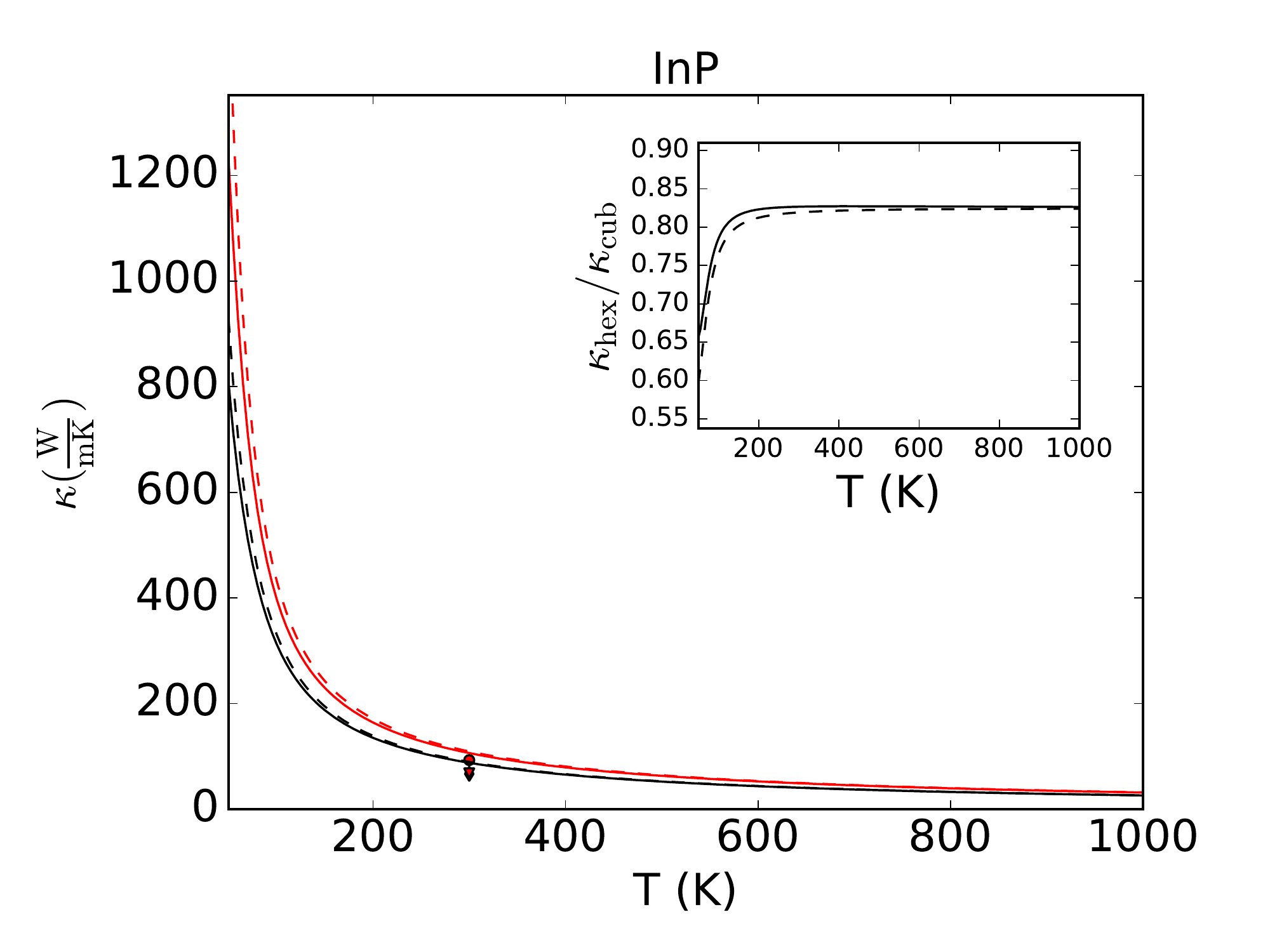}
	\caption{Hexagonal(black) and cubic(red) InP thermal conductivity $\kappa$ (solid) and $\kappa_{pure}$ (dashed) trace mean as a function of temperature. Experimental results for cubic phase are from Ref.~\citenum{SteigmeierPR1966} (triangle), Ref.~\citenum{HTCM_shinde} (circle) and Ref.~\citenum{SteigmeierPRev1963} (diamond). Inset: ratio between hexagonal and cubic thermal conductivity as a function of the temperature with (solid) and without (dashed) isotopic scattering.}
	\label{inp-ktr}
\end{figure}

\begin{figure}
	\centering
	\includegraphics[width=\linewidth]{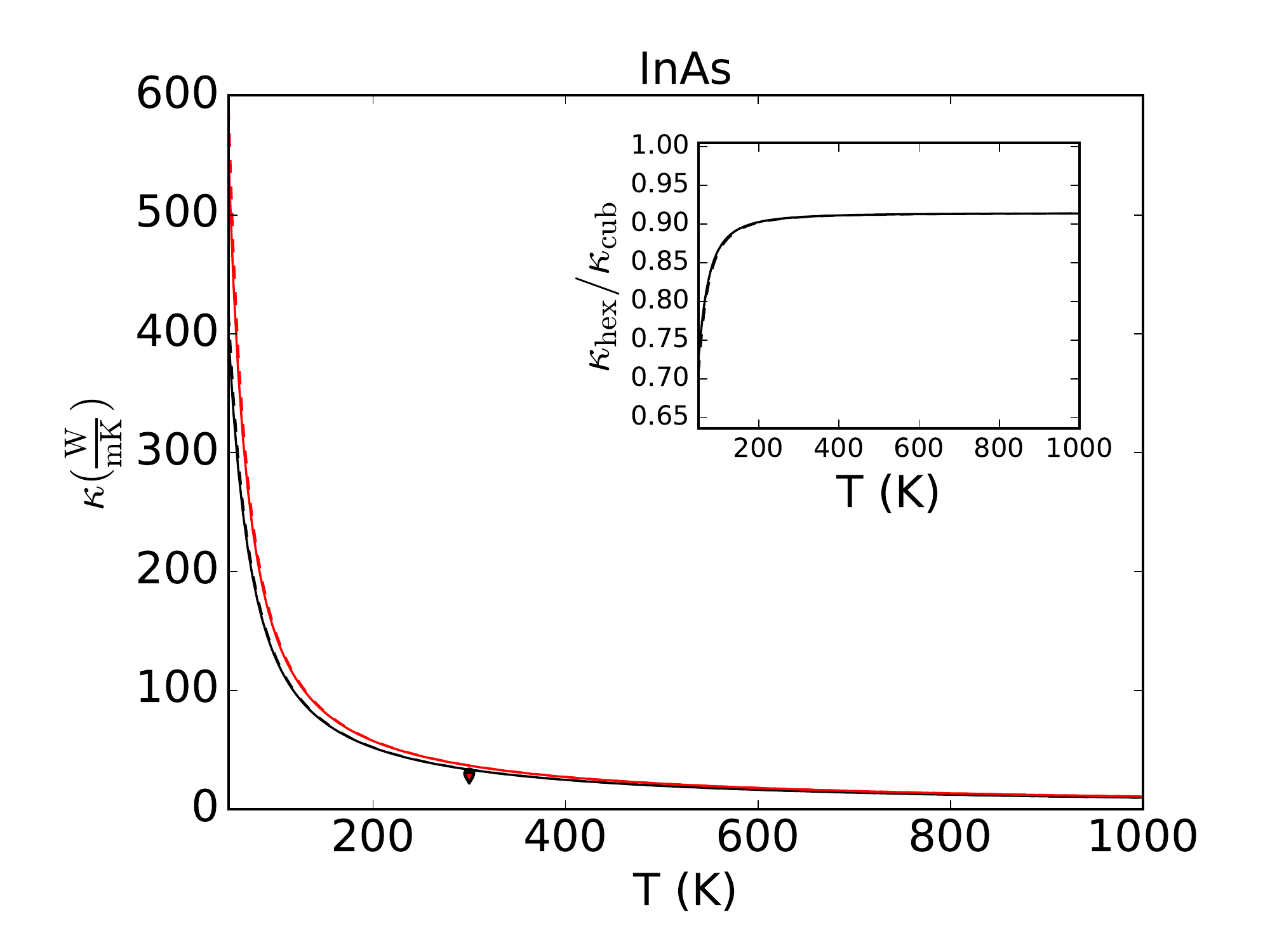}
	\caption{Hexagonal(black) and cubic(red) InAs thermal conductivity $\kappa$ (solid) and $\kappa_{pure}$ (dashed) trace mean as a function of temperature. Experimental results for cubic phase are from Ref.~\citenum{SteigmeierPR1966} (triangle), Ref.~\citenum{HTCM_shinde} (circle) and Ref.~\citenum{SteigmeierPRev1963} (diamond). Inset: ratio between hexagonal and cubic thermal conductivity as a function of the temperature with (solid) and without (dashed) isotopic scattering.}
	\label{inas-ktr}
\end{figure}

\begin{figure}
	\centering
	\includegraphics[width=\linewidth]{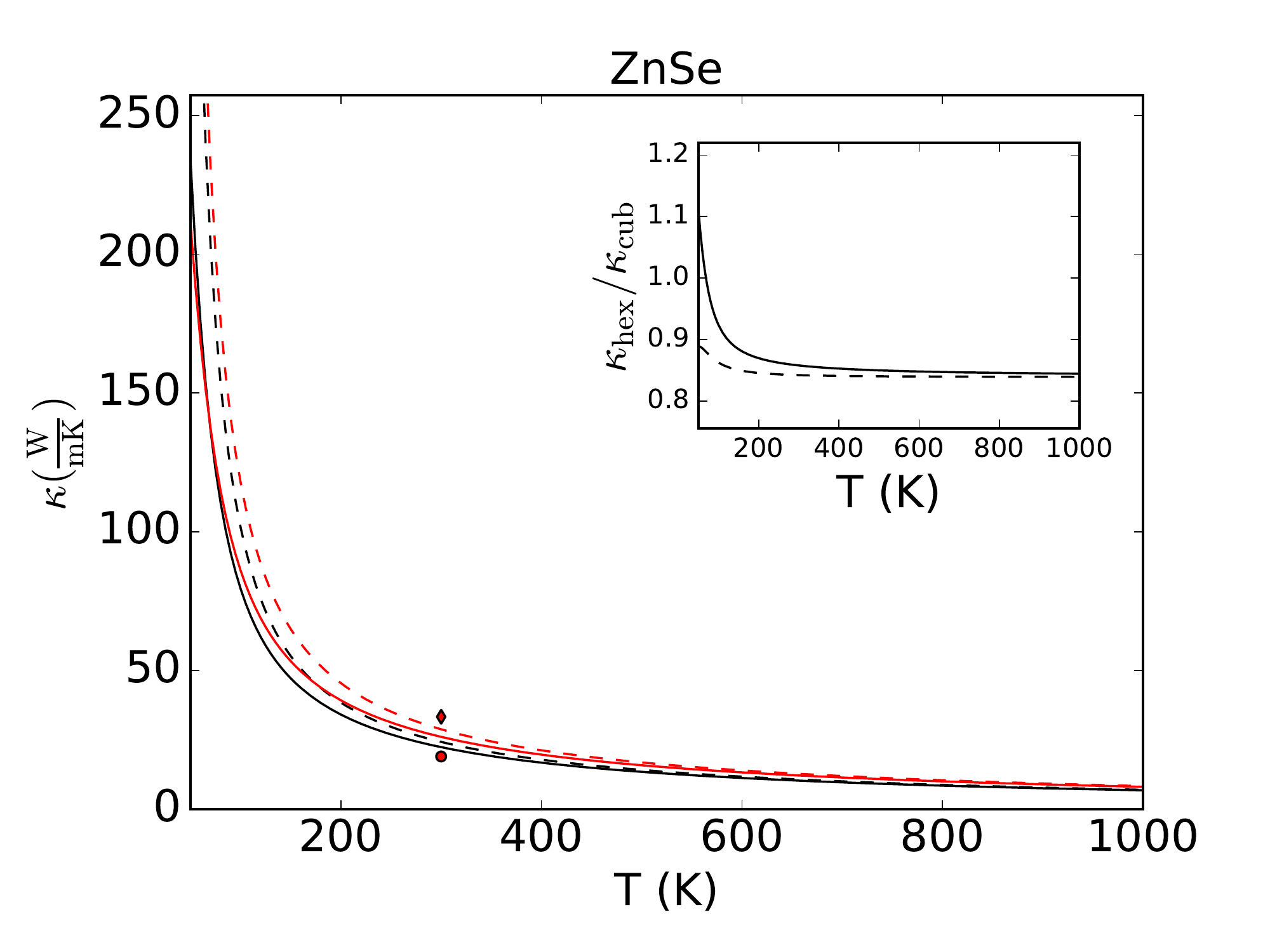}
	\caption{Hexagonal(black) and cubic(red) ZnSe thermal conductivity $\kappa$ (solid) and $\kappa_{pure}$ (dashed) trace mean as a function of temperature. Experimental results for cubic phase are from Ref.~\citenum{HTCM_shinde,Madelung2ZnSe} (circle) and Ref.~\citenum{SLACK19791} (diamond). Inset: ratio between hexagonal and cubic thermal conductivity as a function of the temperature with (solid) and without (dashed) isotopic scattering.}
	\label{znse-ktr}
\end{figure}

We found (see Table~\ref{table-kappas} and Figs.~\ref{bn-ktr}-\ref{znse-ktr}) that most of materials under study (GaAs, GaP, InP, InAs, ZnSe and AlAs) follow the silicon behavior of reducing their $\kappa$ with symmetry~\cite{MartiAPL2017}. However, we observe that is possible for some materials (GaN) to have the opposite behavior, namely $\kappa$ increases when symmetry is reduced, i.e. going from ZB to WZ. Moreover, as previously observed in other materials~\cite{LiJAP2013}, BN can show these two opposite behaviors at different temperatures.


\begin{table*}
	\centering
	\caption{\label{table-kappas} Calculated $\kappa$ and $\kappa_{hex}/\kappa_{cub}$ ratios at 300~K. The experimental $\kappa$ at room temperature is also presented at normal conditions. The values for all $\kappa$'s are given in W/m$\cdot$K.}
	\begin{threeparttable}
		\begin{tabular}{@{}lllllllll@{}}
			\toprule
			\hline \hline \\
			& \makecell{$\kappa^{calc}_{cub}$ \\ Ref.~\citenum{phono3py} } & \makecell{${\kappa^{calc}_{hex}}_{\dagger}$ \\ Ref.~\citenum{phono3py} }
			& \makecell{$\kappa^{calc}_{cub}$ \\ Ref.~\citenum{LindsayPRB2013} } & \makecell{${\kappa^{calc}_{hex}}^{\dagger}$ \\ Ref.~\citenum{LindsayPRB2013} }
			& \makecell{$\kappa^{calc}_{cub}$ \\ This work} & \makecell{${\kappa^{calc}_{hex}}^{\dagger}$ \\ This work}
			& \makecell{${{}^{\kappa_{hex}}\!/_{\kappa_{cub}}}$ \\ \;} & \makecell{${\kappa^{exp}}^{\ddagger}$} \\ \\ \midrule
			\hline \\
			BN & \makecell[tc]{726} & \makecell[tc]{592\\(602/573)$^{\S}$} & \makecell[tc]{940$^{\dagger\dagger}$}  & \makecell[tc]{--} & \makecell[tc]{1071}  & \makecell[tc]{887.7 \\ (906.6/849.9)$^{\S}$} & \makecell[tc]{0.845} & \makecell[tc]{760$^a$ (cub)\\1200$^b$ (cub)} \\
			\\
			AlAs & \makecell[tc]{86.8} & \makecell[tc]{72.9\\(73.9/71.0)} & \makecell[tc]{105} & \makecell[tc]{--} & \makecell[tc]{100.0}  & \makecell[tc]{ 65.84\\ (65.54/66.45)} & \makecell[tc]{0.655} & \makecell[tc]{98$^a$ (cub) \\ 91$^c$ (cub)} \\
			\\
			GaN & \makecell[tc]{181} & \makecell[tc]{ 171 \\ (171/172)}& \makecell[tc]{215}& \makecell[tc]{241\\(242/239)}& \makecell[tc]{290.7}  & \makecell[tc]{304.3 \\ (293.8/325.3)} & \makecell[tc]{1.047} & \makecell[tc]{253$^d$ (hex)\\269$^e$ (hex)\\294$^f$ (hex)\\280$^g$ (hex)\\300$^g$ (hex)\\330$^g$ (hex)\\380$^g$ (hex)} \\
			\\
			GaP & \makecell[tc]{104}  & \makecell[tc]{92.8\\(96.5/85.4)} & \makecell[tc]{131} & \makecell[tc]{--} & \makecell[tc]{157.1}  & \makecell[tc]{144.7\\(148.8/136.4)} & \makecell[tc]{0.915} & \makecell[tc]{77$^h$ (cub)\\100$^a$ (cub)\\110$^i$ (cub)} \\
			\\
			GaAs & \makecell[tc]{32.1} & \makecell[tc]{27.2\\(27.8/25.9)} & \makecell[tc]{54} & \makecell[tc]{--} & \makecell[tc]{47.23}  & \makecell[tc]{39.52\\(38.97/40.61)} & \makecell[tc]{0.837} & \makecell[tc]{45$^a$ (cub) \\45.5$^j$ (cub)\\46$^k$ (cub)} \\
			\\
			InP & \makecell[tc]{85.2} & \makecell[tc]{68.9\\69.3/68.2} & \makecell[tc]{89} & \makecell[tc]{--} & \makecell[tc]{106.2}  & \makecell[tc]{87.82\\(85.46/92.55)} & \makecell[tc]{0.827} & \makecell[tc]{93$^a$ (cub) \\ 67$^k$ (cub) \\ 68$^h$ (cub)} \\
			\\
			InAs & \makecell[tc]{25.2} & \makecell[tc]{18.3\\(18.5/18.0)} & \makecell[tc]{36} & \makecell[tc]{--} & \makecell[tc]{36.63}  & \makecell[tc]{33.29 \\ (33.13/33.61)} & \makecell[tc]{0.909} & \makecell[tc]{30$^a$ (cub) \\ 27.3$^j$ (cub) \\ 26.5$^h$ (cub)} \\
			\\
			ZnSe & \makecell[tc]{15.6} & \makecell[tc]{14.0\\(13.8/14.5)} & \makecell[tc]{--} & \makecell[tc]{--} & \makecell[tc]{26.06}  & \makecell[tc]{22.35 \\ (21.65/23.76)} & \makecell[tc]{0.858} & \makecell[tc]{19$^{a,l}$ (cub)\\33$^m$ (cub)} \\
			\\ \bottomrule
			\hline \hline	
		\end{tabular}
		\begin{tablenotes}
			\small
			\item $^{\dagger}$ Mean of $\kappa$ trace (in-plane $\kappa$ / out-of-plane $\kappa$, along c-crystallographic axis).
			\item $^{\ddagger}$ The material phase of the experimental measurement is indicated between brackets.
			\item $^{\S}$ The hexagonal phase refers to WZ, as opposed to the layered h-BN phase.
			\item $^{\dagger\dagger}$ This value is from Ref. \citenum{LindsayPRL2013} 
			\item $^a$ Ref.~\citenum{HTCM_shinde}
			\item $^b$ Refs.~\citenum{KARIM1993502,SamantarayBN}
			\item $^c$ Ref.~\citenum{AfromowitzJAP1973}
			\item $^d$ Ref.~\citenum{ShibataMRP2007}
			\item $^e$ Ref.~\citenum{JezowskiMRE2015}
			\item $^f$ Ref.~\citenum{RichterPSSC2011}
			\item $^g$ Ref.~\citenum{ShibataUS20090081110A1}. Measurements are done at 298.15 K.
			\item $^h$ Ref. ~\citenum{SteigmeierPR1966}
			\item $^i$ Ref.~\citenum{HandbookElecMat}
			\item $^j$ Ref.~\citenum{POSSzeWiley}	
			\item $^k$ Ref.~\citenum{SteigmeierPRev1963} 
			\item $^l$ Ref. ~\citenum{Madelung2ZnSe}
			\item $^m$ Ref.~\citenum{SLACK19791}

			\end{tablenotes}
	\end{threeparttable}
\end{table*}

\subsubsection{\label{sssec:saok} Standard analysis of $\mathrm{\kappa_{hex}/\kappa_{cub}}$}

Four conditions that a crystal must fulfill to have a high $\kappa$ have been long established~~\cite{SlackJPCS1973}: (I) to be structurally simple, (II) to be composed of light elements, (III) to have strong covalent bonds---represented normally by a high Debye temperature/frequency ($f_D$)---and (IV) to be harmonic, which is normally associated to a low Gr\"uneisen parameter ($\gamma$). 

Recently, three additional conditions for a high $\kappa$ have been introduced by Lindsay \textit{et al.}~\cite{LindsayPRL2013} and Mukhopadhyay \textit{et al.}~\cite{MukhopadhyayPRB2016}. They demonstrated that, in addition to the four previous conditions, a higher $\kappa$ is obtained if the material also has (V) high ``acoustic-optical'' (a-o) gap, (VI) a high ``acoustic bunching'' and (VII) low optical bandwidth.

\begin{table*}[t]
	\centering
	\caption{\label{table-conditions}Seven standard conditions of high $\kappa$ for all materials in both phases. Boldface indicates the characteristic more favorable to a higher $\kappa$.  $M_{avg}$ stands for the average atomic mass of the unit cell, $f_{D}$ for the Debye frequency and $\gamma_{300K}$ for the Gr\"uneisen parameter at room temperature.}
	\begin{threeparttable}
		\begin{tabular}{@{}llllllll@{}}
			\toprule
			\hline \hline \\
			\makecell[tc]{material} &\makecell[tc]{(I)\\crystal\\structure} &\makecell[tc]{(II)\\$\mathrm{M_{avg}}$\\$(amu)$}  &\makecell[tc]{(III)\\$f_D$\\$\mathrm{(THz)}$} &\makecell[tc]{(IV)\\$\gamma_{300K}$} & \makecell[tc]{(V)\\``a-o'' gap\\$\mathrm{(THz)}$} &\makecell[tc]{(VI)\\``acoustic''\\bunching\\$\mathrm{(THz)}$} &\makecell[tc]{(VII)\\optical\\ bandwidth\\$\mathrm{(THz)}$} \\ \midrule
			\hline \\
			\makecell[tc]{BN} 
			& \makecell[tc]{w \\ \textbf{zb}} 
			& \makecell[tc]{12.41\\12.41}
			& \makecell[tc]{39.15\\39.15}
			& \makecell[tc]{\textbf{0.67}\\0.73}  
			& \makecell[tc]{0.00\\0.00}
			& \makecell[tc]{30.11\\\textbf{15.55}}
			& \makecell[tc]{11.41\\\textbf{9.37}}
			\\ \\
			\makecell[tc]{AlAs} 
			& \makecell[tc]{w \\ \textbf{zb}} 
			& \makecell[tc]{50.95\\50.95}
			& \makecell[tc]{\textbf{5.44}\\5.41}
			& \makecell[tc]{\textbf{0.29}\\0.40}  
			& \makecell[tc]{3.54\\\textbf{3.55}}
			& \makecell[tc]{6.32\\\textbf{3.56}}
			& \makecell[tc]{1.93\\1.93}
			\\ \\
			\makecell[tc]{GaN} 
			& \makecell[tc]{w \\ \textbf{zb}} 
			& \makecell[tc]{41.86\\41.86}
			& \makecell[tc]{\textbf{9.75}\\9.71}
			& \makecell[tc]{\textbf{0.64}\\0.70}  
			& \makecell[tc]{6.18\\\textbf{6.50}}
			& \makecell[tc]{10.27\\\textbf{6.43}}
			& \makecell[tc]{5.87\\\textbf{5.36}}
			\\ \\
			\makecell[tc]{GaP} 
			& \makecell[tc]{w \\ \textbf{zb}} 
			& \makecell[tc]{50.35\\50.35}
			& \makecell[tc]{\textbf{6.66}\\5.57}
			& \makecell[tc]{\textbf{0.51}\\0.57}  
			& \makecell[tc]{\textbf{2.67}\\2.49}
			& \makecell[tc]{7.01\\\textbf{4.53}}
			& \makecell[tc]{2.10\\\textbf{1.99}}
			\\ \\
			\makecell[tc]{GaAs} 
			& \makecell[tc]{w \\ \textbf{zb}} 
			& \makecell[tc]{72.32\\72.32}
			& \makecell[tc]{4.53\\\textbf{4.67}}
			& \makecell[tc]{\textbf{0.55}\\0.60}  
			& \makecell[tc]{0.00\\0.00}
			& \makecell[tc]{6.21\\\textbf{4.35}}
			& \makecell[tc]{2.89\\\textbf{2.53}}
			\\ \\				
			\makecell[tc]{InP} 
			& \makecell[tc]{w \\ \textbf{zb}} 
			& \makecell[tc]{72.90\\72.90}
			& \makecell[tc]{\textbf{3.51}\\3.48}
			& \makecell[tc]{\textbf{0.38}\\0.48}  
			& \makecell[tc]{\textbf{3.88}\\3.69}
			& \makecell[tc]{4.92\\\textbf{3.41}}
			& \makecell[tc]{1.42\\\textbf{1.35}}
			\\ \\
			\makecell[tc]{InAs} 
			& \makecell[tc]{w \\ \textbf{zb}} 
			& \makecell[tc]{94.87\\94.87}
			& \makecell[tc]{2.92\\\textbf{3.07}}
			& \makecell[tc]{\textbf{0.36}\\0.38}  
			& \makecell[tc]{\textbf{0.82}\\0.51}
			& \makecell[tc]{4.56\\\textbf{3.62}}
			& \makecell[tc]{1.55\\\textbf{1.39}}
			\\ \\	
			\makecell[tc]{ZnSe} 
			& \makecell[tc]{w \\ \textbf{zb}} 
			& \makecell[tc]{72.17\\72.17}
			& \makecell[tc]{4.70\\\textbf{4.47}}
			& \makecell[tc]{\textbf{0.67}\\0.71}  
			& \makecell[tc]{0.00\\0.00}
			& \makecell[tc]{5.29\\\textbf{4.00}}
			& \makecell[tc]{\textbf{2.03}\\2.06}
			\\ \\	
			\\ \bottomrule
			\hline \hline	
		\end{tabular}
	\end{threeparttable}
\end{table*}

We have collected those seven conditions for all materials under study, with values from our first-principles calculations, in Table~\ref{table-conditions}. Interestingly, by observing those conditions together with the $\kappa$ values, we see that they do not clarify which of the phases---cubic or hexagonal---is the most conductive: for instance, if one looks at GaN four out of seven criteria suggest that zinc-blende phase should be the most conductive one, while one of them is neutral and only two of them predict a larger $\kappa$ for the wurtzite. Yet, the calculations predict the latter to be more conductive. Additionally, those conditions cannot either explain the reason for some materials, like BN, to change their $\kappa_{hex}/\kappa_{cub}$ ratio behavior with temperature. Such a shortcoming is due to being based on arguments of a qualitative character that do not quantify the relative importance of each condition, thus the necessity of a more quantitative viewpoint to predict/understand which phase is the most conductive at a given temperature.

\subsubsection{\label{sssec:aps} Effective anharmonicity and accessible phase space }

To gain insight on the $\kappa_{hex}/\kappa_{cub}$ ratio at different temperatures, we focus our analysis on two quantities that together contain all conditions: the three-phonon scattering matrix elements or anharmonicity (high-$\kappa$ conditions I,II, III and IV) and phase space, i.e. all the energy conserving three-phonon combinations (high-$\kappa$ conditions I,II,V,VI and VII). In fact, these two quantities directly contribute to $\kappa$ via the three-phonon scattering rate ($\Gamma_{\lambda\lambda'\lambda''}^{\pm}$)~\cite{ShengBTE}:
\begin{equation}
\Gamma_{\lambda\lambda'\lambda''}^{\pm} = \frac{\hbar\pi|V^{\pm}_{\lambda\lambda'\lambda''}|^2}{4\omega_{\lambda}\omega_{\lambda'}\omega_{\lambda''}} \begin{Bmatrix}
n_{0}'-n_{0}''\\
n_{0}'+n_{0}''+1
\end{Bmatrix} \delta(\omega_{\lambda}\pm\omega_{\lambda'}-\omega_{\lambda''}) ,
\label{3ph_eq}
\end{equation}
where $V^{\pm}_{\lambda\lambda'\lambda''}$ stands for the three-phonon scattering matrix element (+ for absorption processes and - for emission) and $n_{0}'$ is a shorthand for $n_{0}(\omega_{\lambda'})$ and similarly for $n_{0}''$.

As it can be seen from Eq.~(\ref{3ph_eq}), an increment in anharmonicity ($V^{\pm}_{\lambda\lambda'\lambda''}$) produces an increment in the scattering rate, therefore a reduction in $\kappa$. In the same way, an increment in phase space, represented in Eq.~(\ref{3ph_eq}) by the energy conservation delta~$\delta(\omega_{\lambda}\pm\omega_{\lambda'}-\omega_{\lambda''})$, also reduces $\kappa$. 

Comparing the phonon dispersion of both phases for all materials (see Figs.~\ref{disp1}-\ref{disp3}~\cite{SI}) it becomes obvious that symmetry reduction, which causes the appearance of new low/medium optical eigenmodes with non-vanishing scattering matrix elements, also increases the phase space for transitions at a given temperature. In order to obtain an actual measurement of accessible phase space while taking into account the effect of temperature, we calculate it ($\delta_{occ,T}$) defined as follows:
\begin{equation}
\delta_{occ,T} = \frac{2}{3}(\delta^{+}_{occ,T} + \frac{1}{2}\delta^{-}_{occ,T})
\label{delta_eq}
\end{equation}
\begin{multline}
\delta^{+}_{occ,T} = 2 \pi \left[\frac{V_{BZ}}{( 2\pi )^{3}}\right]^{3} \sum_{n,n',n''} \iiint\limits_{BZ}  \delta(\omega_{\lambda}+\omega_{\lambda'}-\omega_{\lambda''}) \times \\
   n_{0} n_{0}' (1+n_{0}'') \; \delta_{q+q',q''+ G} \; d^3q''d^3q'd^3q ,
\label{delta_eq+}
\end{multline}
\begin{multline}
\delta^{-}_{occ,T} = 2 \pi \left[\frac{V_{BZ}}{( 2\pi )^{3}}\right]^{3} \sum_{n,n',n''} \iiint\limits_{BZ} \delta(\omega_{\lambda}-\omega_{\lambda'}-\omega_{\lambda''}) \times \\
  n_{0} (1+n_{0}') (1+n_{0}'') \; \delta_{q-q',q''+ G} \; d^3q''d^3q'd^3q ,
\label{delta_eq-}
\end{multline}
where $\delta_{q\pm q',q''+ G}$ is the momentum conservation condition; $\delta^{+}_{occ,T}$ and $\delta^{-}_{occ,T}$ are the contribution to the accessible phase space of absorption and emission processes, and the $2/3$ and $1/2$ weighting factors ensure the normalization and non double counting of processes~\cite{LindsayJPCM2008}.
Therefore, $\delta_{occ,T}$  gives an idea for a given temperature of the accessible part of the available phase space. Certainly, the attainable phase space for the different materials (see Table ~\ref{table-deltavpp} and Figs.~\ref{DELTAS1}-\ref{DELTAS3}~\cite{SI}) confirms that, as inferred from dispersion relations, the hexagonal phase has a greater accessible phase space which contributes to reducing the $\kappa_{hex}/\kappa_{cub}$ ratio. This is a general feature of all the studied materials.

On the other hand, from the scattering matrix elements ($|V^{\pm}_{\lambda\lambda'\lambda''}|^2$) of energy-conserving three-phonon processes one obtains direct information of the material anharmonicity. Notwithstanding that this bare anharmonicity is an interesting quantity by itself, it is not useful for us because it gives the same importance to processes in which the involved modes are occupied to those in which they are not. Therefore, analogously to what we did with the accessible phase space, we define the temperature-dependent mean effective anharmonicity as:
\begin{equation}
\overline{{|V_{\lambda\lambda'\lambda''}|^2_{occ,T}}} = \frac{2}{3}(\overline{|V^{+}_{\lambda\lambda'\lambda''}|^2_{occ,T}} + \frac{1}{2} \overline{|V^{-}_{\lambda\lambda'\lambda''}|^2_{occ,T}})
\label{vpp_eq}
\end{equation}
where $\overline{|V^{+}_{\lambda\lambda'\lambda''}|^2_{occ,T}}$ and $\overline{|V^{-}_{\lambda\lambda'\lambda''}|^2_{occ,T}}$ are the arithmetic means of the population weighted three-phonon matrix elements squared modulus for absorption and emission processes:
\begin{alignat}{1}
|V^{+}_{\lambda\lambda'\lambda''}|^2_{occ,T} &= |V^{+}_{\lambda\lambda'\lambda''}|^2·n_{0}·n_{0}'·(1+n_{0}'')
\label{vpp_eq+} \\
|V^{-}_{\lambda\lambda'\lambda''}|^2_{occ,T} &=  |V^{-}_{\lambda\lambda'\lambda''}|^2·n_{0}·(1+n_{0}')·(1+n_{0}'') .
\label{vpp_eq-}
\end{alignat}

From the mean effective anharmonicity, it can be seen that, for all materials under study, the cubic phase is more anharmonic than the hexagonal phase both at 77~K (a representative value of the low temperature regime) and at 300~K (see Table~\ref{table-deltavpp} and Figs.~\ref{vpp_fulls1}-\ref{vpp_fulls3}~\cite{SI}).

\begin{table*}[t]
	\centering
	\caption{\label{table-deltavpp}Mean of effective anharmonicity and accessible phase space for zinc-blende and wurtzite phases at 77 and 300K, together with $\kappa$ ratios. Boldface indicates the characteristic more favorable to a higher $\kappa$. Mean of effective anharmonicities and accessible phase spaces are given in eV$^2$/(amu$^3\cdot\mathrm{\AA}^6$) and ps, respectively.}
	\begin{threeparttable}
		\begin{tabular}{@{}lllllll@{}}
			\toprule
			\hline \hline \\
			\makecell[tc]{material} 
			&\makecell[tc]{$\overline{|V_{\lambda\lambda'\lambda''}|^{ 2 }_{occ,77K}}$} 
			&\makecell[tc]{$\delta_{occ,77K}$}
			&\makecell[tc]{$\kappa_{hex}^{77K}/\kappa_{cub}^{77K}$}
			&\makecell[tc]{$\overline{|V_{\lambda\lambda'\lambda''}|^{ 2 }_{occ,300K}}$} 
			&\makecell[tc]{$\delta_{occ,300K}$ } 
			&\makecell[tc]{$\kappa_{hex}^{300K}/\kappa_{cub}^{300K}$}\\ \\
			\midrule
			\hline \\
			\makecell[tc]  {\textbf{BN} \\ ZB \\ WZ \\ WZ/ZB} 
			& \makecell[tc]{~\\$3.387\e{-10}$\\$\mathbf{7.904\e{-11}}$\\0.233}
			& \makecell[tc]{~\\$\mathbf{2.324\e{-6}}$\\$1.231\e{-5}$\\5.297}
			& \makecell[tc]{1.284\\~\\~\\~}
			& \makecell[tc]{~\\$1.350\e{-4}$\\$\mathbf{3.422\e{-5}}$\\0.253}
			& \makecell[tc]{~\\$\mathbf{8.757\e{-3}}$\\$7.284\e{-2}$\\8.318} 
			& \makecell[tc]{0.829\\~\\~\\~}
			\\ \\
			\makecell[tc]  {\textbf{AlAs} \\ ZB \\ WZ\\ WZ/ZB} 
			& \makecell[tc]{~\\$1.538\e{-7}$\\$\mathbf{4.457\e{-8}}$\\0.290}
			& \makecell[tc]{~\\$\mathbf{1.100\e{-1}}$\\$9.662\e{-1}$\\8.784}
			& \makecell[tc]{0.617\\~\\~}
			& \makecell[tc]{~\\$4.378\e{-5}$\\$\mathbf{1.051\e{-5}}$\\0.240}
			& \makecell[tc]{~\\$\mathbf{1.186\e{1}}$\\$1.018\e{2}$\\8.583}
			& \makecell[tc]{0.658\\~\\~}
			\\ \\
			\makecell[tc]  {\textbf{GaN} \\ ZB \\ WZ \\ WZ/ZB} 
			& \makecell[tc]{~\\$7.227\e{-8}$\\$\mathbf{1.522\e{-8}}$\\0.210}
			& \makecell[tc]{~\\$\mathbf{5.877\e{-3}}$\\$4.088\e{-2}$\\6.956}
			& \makecell[tc]{1.359\\~\\~\\~}
			& \makecell[tc]{~\\$1.157\e{-4}$\\$\mathbf{2.424\e{-5}}$\\0.210}
			& \makecell[tc]{~\\$\mathbf{1.217\e{0}}$\\$9.302\e{0}$\\7.643}
			& \makecell[tc]{1.047\\~\\~\\~} 
			\\ \\
			\makecell[tc]  {\textbf{GaP} \\ ZB \\ WZ\\ WZ/ZB}
			& \makecell[tc]{~\\$1.631\e{-7}$\\$\mathbf{3.797\e{-8}}$\\0.233}
			& \makecell[tc]{~\\$\mathbf{6.190\e{-2}}$\\$5.566\e{-1}$\\8.992}
			& \makecell[tc]{0.882\\~\\~}
			& \makecell[tc]{~\\$7.073\e{-5}$\\$\mathbf{1.471\e{-5}}$\\0.208}
			& \makecell[tc]{~\\$\mathbf{8.147\e{0}}$\\$7.103\e{1}$\\8.719}
			& \makecell[tc]{0.921\\~\\~} 
			\\ \\
			\makecell[tc]  {\textbf{GaAs} \\ ZB \\ WZ\\ WZ/ZB} 
			& \makecell[tc]{~\\$1.914\e{-7}$\\$\mathbf{4.817\e{-8}}$\\0.252}
			& \makecell[tc]{~\\$\mathbf{2.587\e{-1}}$\\$2.230\e{0}$\\8.620}
			& \makecell[tc]{0.916\\~\\~\\~}
			& \makecell[tc]{~\\$3.777\e{-5}$\\$\mathbf{9.459\e{-6}}$\\0.250}
			& \makecell[tc]{~\\$\mathbf{3.019\e{1}}$\\$2.559\e{2}$\\8.476}
			& \makecell[tc]{0.837\\~\\~} 
			\\ \\				
			\makecell[tc]  {\textbf{InP} \\ ZB \\ WZ\\ WZ/ZB}
			& \makecell[tc]{~\\$1.029\e{-7}$\\$\mathbf{2.501\e{-8}}$\\0.243}
			& \makecell[tc]{~\\$\mathbf{3.581\e{-1}}$\\$3.030\e{0}$\\8.461}
			& \makecell[tc]{0.741\\~\\~\\~}
			& \makecell[tc]{~\\$2.115\e{-5}$\\$\mathbf{4.692\e{-6}}$\\0.222}
			& \makecell[tc]{~\\$\mathbf{3.114\e{1}}$\\$2.610\e{2}$\\8.382}
			& \makecell[tc]{0.827\\~\\~\\~} 
			\\ \\
			\makecell[tc]  {\textbf{InAs} \\ ZB \\ WZ\\ WZ/ZB} 
			& \makecell[tc]{~\\$2.226\e{-7}$\\$\mathbf{5.220\e{-8}}$\\0.235}
			& \makecell[tc]{~\\$\mathbf{8.224\e{-1}}$\\$6.651\e{0}$\\8.087}
			& \makecell[tc]{0.832\\~\\~\\~}
			& \makecell[tc]{~\\$2.966\e{-5}$\\$\mathbf{6.776\e{-6}}$\\0.228}
			& \makecell[tc]{~\\$\mathbf{6.909\e{1}}$\\$5.548\e{2}$\\8.030}
			& \makecell[tc]{0.909\\~\\~\\~}
			\\ \\	
			\makecell[tc]  {\textbf{ZnSe} \\ ZB \\ WZ\\ WZ/ZB}
			& \makecell[tc]{~\\$3.025\e{-7}$\\$\mathbf{7.207\e{-8}}$\\0.238}
			& \makecell[tc]{~\\$\mathbf{4.817\e{-1}}$\\$3.548\e{0}$\\7.366}
			& \makecell[tc]{0.971\\~\\~\\~}
			& \makecell[tc]{~\\$4.648\e{-5}$\\$\mathbf{1.105\e{-5}}$\\0.238}
			& \makecell[tc]{~\\$\mathbf{4.902\e{1}}$\\$3.689\e{2}$\\7.25}
			& \makecell[tc]{0.857\\~\\~\\~} 
			\\ \\	
			\\ \bottomrule
			\hline \hline	
		\end{tabular}
	\end{threeparttable}
\end{table*}

Therefore, we have two antagonistic processes occurring when reducing the symmetry from cubic to hexagonal: an increment in the available phase space for phonon-phonon scattering events and a lowering of the anharmonicity, which makes the strength of those events weaker when compared to the cubic ones, with no indication of their relative importance as regards to $\kappa$.

\subsubsection{\label{sssec:discussion} Discussion}

As we already mentioned, despite being widely used to predict the relationship between thermal conductivity for different materials~\cite{MukhopadhyayPRB2016}, the criteria listed in Sec.~\ref{sssec:saok} lack the capacity to discern the relative importance of opposed processes.

\begin{figure*}
	\centering
	\includegraphics[width=1.00\textwidth]{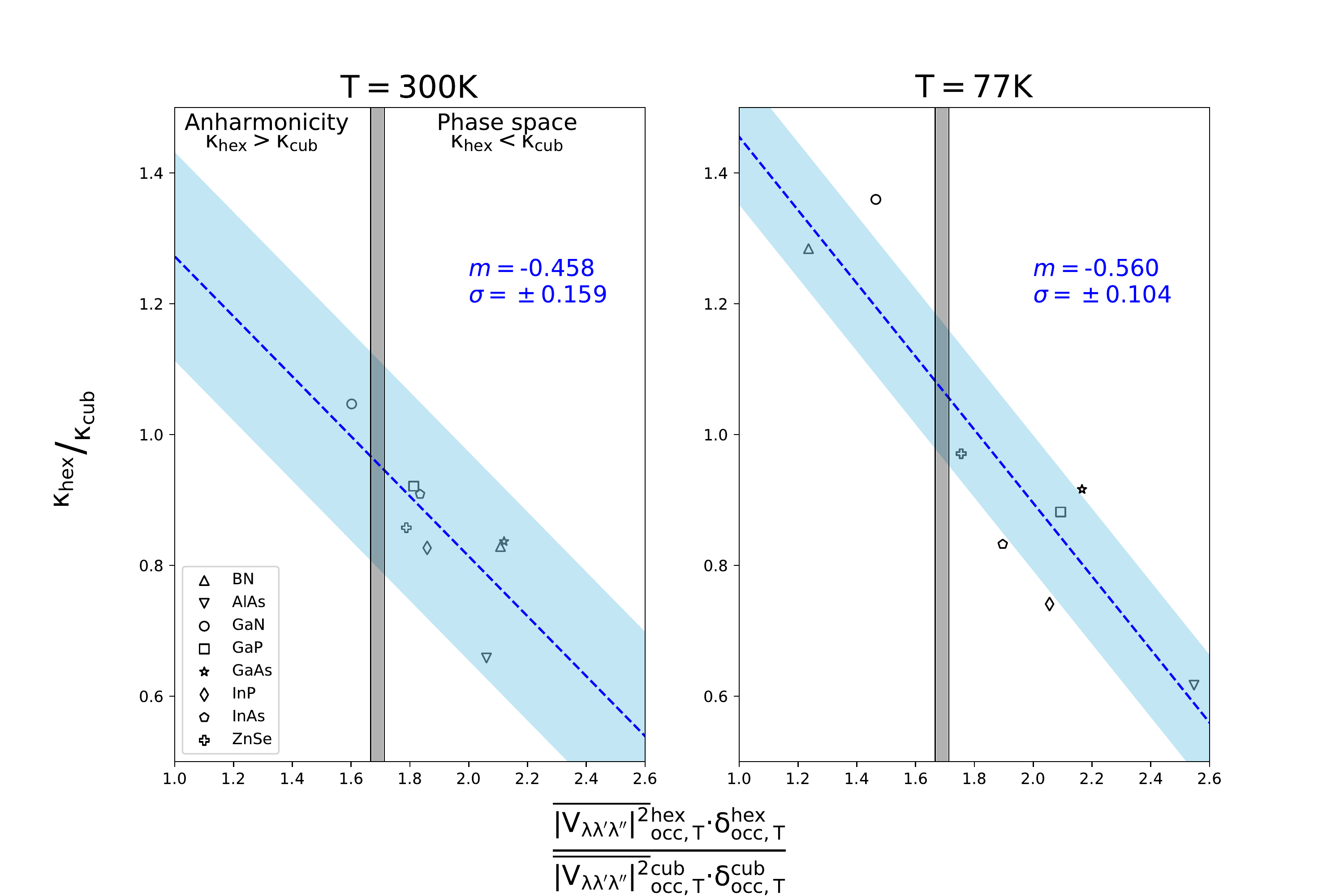}
	\caption{$\kappa_{hex}/\kappa_{cub}$ ratio (with isotopic scattering) as a function of the hexagonal-cubic ratio of the mean effective anharmonicity and the accessible phase space product at 77K (left) and 300K (right) for different materials. The factor that controls the ratio $\mathrm{\kappa_{hex}/\kappa_{cub}}$ (anharmonicity or phase space) for a given REAAPS is indicated together with which phase is the most conducting for that value. The grey filled region corresponds to the limiting region. The dashed blue line corresponds to a linear regression of the data, with the corresponding slope $m$ and standard deviation, $\sigma$ provided in the figure. The cyan area corresponds to an interval $\pm \sigma$.
		\label{ratios_anhdel}
	}
\end{figure*}

\begin{figure*}
	\centering
	\includegraphics[width=1.00\textwidth]{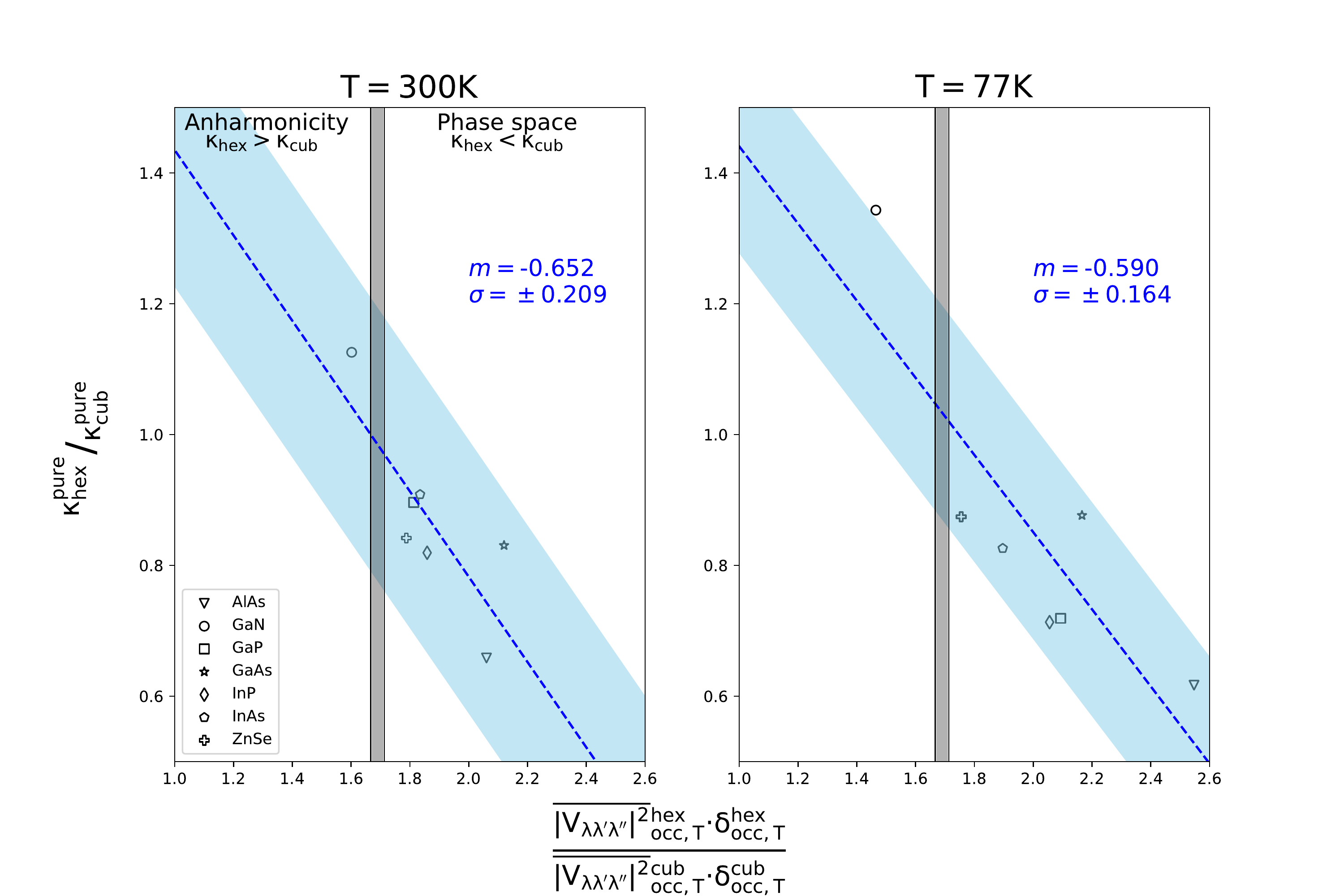}
	\caption{$\kappa_{hex}^{pure}/\kappa_{cub}^{pure}$ ratio (without isotopic scattering) as a function of the hexagonal-cubic ratio of the mean effective anharmonicity and the accessible phase space product at 77K (left) and 300K (right) for different materials. The factor that controls the ratio $\mathrm{\kappa_{hex}^{pure}/\kappa_{cub}^{pure}}$ (anharmonicity or phase space) for a given REAAPS is indicated together with which phase is the most conducting for that value. The grey filled region corresponds to the limiting region. The dashed blue line corresponds to a linear regression of the data, with the corresponding slope $m$ and standard deviation, $\sigma$ provided in the figure. The cyan area corresponds to an interval $\pm \sigma$.
		\label{ratios_anhdel_pure}
	}
\end{figure*}

To overcome such a limitation and owing to the fact that scattering rates are a product of the anharmonicity with phase space [see Eq.~(\ref{3ph_eq})], we represent in Fig.~\ref{ratios_anhdel} $\kappa_{hex}/\kappa_{cub}$  versus the hexagonal-cubic ratio of the mean effective anharmonicity, Eq.~(\ref{vpp_eq}), and the accessible phase space product, Eq.~(\ref{delta_eq}) at 77K and 300K. We call this ratio, which is a central magnitude in our discussion, REAAPS, standing for Ratio of Effective Anharmonicity and Accessible Phase Space product. Isotopic scattering, despite being the clear mechanism determining the ratio at low temperature for some materials like GaP, is not accounted in such analysis. To avoid the influence of the isotope effect, the same procedure, but with the ratios of isotopically pure materials, can be repeated (see Fig.~\ref{ratios_anhdel_pure}).

By this analysis of the relative importance of the accessible phase space and effective anharmonicity at a given temperature, it can be observed from Fig.~\ref{ratios_anhdel_pure} that the $\kappa_{hex}^{pure}/\kappa_{cub}^{pure}$ ratio shows a strong correlation with the REAAPS, and that $\kappa_{hex}^{pure} > \kappa_{cub}^{pure}$ when REAAPS is lower than $\sim1.66$, indicating that in these cases the effective anharmonicity increment in the zinc-blende phase compensates the wurtzite accessible phase space increment, thus giving a higher conductivity in wurtzite phase. For higher values of REAAPS, the dominant factor is the increment in the accessible phase space of the wurtzite phase, hence the higher conductivity of the zinc-blende phase. Interestingly, the plots with (Fig.~\ref{ratios_anhdel}) and without (Fig.~\ref{ratios_anhdel_pure}) isotopic scattering are quite similar at both temperatures, indicating that three-phonon processes are the ones dominating the ratio behavior at the studied temperatures. We note that both the threshold of $\sim1.66$ and the slopes of the linear regressions are independent, to a large extent, of temperature (see Fig.~\ref{reaaps700}~\cite{SI} for a plot at a high temperature, 700~K).

Our analysis helps to explain some of the ratio behaviors such as in GaN, which has $\kappa_{hex} > \kappa_{cub}$ over all the temperature range considered. As observed in Table~\ref{table-deltavpp}, the anharmonicity of the hexagonal phase is particularly small when compared to that of the cubic phase. Additionally, the ratio of the accessible phase space is somewhat lower than in other materials. Both factors favor a higher $\kappa$ for the WZ phase, as it is indeed the case. This is in agreement with previous works by Lindsay~\textit{et al.}~\cite{LindsayPRB2013} but in opposition to the results by Togo~\textit{et al.}~\cite{phono3py}, which, as mentioned before, are obtained by a different approach to the LBTE that yields different values for the thermal conductivity in other materials as well~\cite{Torres2DMAT2019}.

The case of BN is also interesting. Both Table~\ref{table-deltavpp} and Fig.~\ref{ratios_anhdel} show that there is nothing particular to BN at 300K, having $\kappa_{hex}^{300K} < \kappa_{cub}^{300K}$. However, when the temperature is lowered to 77K, the accessible phase space decreases much more than in the other materials (a consequence of the hard phonon modes) and the ratio of accessible phase space takes a low value of 5.297, significantly different from the rest of the materials. From this it can be concluded that the change from $\kappa_{hex}^{300K} < \kappa_{cub}^{300K}$ to $\kappa_{hex}^{77K} > \kappa_{cub}^{77K}$ is due to the abnormally large increase of $\kappa_{hex}$ as temperature is decreased because of the quicker decrease of the accessible phase space in the WZ phase.

Moreover, one can also see from Table~\ref{table-deltavpp} that for the vast majority of materials the behavior of the ratio with temperature is the opposite to the behavior of REAAPS with temperature. Therefore, when REAAPS decreases (increases) with the temperature, the ratio increases (decreases) for almost all materials, explaining the temperature dependence of $\kappa_{hex}^{pure} / \kappa_{cub}^{pure}$. However, this does not occur in the case of GaAs, where contrarily to the rest of the materials, the evolution of the REAAPS seems to indicate that the ratio should increase with temperature. This disagreement may show a limitation of our analysis, where we have not taken into account the correlations between anharmonicity and accessible phase space, which can play an important role in borderline cases. When isotopic scattering is added into consideration, it can significantly alter the monotonic dependence of the ratio {\it vs.} the temperature at very low temperatures, see the case of GaP for an example.

\subsubsection{\label{sssec:4phonon} Four-phonon scattering for GaN}

Feng {\it et al.}~\cite{FengPRB2017} have conducted a rigorous study of four-phonon scattering on three representative materials, showing that a significant reduction of $\kappa$ is caused at high temperatures. This effect was shown to be particularly strong in BAs, a material with a large a-o gap, where the inclusion of four-phonon processes opened up scattering channels that were forbidden in a three-phonon event. GaN also presents a large a-o gap, and thus it is interesting to estimate how four-phonon scattering can affect the predicted values for $\kappa$.

Although a full study is out of scope of the present work, Ref.~\citenum{FengPRB2017} provides some guidelines on how to estimate the effect of four-phonon scattering on the thermal conductivity. We have followed the procedure there reported for the estimation, computing the anharmonicity ratio $\left| \Phi_4 / \Phi_3 \right|^2 / \left| \Phi_2 \right|$, where $\Phi_n$ is the $n$-th order on-site force constant for Ga along the stacking direction ([111] or [0001]), e.g. $\Phi_3 = \Phi^{\rm WZ}_{0,{\rm Ga},z;0,{\rm Ga},z;0,{\rm Ga},z} $, and comparing the cubic and hexagonal phases. These directions are chosen because (a) they correspond to the directions of the cation-anion bond, (b) they are the directions maximum structural difference between ZB and WZ, and (c) the IFCs do not take zero values. The results, shown in Table~\ref{table-anharmRatio}, indicate that the inclusion of four-order processes maintains that the anharmonicity of GaN-ZB is stronger than that of GaN-WZ. Also, those anharmonicity ratios, although the IFCs were computed along different directions than in Ref.~\citenum{FengPRB2017} and we do not know up to what point they can be directly compared, have a higher numerical value than those provided for diamond, BAs and Si~\cite{FengPRB2017}. Furthermore, given that GaN-ZB has a larger a-o gap than its WZ counterpart, we expect that at high temperatures ZB will be more affected by the inclusion of four-phonon processes than the hexagonal phase. From all these considerations, we can predict that $\kappa_{hex} / \kappa_{cub}$ will increase at high temperature once four-phonon processes are included in the analysis.

\begin{table}[t]
  \centering
	\caption{\label{table-anharmRatio} Anharmonicity ratio $\left| \Phi_4 / \Phi_3 \right|^2 / \left| \Phi_2 \right|$ to get an insight of the relative importance of four-phonon processes in GaN-ZB and GaN-WZ. IFCs have been calculated along the stacking direction ([111] or [0001]), and they are given in eV/\AA$^n$.}
		\begin{tabular}{@{}lllllll@{}}
			\toprule
			\hline \hline \\
			& \makecell{phase}
			& \makecell{$\Phi_2$} 
			& \makecell{$\Phi_3$}
			& \makecell{$\Phi_4$}
			& \makecell{$\left| \Phi_4 / \Phi_3 \right|^2 / \left| \Phi_2 \right|$}
			\\ \\ \midrule
			\hline \\
			& \makecell{GaN-ZB}
			& \makecell{20.1} 
			& \makecell{70.0}
			& \makecell{244}
			& \makecell{0.604}
			\\ \\
			& \makecell{GaN-WZ}
			& \makecell{19.7} 
			& \makecell{69.3}
			& \makecell{184}
			& \makecell{0.358}
			\\
			\\ \bottomrule
			\hline \hline	
  \end{tabular}
\end{table}

\subsection{\label{ssec:NWs} Nanowires}

As previously mentioned, phases that are not thermodynamically stable at room temperature and atmospheric pressure in bulk form can naturally occur when the materials are grown as nanowires, and thus it is in nanowires that the zinc-blende and wurtzite phase can both be easily accessed. Therefore, in this section we discuss the conductivity along the stacking direction, which is also the common growth direction ([111] for cubic and [0001] for hexagonal) for nanowires of different materials and diameters.

Despite being one dimensional structures, nanowire phonon dispersions can be approximated to the bulk ones for nanowires with diameters $\gtrsim 60-70$~nm~\cite{ShengBTE}. However, the reduction in the symmetry caused by the nanowire boundaries make it unrealistic to use bulk-like scattering rates, as they become position dependent in the direction perpendicular to the nanowire. To solve this problem, an iterative solution under the diffusive regime as proposed by Li \textit{et al.}~\cite{BTEnanowires} is used as implemented in the ShengBTE code~\cite{ShengBTE}. 

\begin{figure*}
	\begin{subfigure}{0.43\linewidth}
		\centering
		\includegraphics[width=\linewidth]{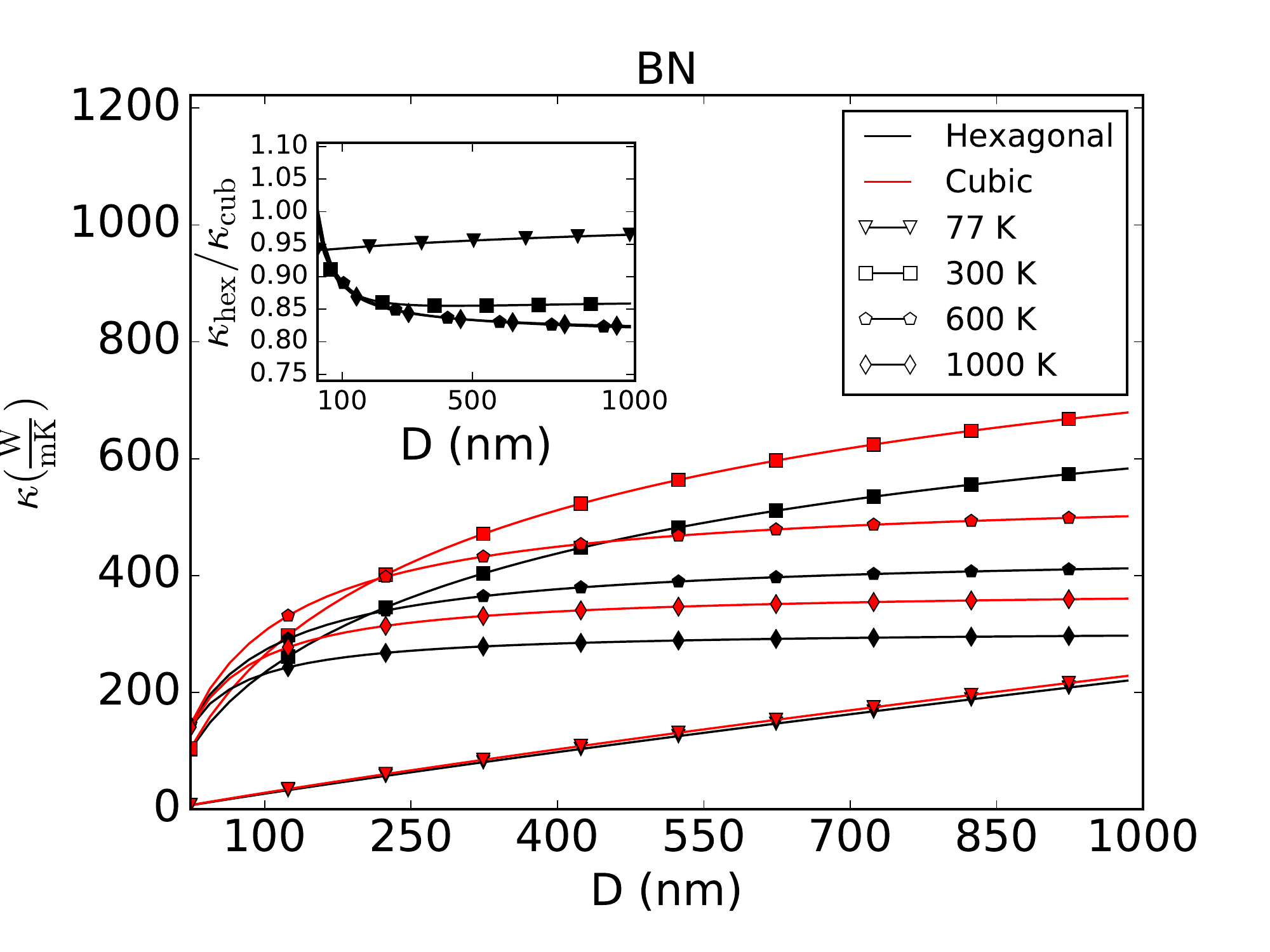}
	\end{subfigure}
	\begin{subfigure}{0.43\linewidth}
		\centering
		\includegraphics[width=\linewidth]{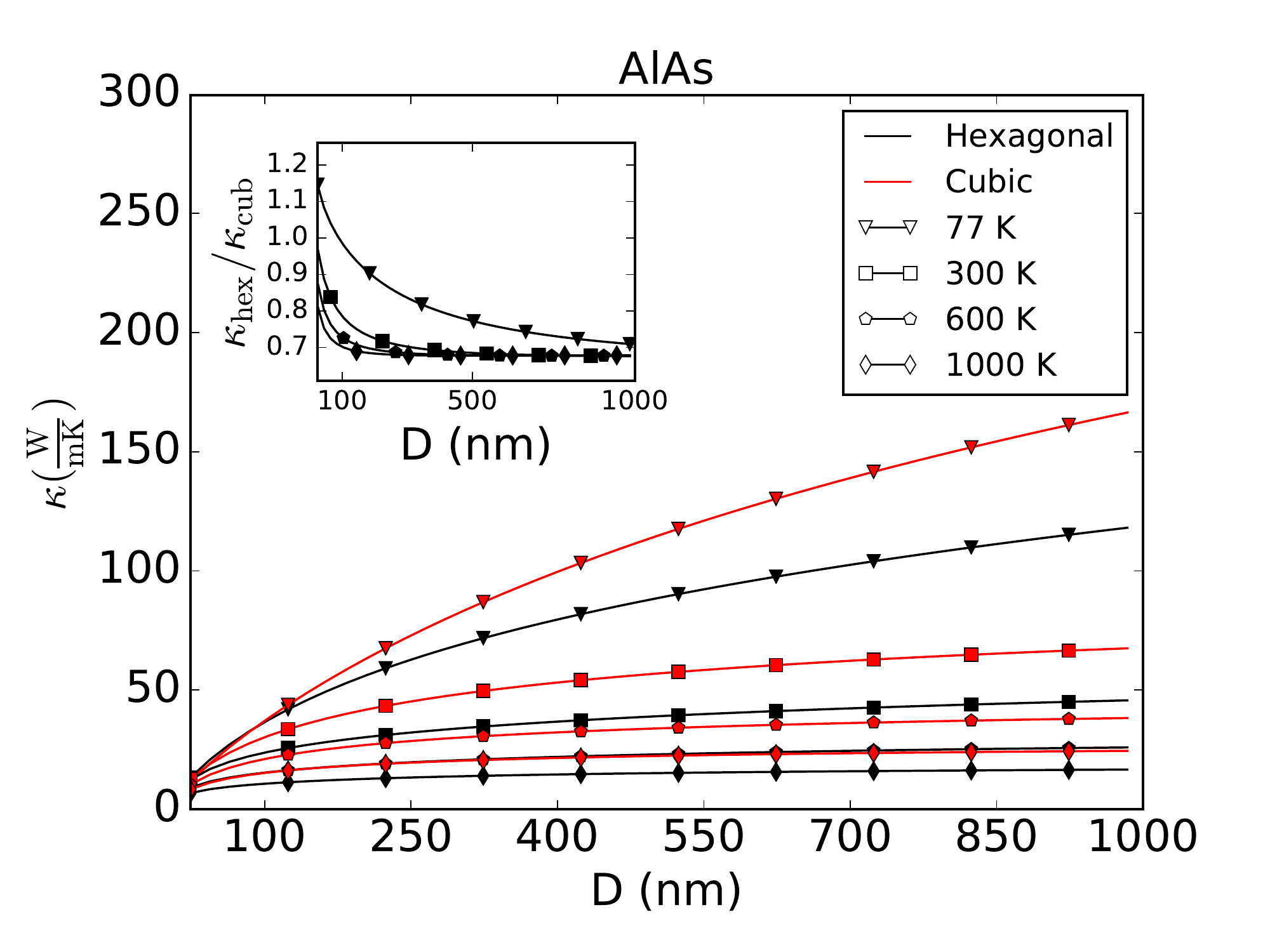}
	\end{subfigure}
	\begin{subfigure}[t]{0.43\textwidth}
		\centering
		\includegraphics[width=\linewidth]{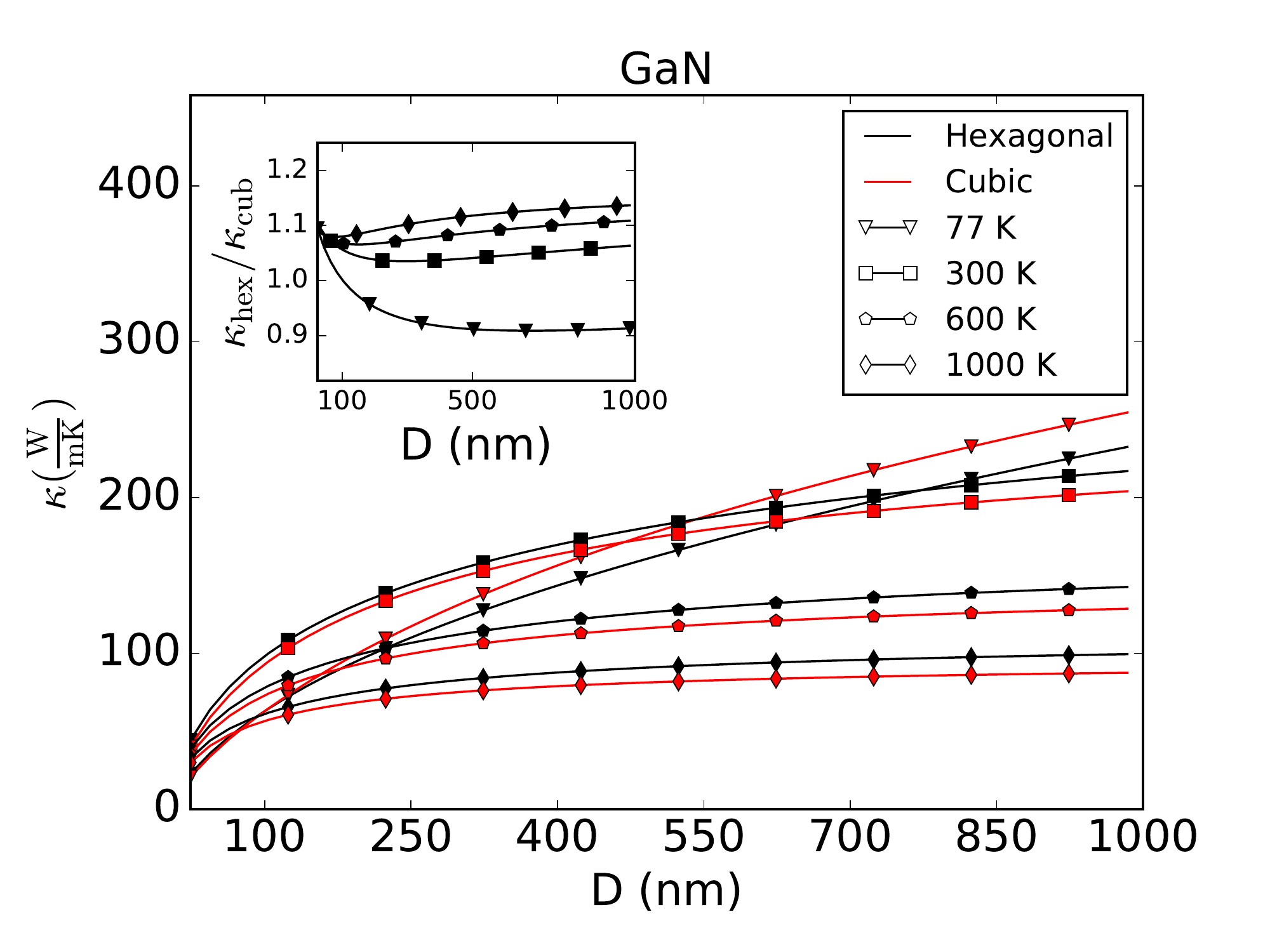}
	\end{subfigure}
	\begin{subfigure}[t]{0.43\textwidth}
		\centering
		\includegraphics[width=\linewidth]{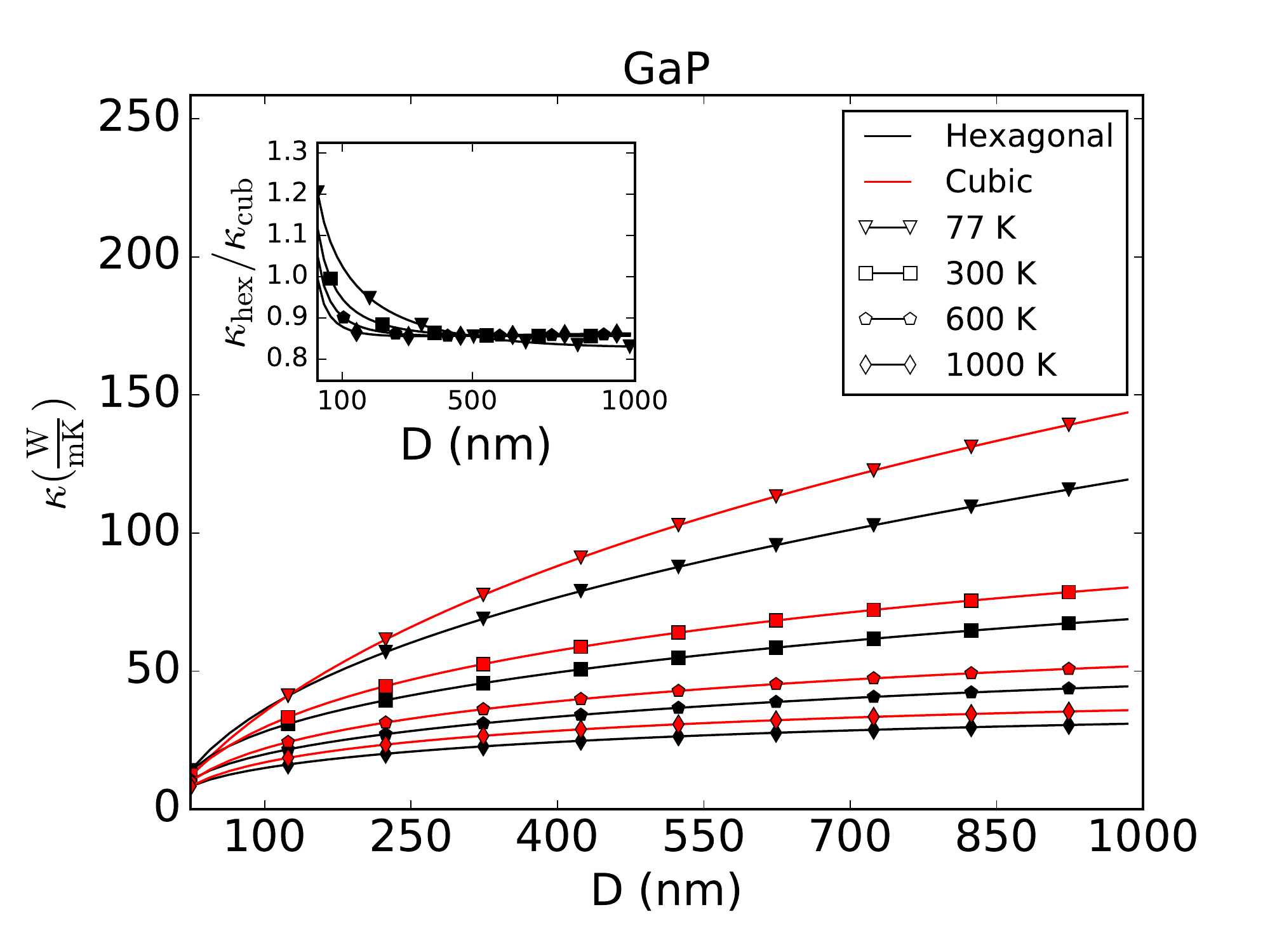}
	\end{subfigure}
	\begin{subfigure}[t]{0.43\textwidth}
		\centering
		\includegraphics[width=\linewidth]{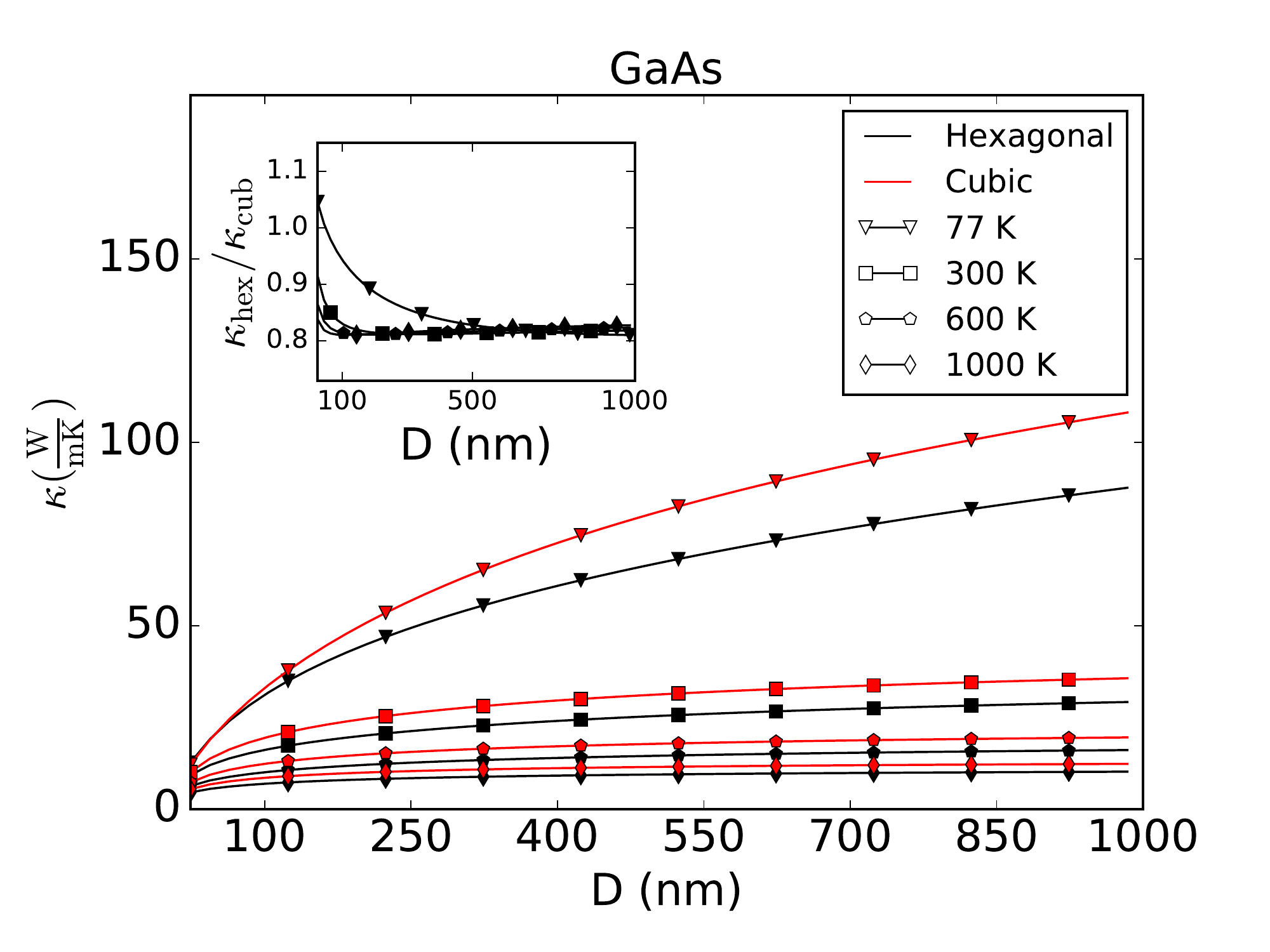}
	\end{subfigure}
	\begin{subfigure}[t]{0.43\textwidth}
		\centering
		\includegraphics[width=\linewidth]{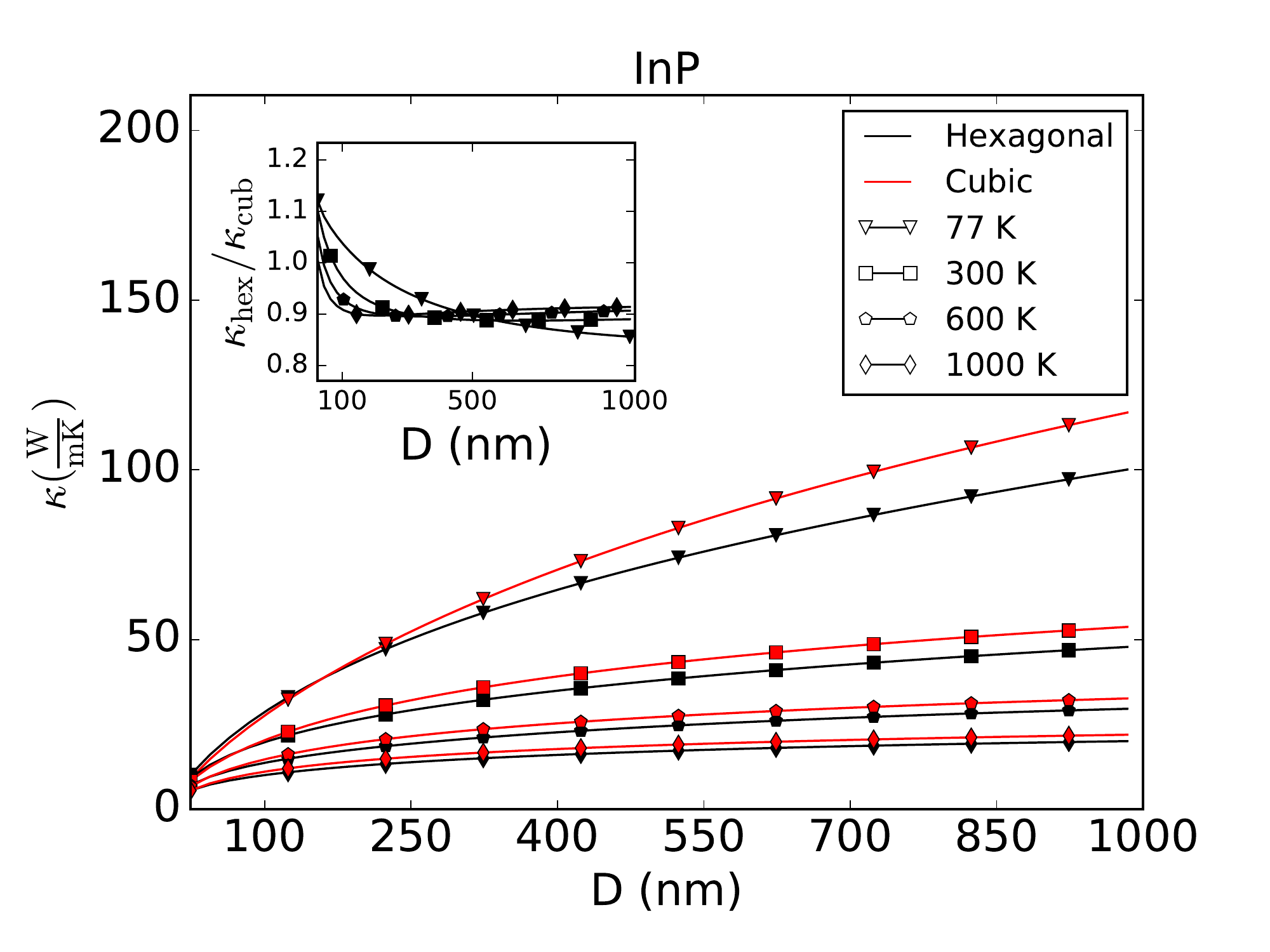}
	\end{subfigure}
	\begin{subfigure}[t]{0.43\textwidth}
		\centering
		\includegraphics[width=\linewidth]{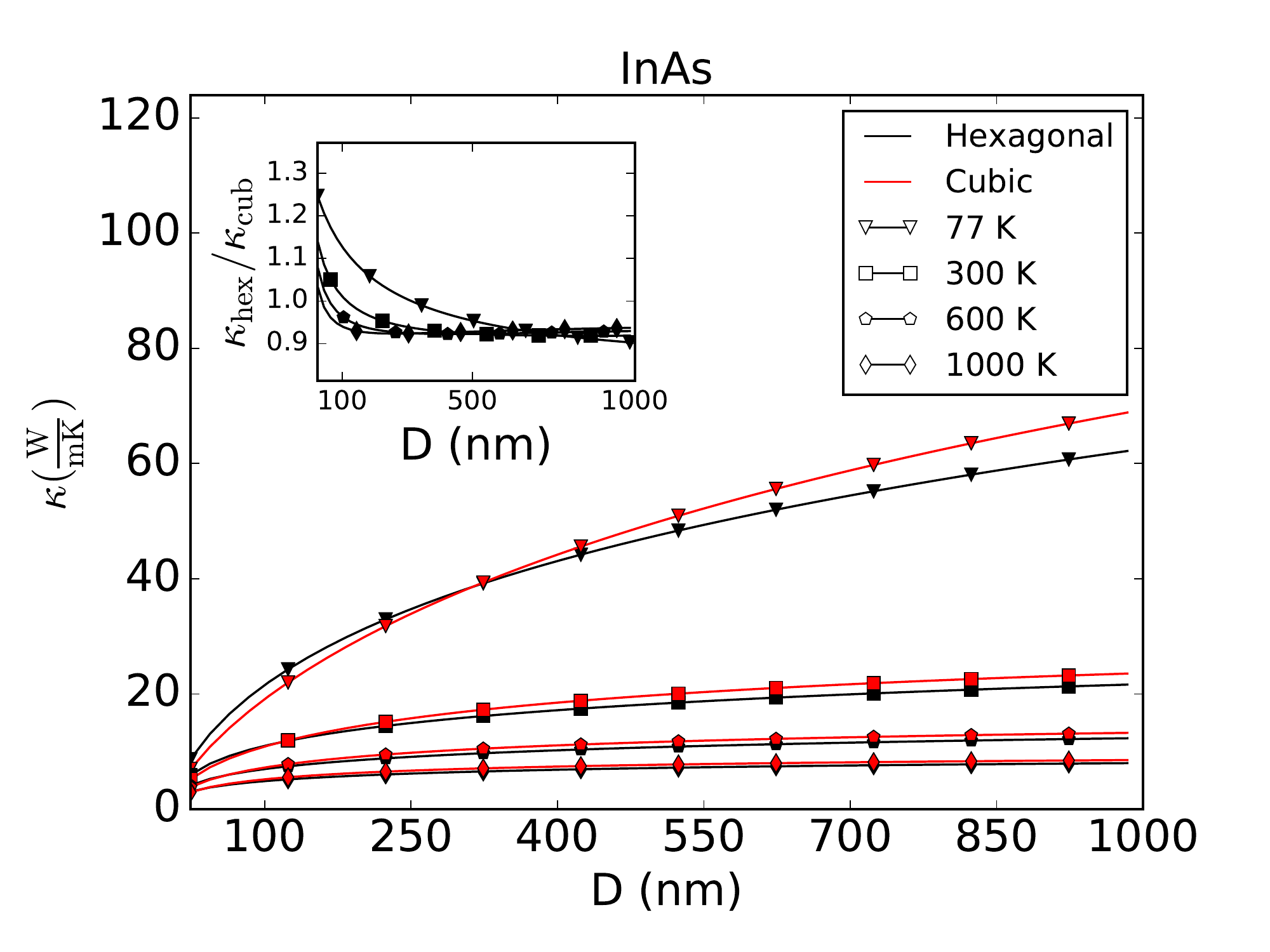}
	\end{subfigure}
	\begin{subfigure}[t]{0.43\textwidth}
		\centering
		\includegraphics[width=\linewidth]{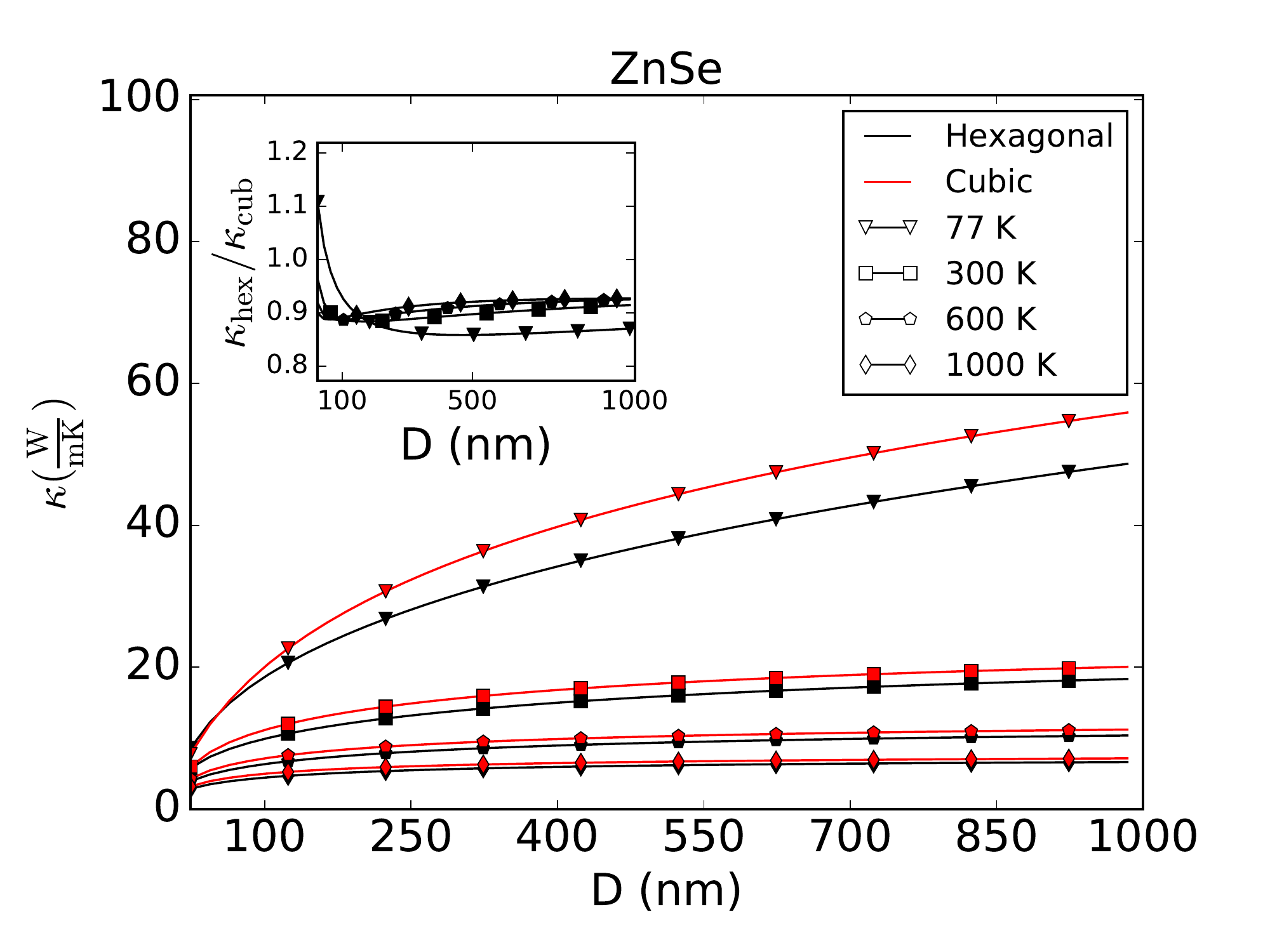}
	\end{subfigure}
	\caption{Thermal lattice conductivity of nanowires along [111] for cubic (red) and [0001] for hexagonal (black) at 77K (triangles), 300K (squares),  600K (pentagons) and 1000K (diamonds) as function of the nanowire diameter. Inset: hexagonal-cubic ratios for nanowires.}
	\label{nanos}
\end{figure*}

\begin{figure*}
	\begin{subfigure}{0.42\linewidth}
		\centering
		\includegraphics[width=\linewidth]{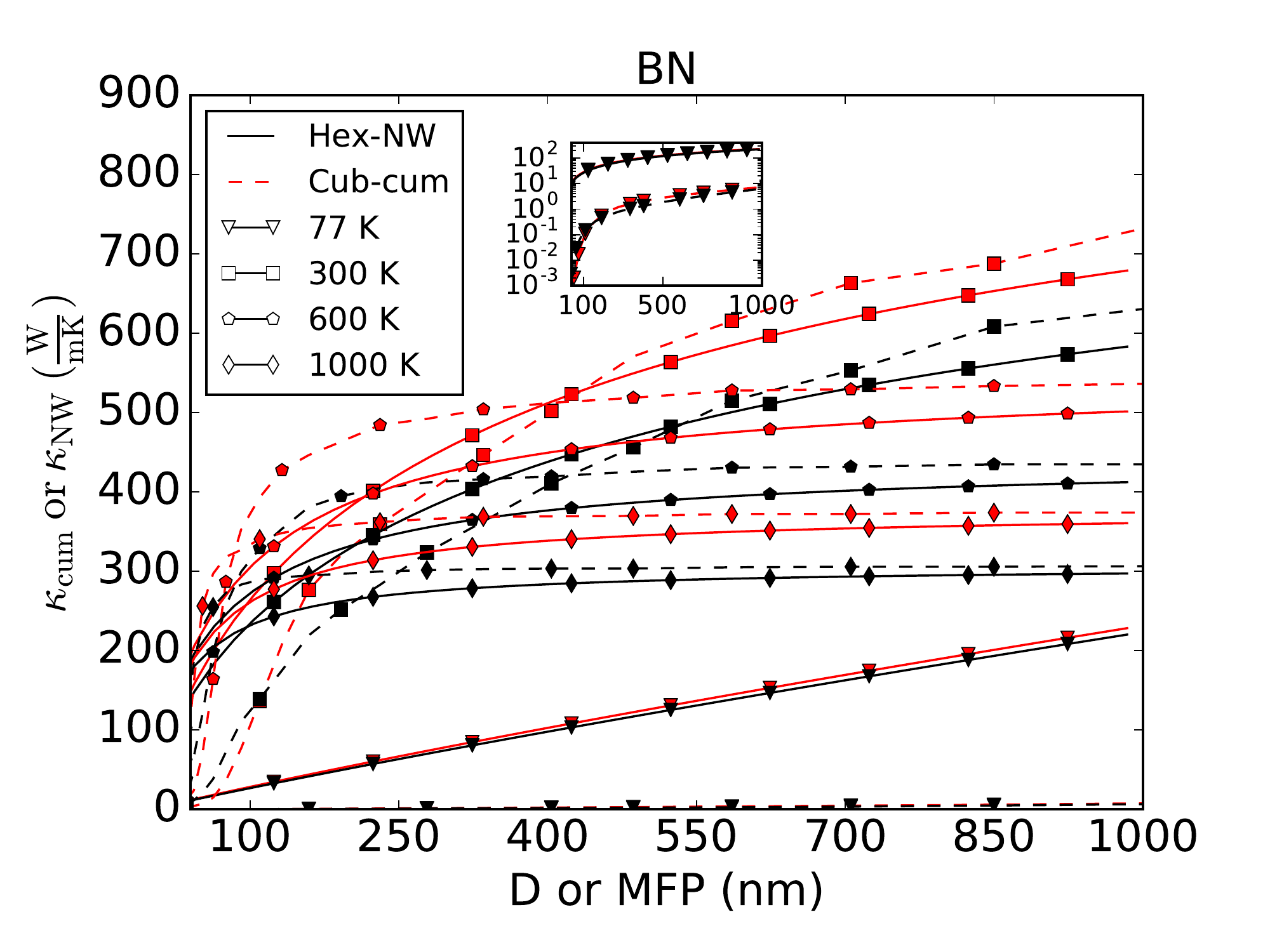}
	\end{subfigure}
	\begin{subfigure}{0.42\linewidth}
		\centering
		\includegraphics[width=\linewidth]{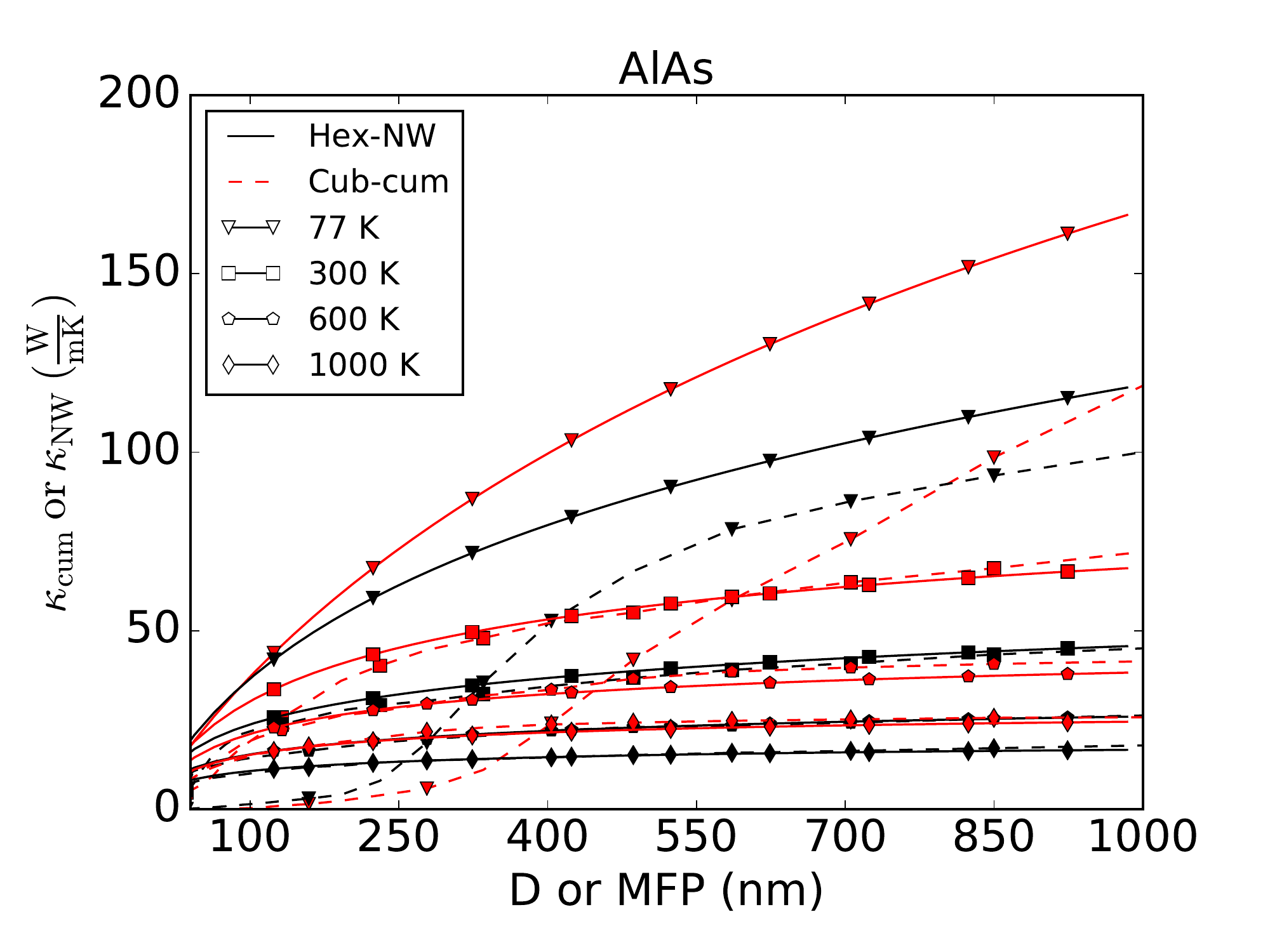}
	\end{subfigure}
	\begin{subfigure}[t]{0.42\textwidth}
		\centering
		\includegraphics[width=\linewidth]{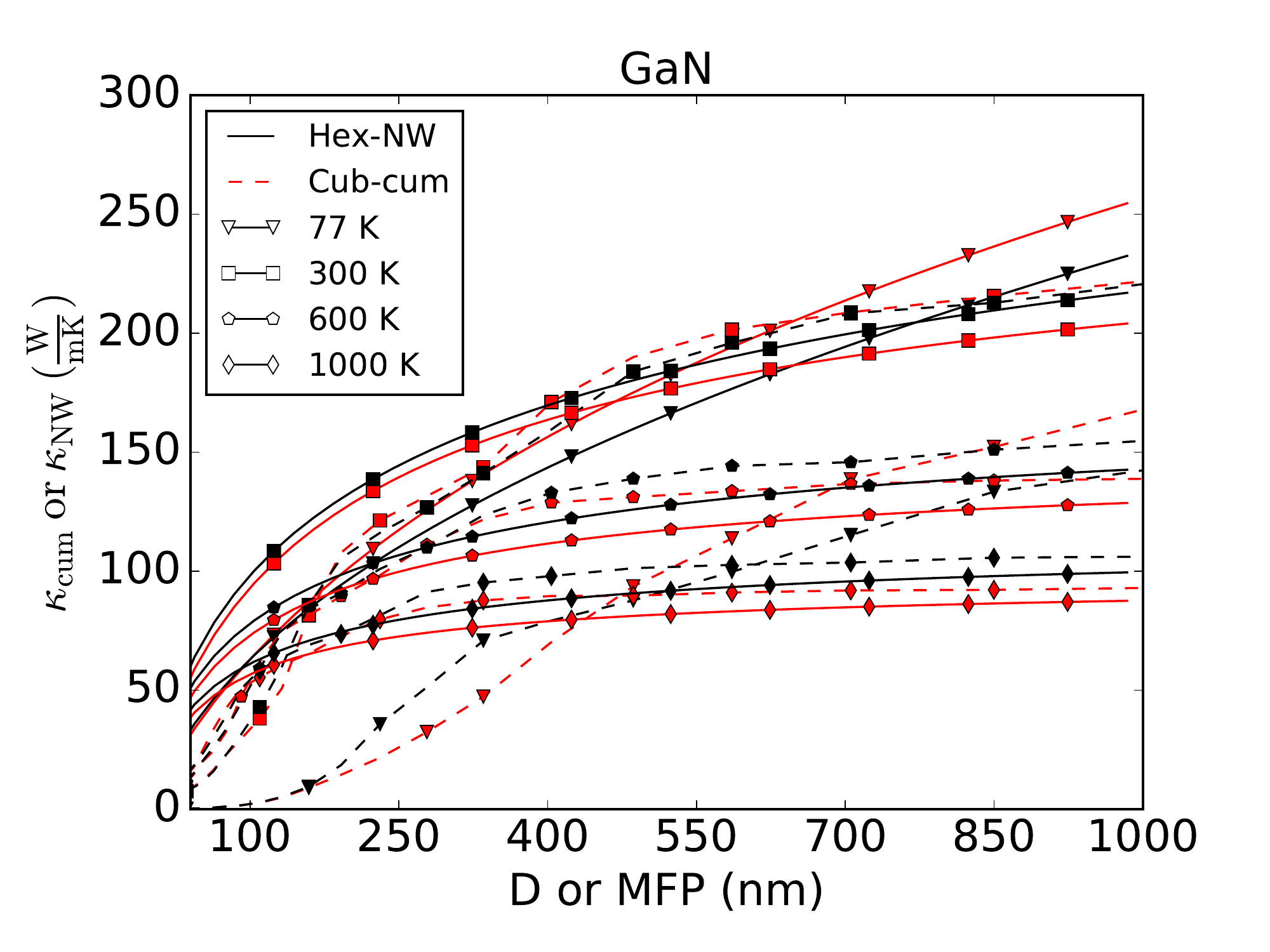}
	\end{subfigure}
	\begin{subfigure}[t]{0.42\textwidth}
		\centering
		\includegraphics[width=\linewidth]{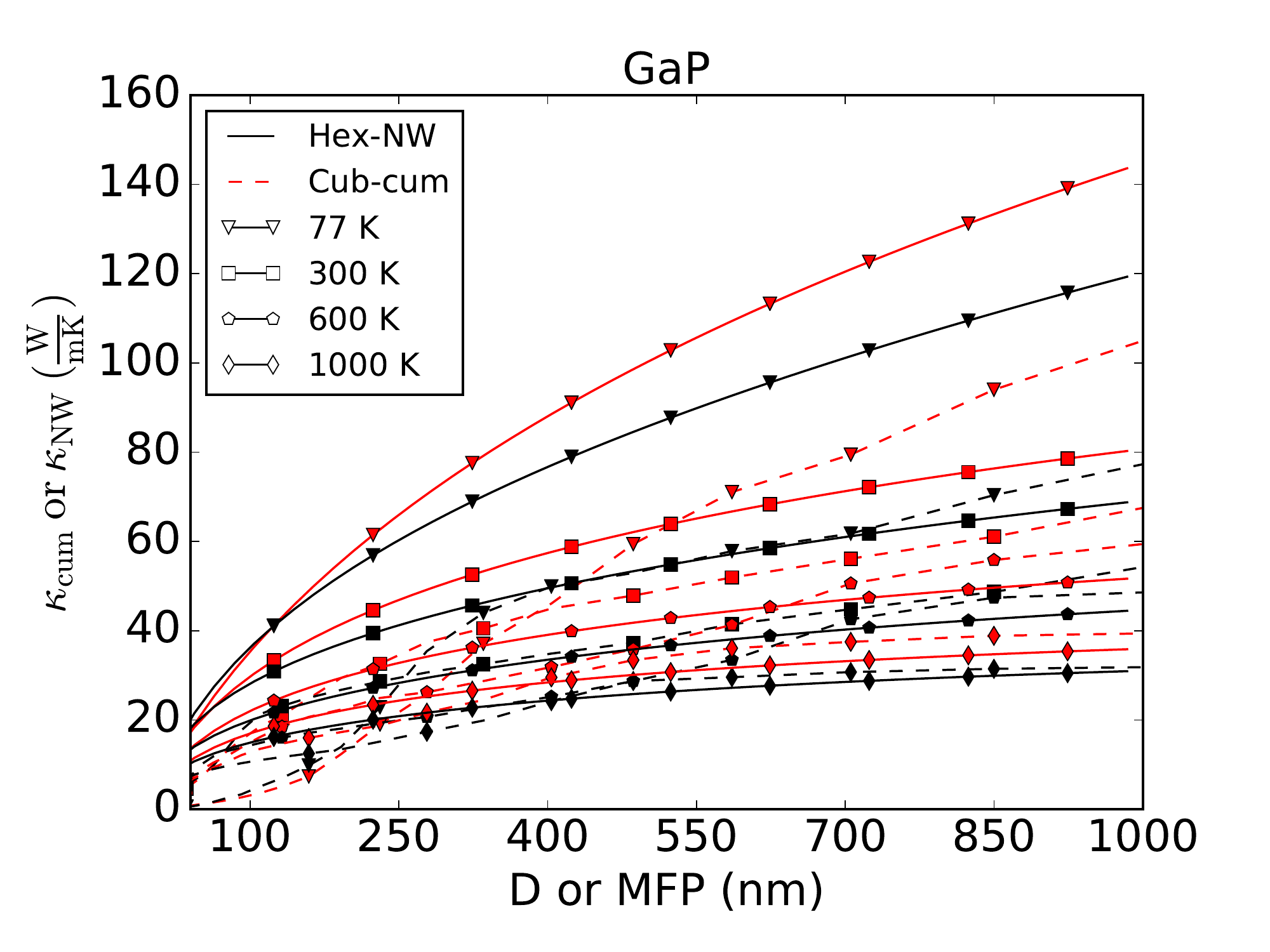}
	\end{subfigure}
	\begin{subfigure}[t]{0.42\textwidth}
		\centering
		\includegraphics[width=\linewidth]{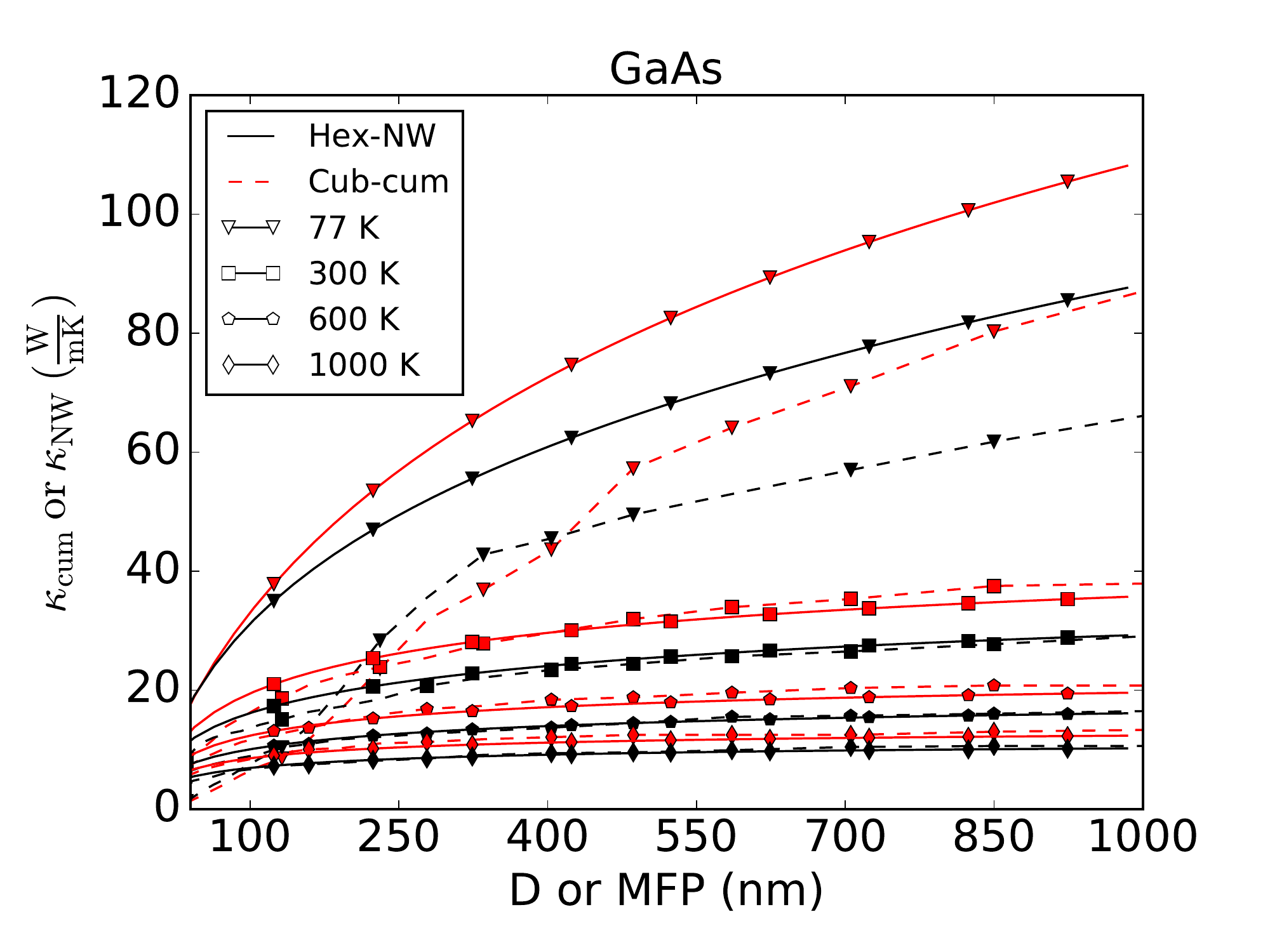}
	\end{subfigure}
	\begin{subfigure}[t]{0.42\textwidth}
		\centering
		\includegraphics[width=\linewidth]{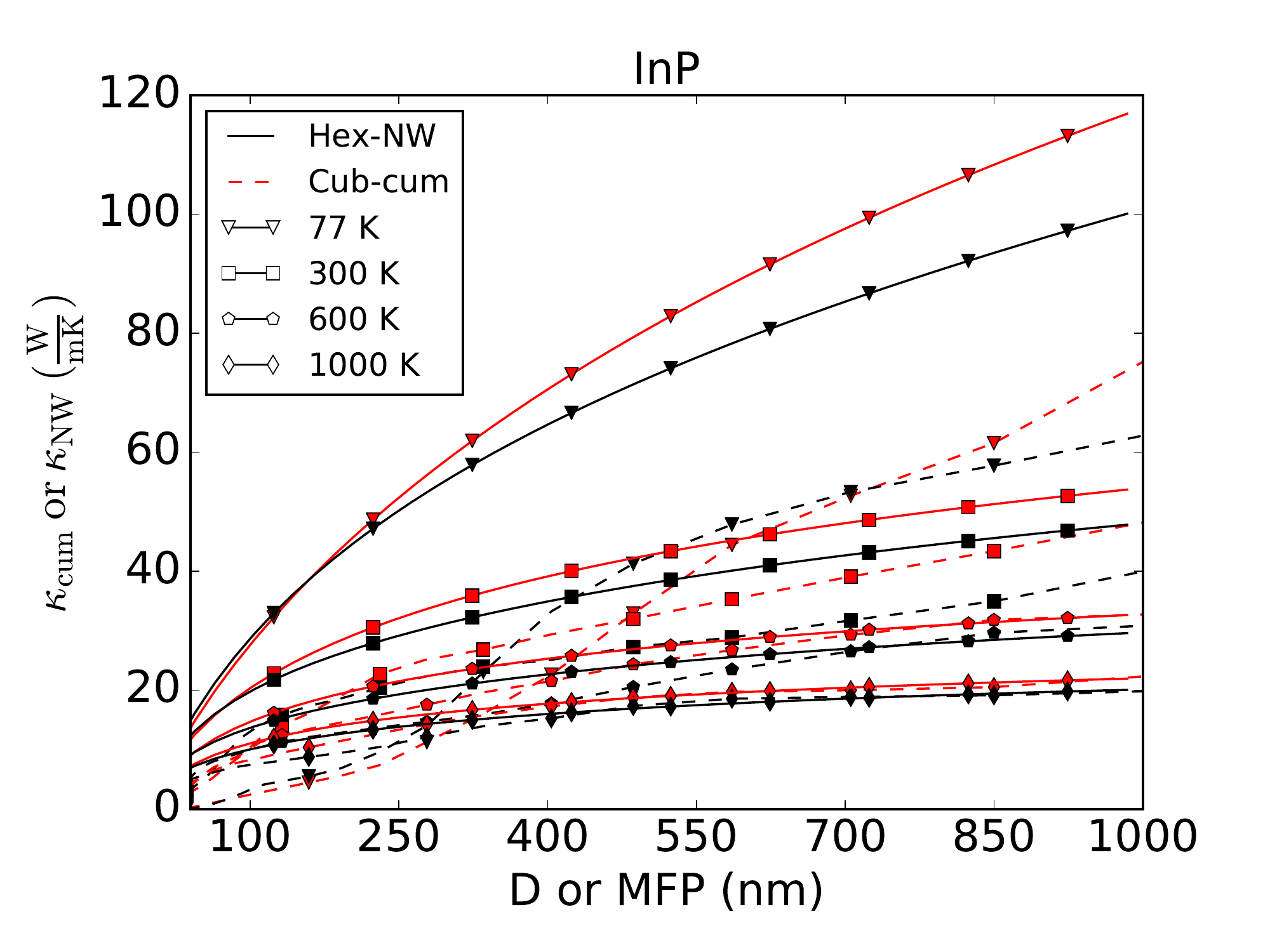}
	\end{subfigure}
	\begin{subfigure}[t]{0.42\textwidth}
		\centering
		\includegraphics[width=\linewidth]{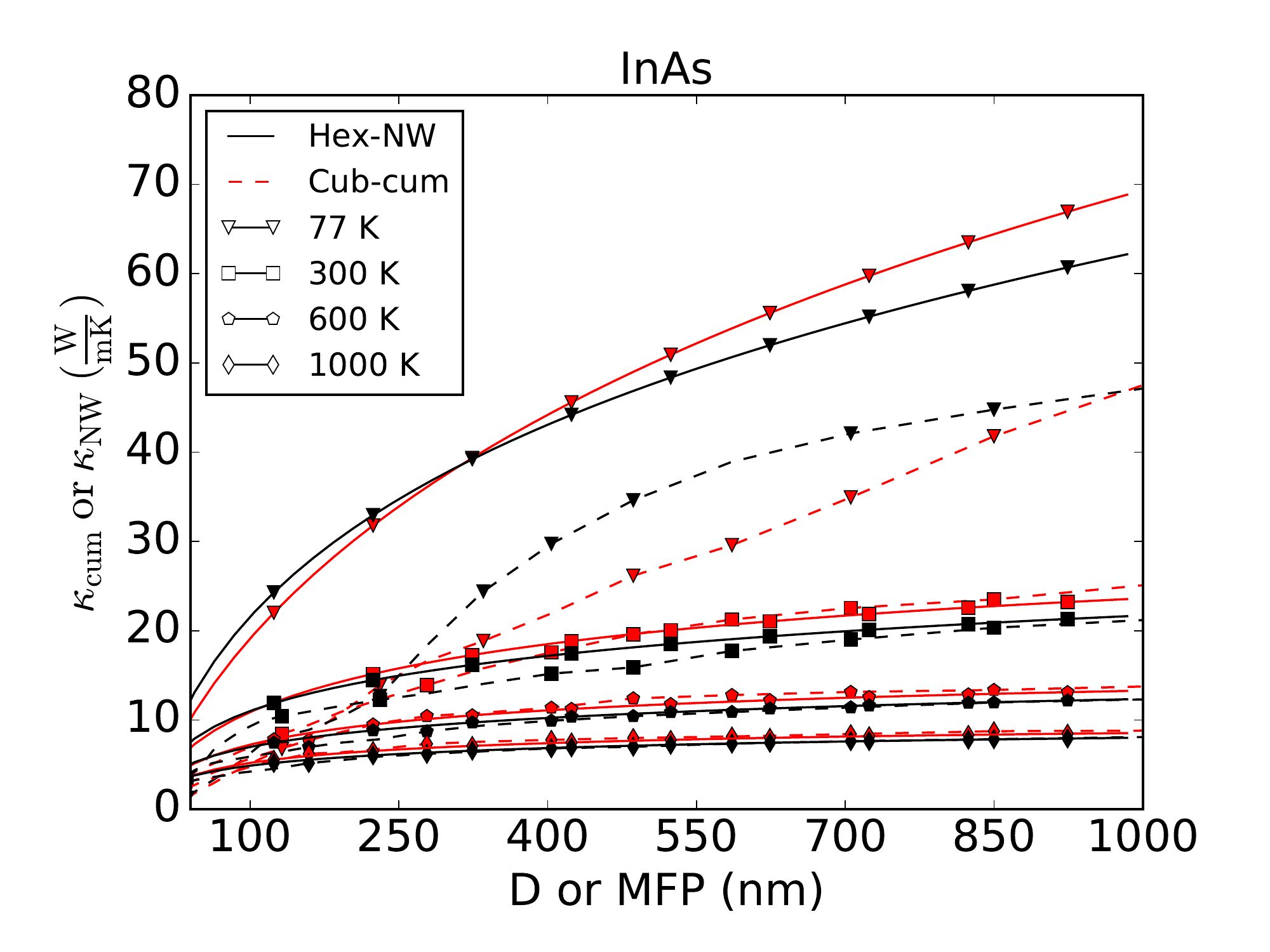}
	\end{subfigure}
	\begin{subfigure}[t]{0.42\textwidth}
		\centering
		\includegraphics[width=\linewidth]{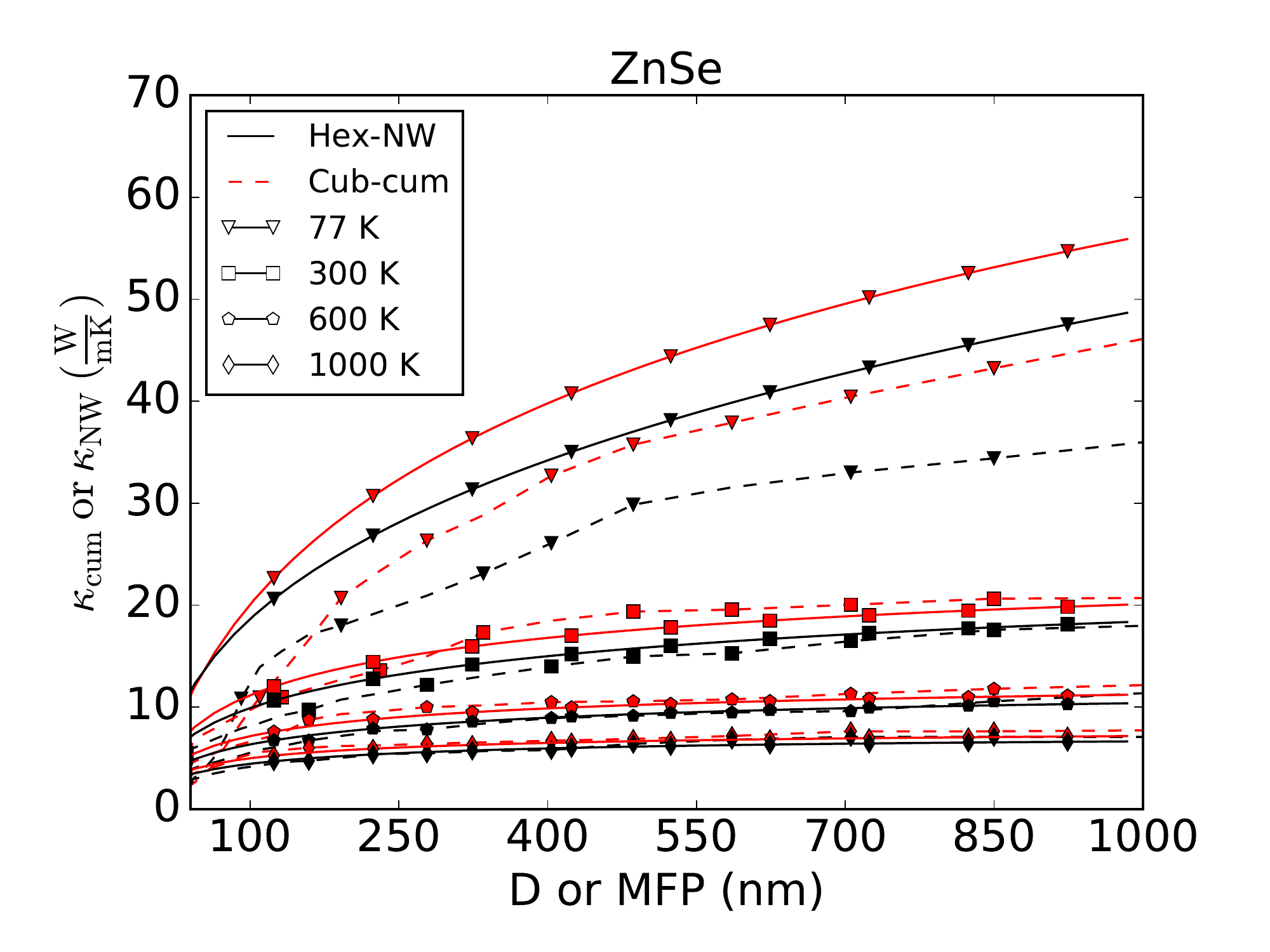}
	\end{subfigure}
	\caption{Dashed lines: Bulk-cumulative thermal lattice conductivity at 77K (triangles), 300K (squares),  600K (pentagons) and 1000K (diamonds) as function of the phonons mean free path for zinc-blende (red) and wurtzite (black). Solid lines: Bulk normalized nanowire thermal lattice conductivity at 77K (triangles), 300K (squares),  600K (pentagons) and 1000K (diamonds) as function of nanowire diameter for zinc-blende (red) along [111] axis and wurtzite (black) along [0001] axis. BN-Inset: Zoom for 77K.}
	\label{MFPS}
\end{figure*}

$\kappa$ along [111] (cubic) and [0001] (hexagonal) is plotted as function of the nanowire diameter for several temperatures in Fig.~\ref{nanos} together with $\kappa_{hex}/\kappa_{cub}$. In order to understand the ratio behavior in NWs, it becomes essential to obtain an insight of the size effects (boundary scattering). To do so, the cumulative thermal conductivity with the phonon mean free path (MFP) was plotted together with the nanowire $\kappa$ for both phases at different temperatures (see Fig.~\ref{MFPS}). Thus, from Fig.~\ref{MFPS} we can confirm that the ratio in NWs is mostly controlled by size effects. Although the cumulative thermal conductivity with MFP do not hold accurate predictive power~\cite{BTEnanowires}, they overall reproduce the behavior of the ratios as a function of diameter (Fig.~\ref{MFP-ratios}~\cite{SI}), especially at high temperatures~\footnote{The lack of agreement at 77K might be due to the larger number of phonons (lower temperatures mean longer MFPs) affected by the different cutoff behaviors for a given MFP/diameter.}. For instance, InAs having a ratio larger than 1 at small diameters is due to the behavior of the ratio of the cumulative functions. Notwithstanding its utility for determining size effects (i.e: boundary scattering) trends, it is also clear that MFP plot cannot be used to perfectly predict the detailed ratio behavior as the cumulative and NW function differ, due to the cumulative function not accounting for phonon propagation axis or the expression of boundary scattering in NWs having an $\exp(-d/\lambda_{\rm MFP})$ behavior---where $d$ is the distance traveled by the phonon hitting the NW surface and $\lambda_{\rm MFP}$ is its MFP---as opposed to an abrupt cutoff~\cite{BTEnanowires}.

Moreover, from the nanowire $\kappa$s we can also observe that BN and GaN show a higher $\kappa$ at 300K than at 77K, thus indicating a displacement in the $\kappa(T)$ peak to higher temperatures when compared to the bulk. Such results are coherent with experimental observations for silicon NWs~\cite{LiAPL2003}, and they are associated to the domination of boundary scattering over Umklapp scattering in the NW geometry when compared to the bulk one. Finally, it is worth mentioning that the available tuning of $\kappa$ ratios by modifying their diameter, especially for AlAs with a range between 0.7 and 1.1, makes these NWs interesting blockpieces for complex thermoelectric and/or phononic systems.

\section{\label{sec:Conclusions} Summary and conclusions}

In this work we have presented {\it ab initio} calculations of the lattice thermal conductivity for the cubic and hexagonal phases of GaAs, GaN, GaP, InAs, InP, AlAs, BN and ZnSe using density functional theory and iteratively solving the phonon Boltzmann Transport Equation. We find that which phase is the most conductive one depends on the relative strength between effective anharmonicity and accessible phase space. Such factors are shown to be antagonistic for all materials due to the higher effective anharmonicity of the cubic phase when compared to the hexagonal one, which, on the other hand, has a higher accessible phase space. Furthermore, we carry out an analysis of which factor is dominant when three phonon processes are the only ones present in each material at 77 and 300K, showing that, when anharmonicity (phase space) dominates, it leads to a higher (lower) conductivity in the hexagonal phase compared to the cubic one. Moreover, we have observed that when the hexagonal-cubic ratio of temperature-weighted anharmonicity and accessible phase space product is less than $\sim1.66$, the dominating factor determining $\kappa$ is anharmonicity~($\kappa_{hex}>\kappa_{cub}$). On the contrary, when that product is higher than $\sim1.66$, the dominating factor determining $\kappa$ is the accessible phase space($\kappa_{hex}<\kappa_{cub}$), thereby making such quantity an excellent tool to predict which is the most conductive phase at a given temperature when other more qualitative analysis fail. We also present results for NWs, showing the effect of boundary scattering on ($\kappa_{hex}/\kappa_{cub}$) and its relation to the phonon mean free paths. Moreover, we find that NWs have the ability to have their $\kappa$ ratio tuned over a wide range with diameter hence making them appealing materials for phononic and thermoelectric applications. The case of AlAs, with a $\kappa_{hex}/\kappa_{cub}$ range between 0.7 and 1.1, is of special interest.

\begin{acknowledgments}
We acknowledge financial support by the Ministerio de Econom\'ia, Industria y Competitividad (MINECO) under grants TEC2015-67462-C2-1-R (MINECO/FEDER) and RTI2018-097876-B-C21 (MICINN/FEDER) and FEDER-MAT2017-90024-P, the Severo Ochoa Centres of Excellence Program under Grant SEV-2015-0496, the Generalitat de Catalunya under grant no. 2017 SGR 1506 and the Ministerio de Educaci\'on, Cultura y Deporte programme of Formaci\'on de Profesorado Universitario under Grant Nos. FPU2016/02565. The authors thankfully acknowledge the computer resources, technical expertise and assistance provided by the Supercomputing and Visualization Center of Madrid (CeSViMa) and the Spanish Supercomputing Network (RES) under project FI-2018-1-0027.\end{acknowledgments}

\bibliography{references}

\end{document}


\title{Supplemental Material for:\\Thermal conductivity for III-V and II-VI semiconductor wurtzite and zinc-blende polytypes: the role of anharmonicity and phase space}

\maketitle

\section{Thermal conductivity convergence with \MakeLowercase{\it q}{\bf-mesh}}
\begin{figure*}
	\begin{subfigure}{0.42\linewidth}
		\centering
		\includegraphics[width=\linewidth]{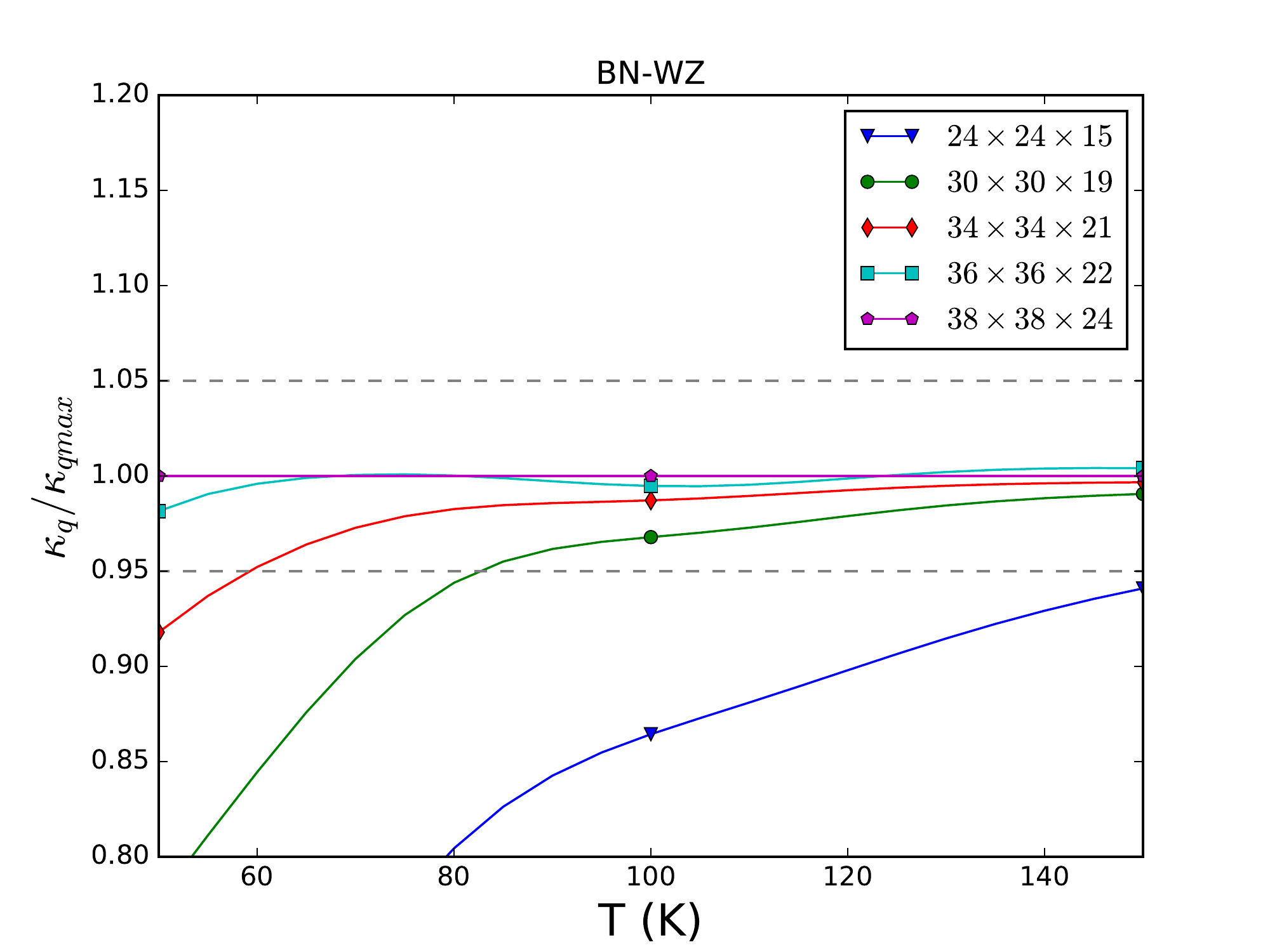}
	\end{subfigure}
	\begin{subfigure}{0.42\linewidth}
		\centering
		\includegraphics[width=\linewidth]{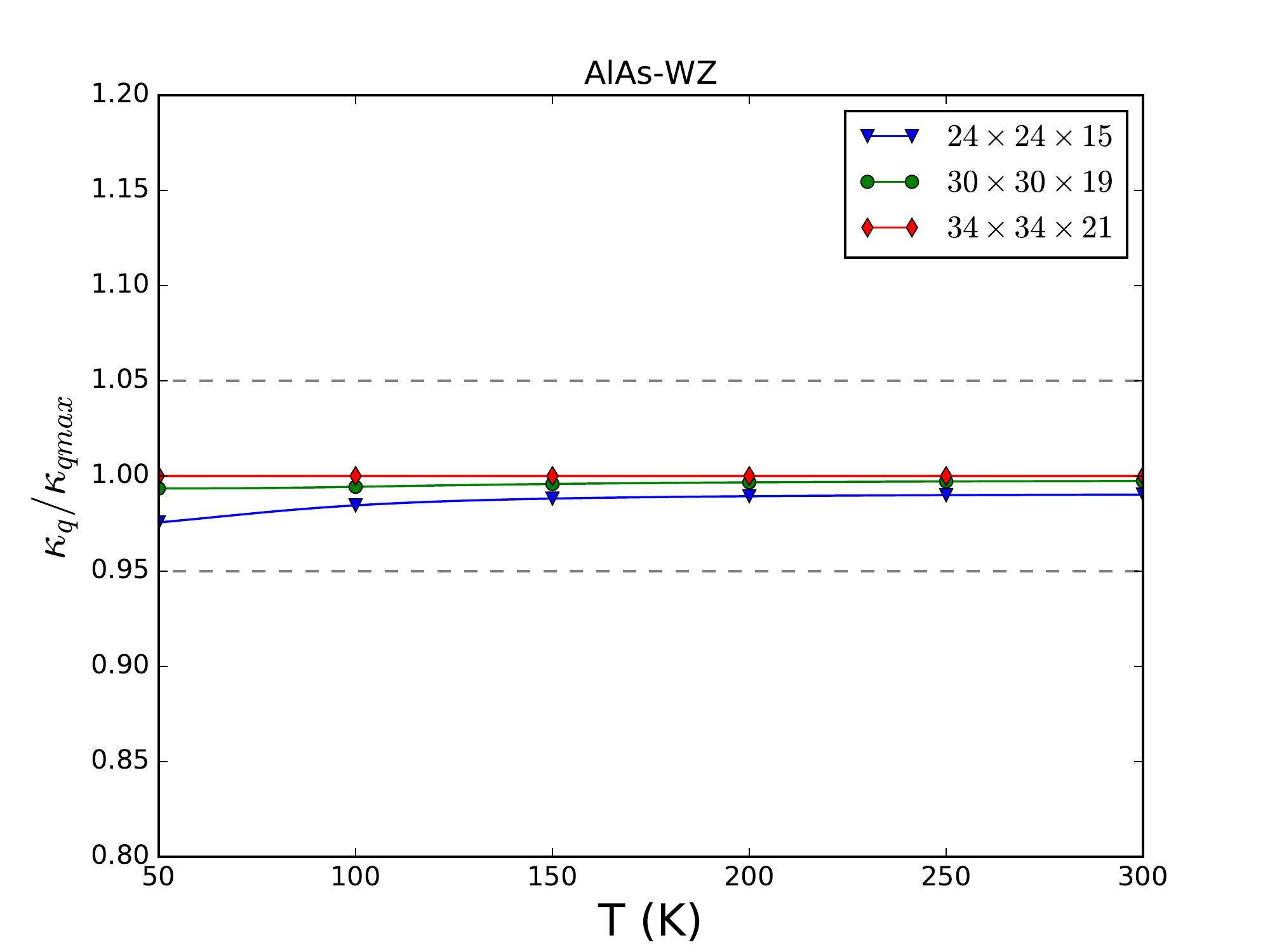}
	\end{subfigure}
	\begin{subfigure}[t]{0.42\textwidth}
		\centering
		\includegraphics[width=\linewidth]{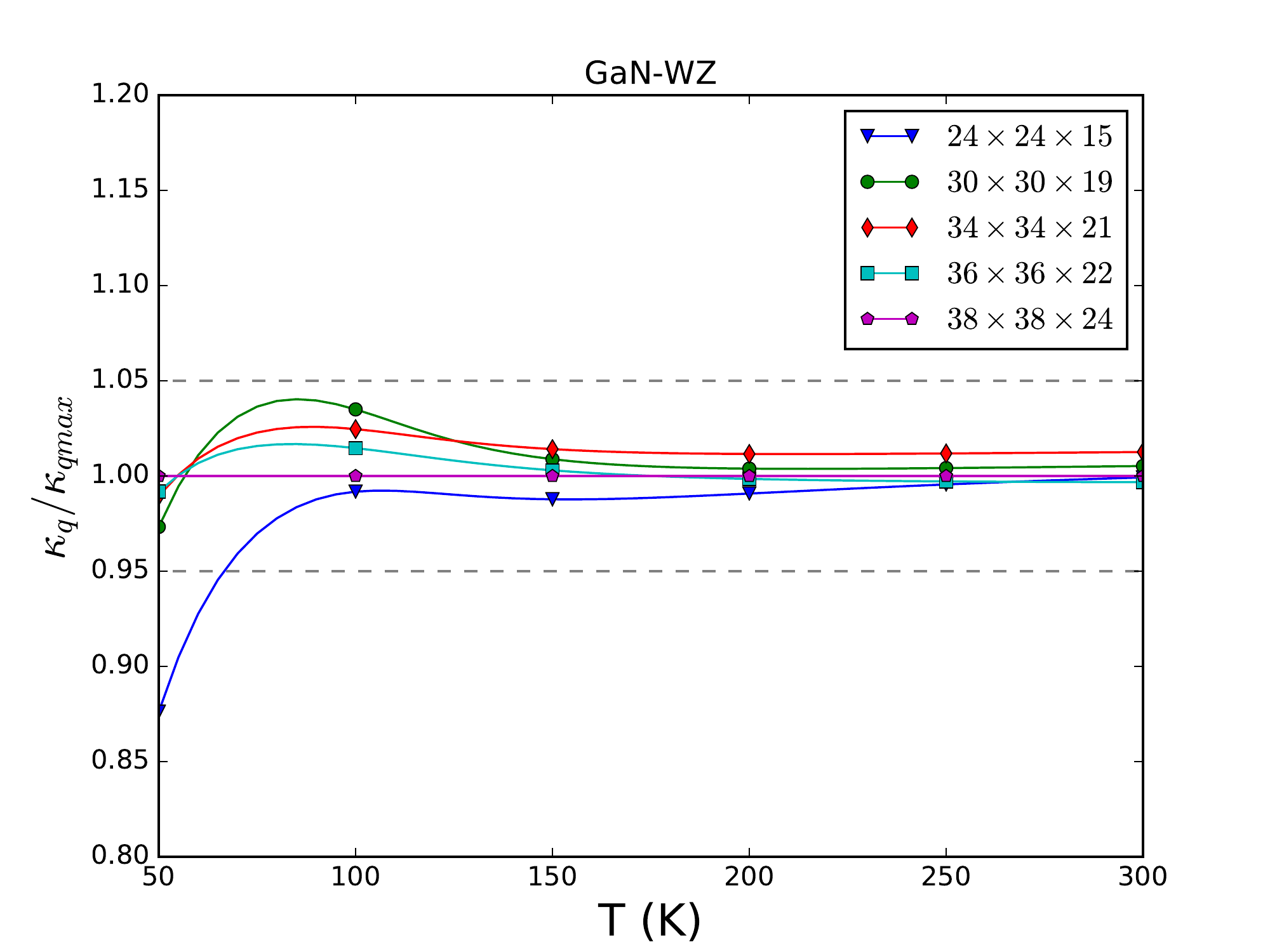}
	\end{subfigure}
	\begin{subfigure}[t]{0.42\textwidth}
		\centering
		\includegraphics[width=\linewidth]{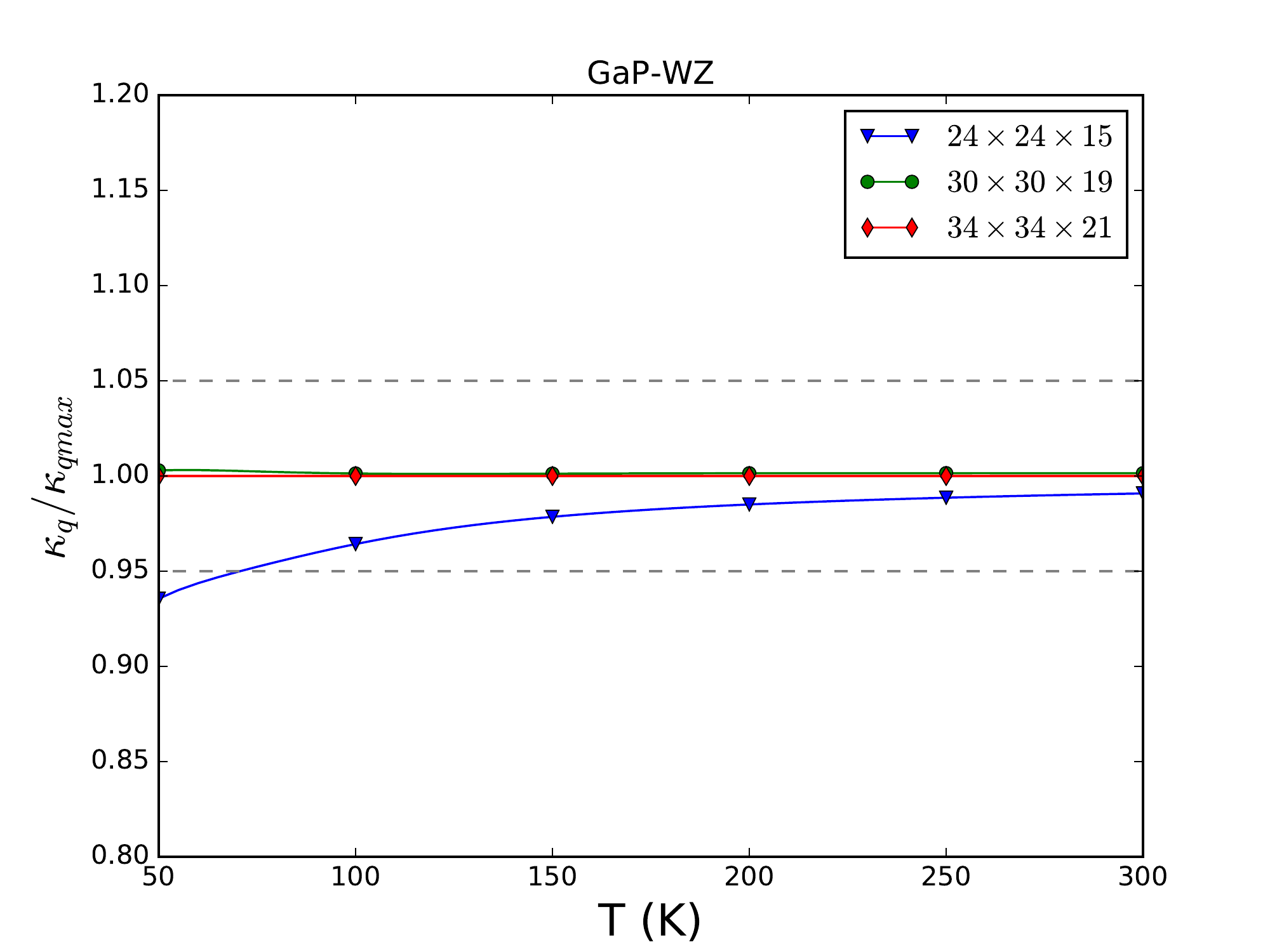}
	\end{subfigure}
	\begin{subfigure}[t]{0.42\textwidth}
		\centering
		\includegraphics[width=\linewidth]{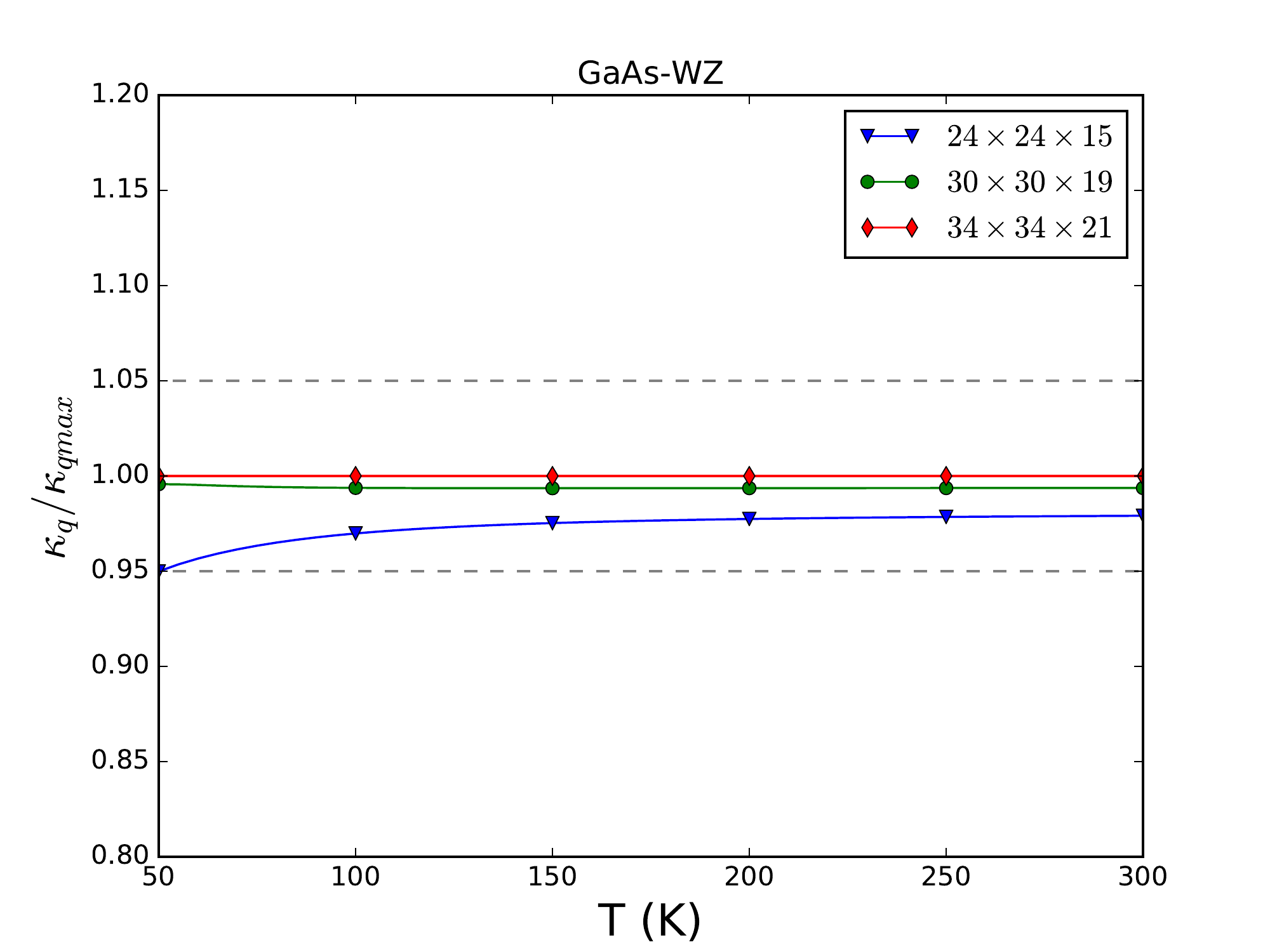}
	\end{subfigure}
	\begin{subfigure}[t]{0.42\textwidth}
		\centering
		\includegraphics[width=\linewidth]{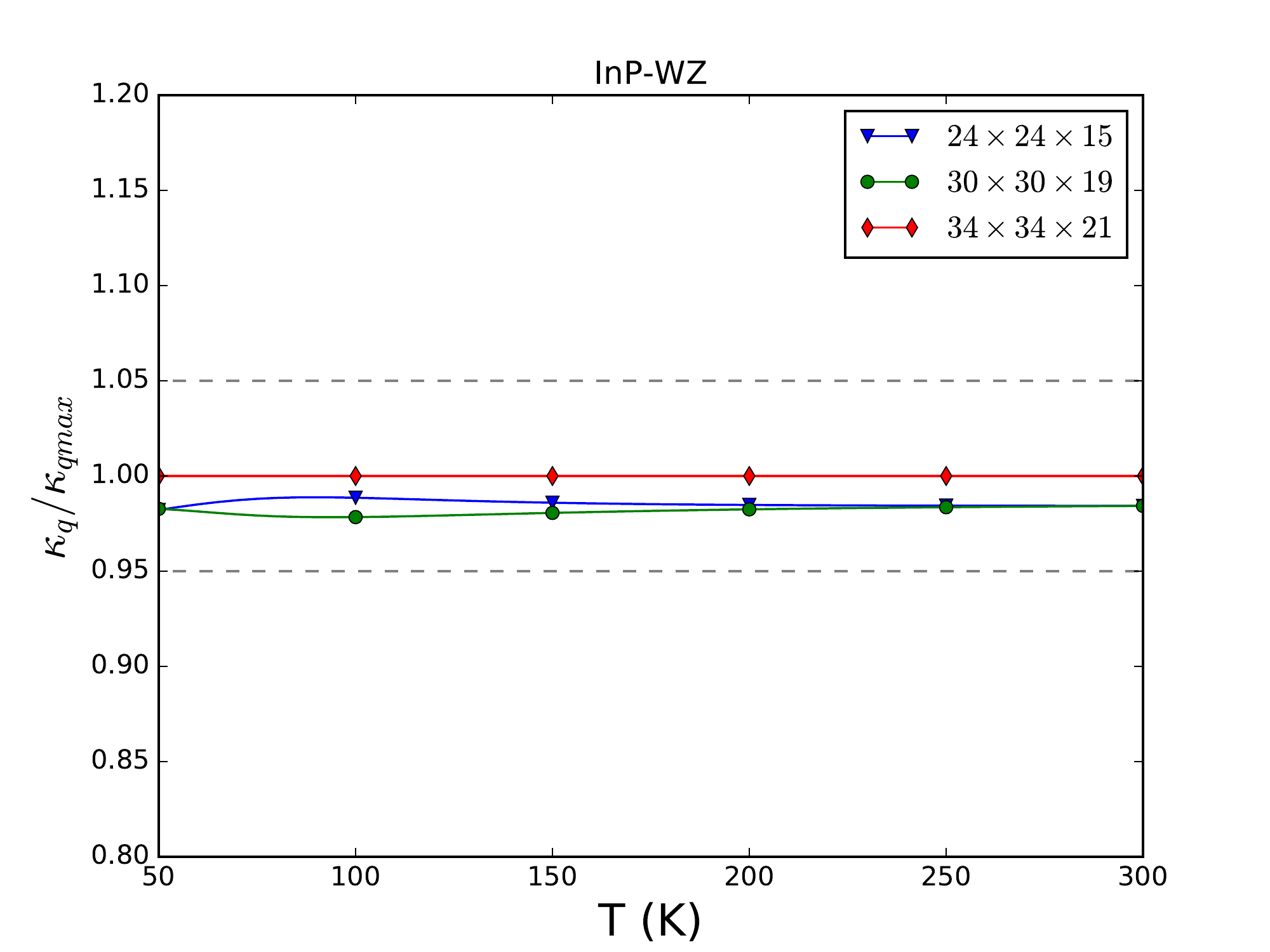}
	\end{subfigure}
	\begin{subfigure}[t]{0.42\textwidth}
		\centering
		\includegraphics[width=\linewidth]{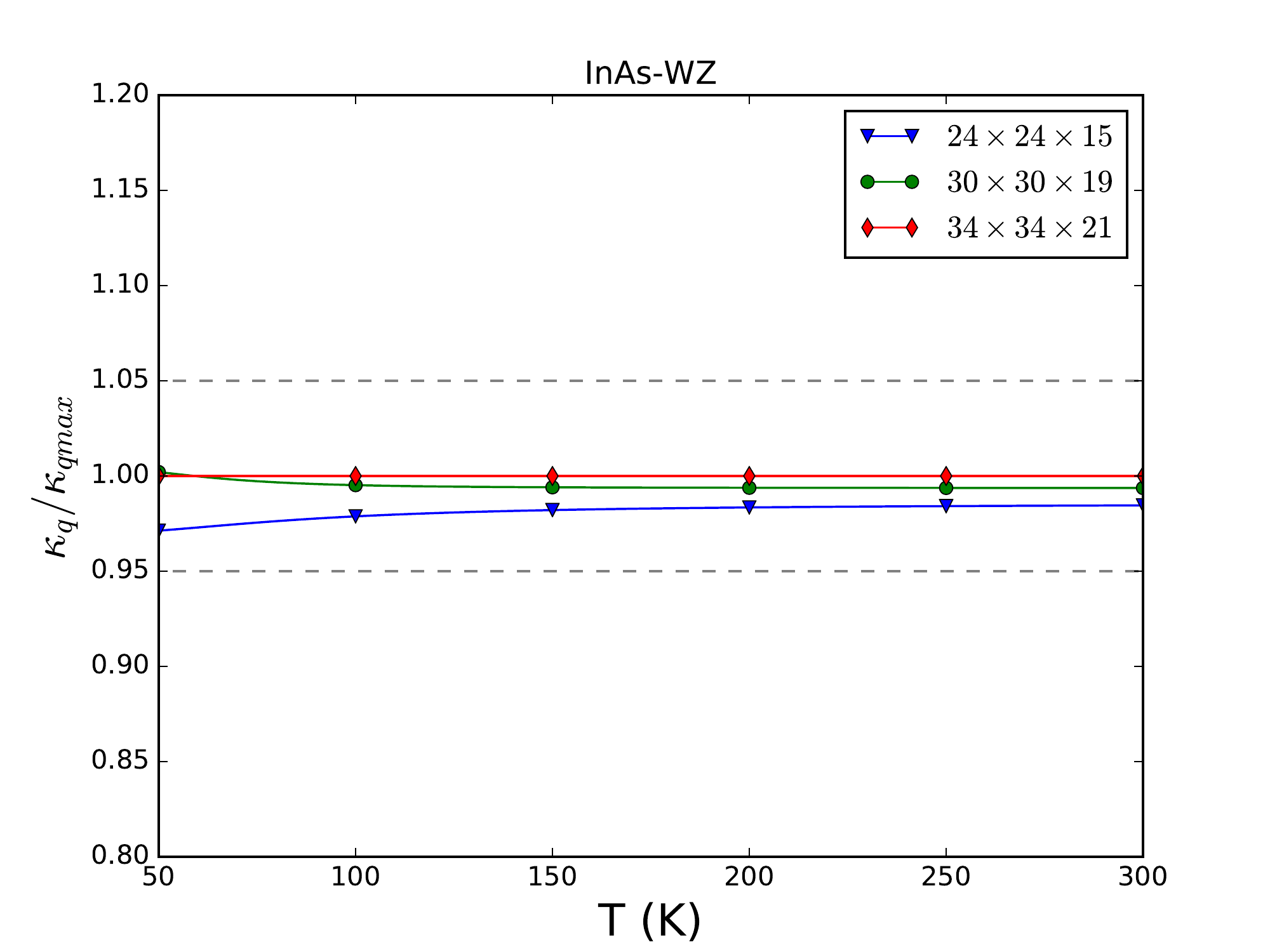}
	\end{subfigure}
	\begin{subfigure}[t]{0.42\textwidth}
		\centering
		\includegraphics[width=\linewidth]{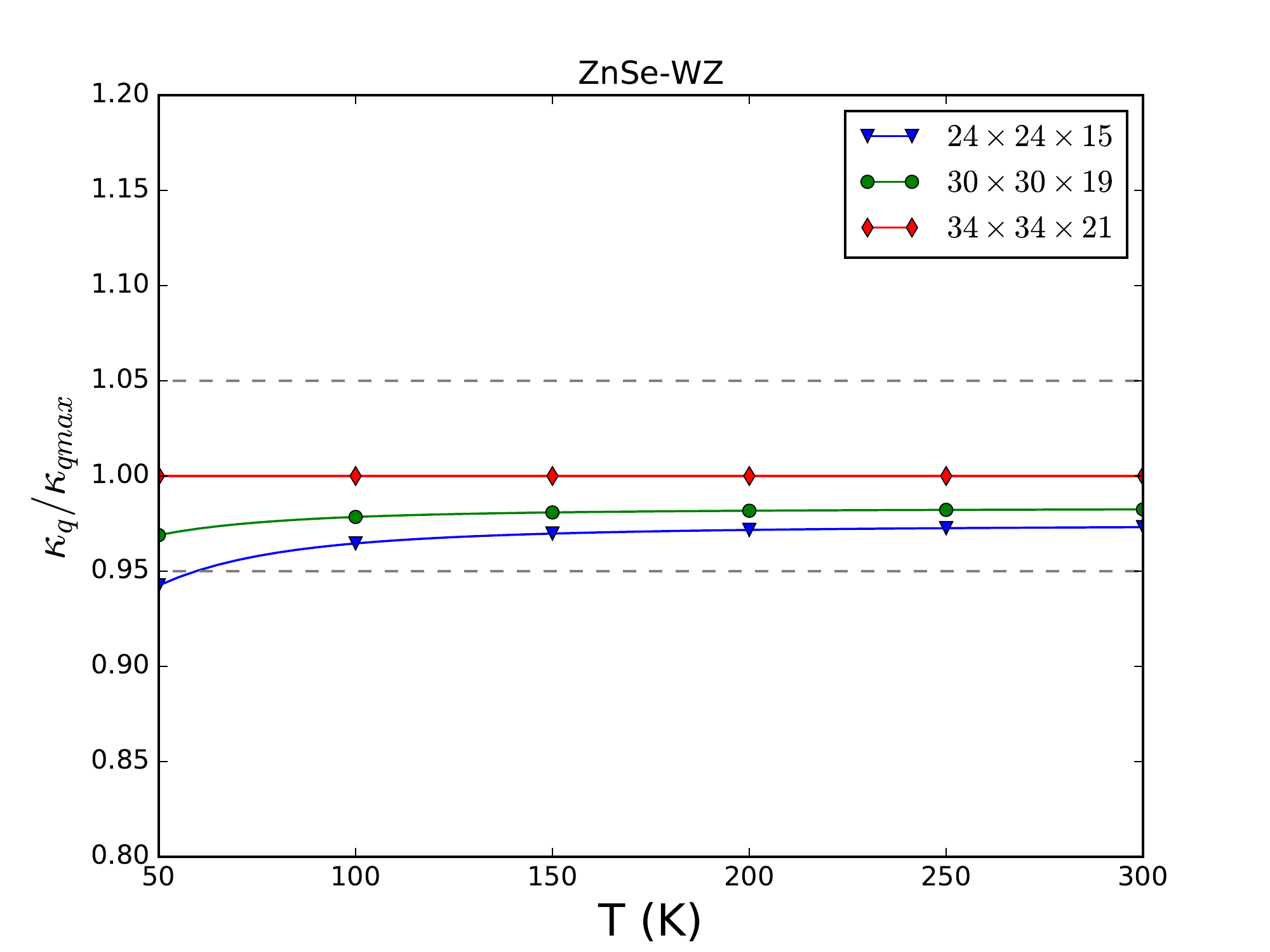}
	\end{subfigure}
	\caption{Convergence of thermal conductivity with respect to the maximum $q$-mesh value for the corresponding material, as function of temperature for the wurtzite phase. Dashed lines indicate the 5\% threshold which we have taken as our convergence criterion.}
	\label{conv-WZ}
\end{figure*}

\begin{figure*}
	\begin{subfigure}{0.42\linewidth}
		\centering
		\includegraphics[width=\linewidth]{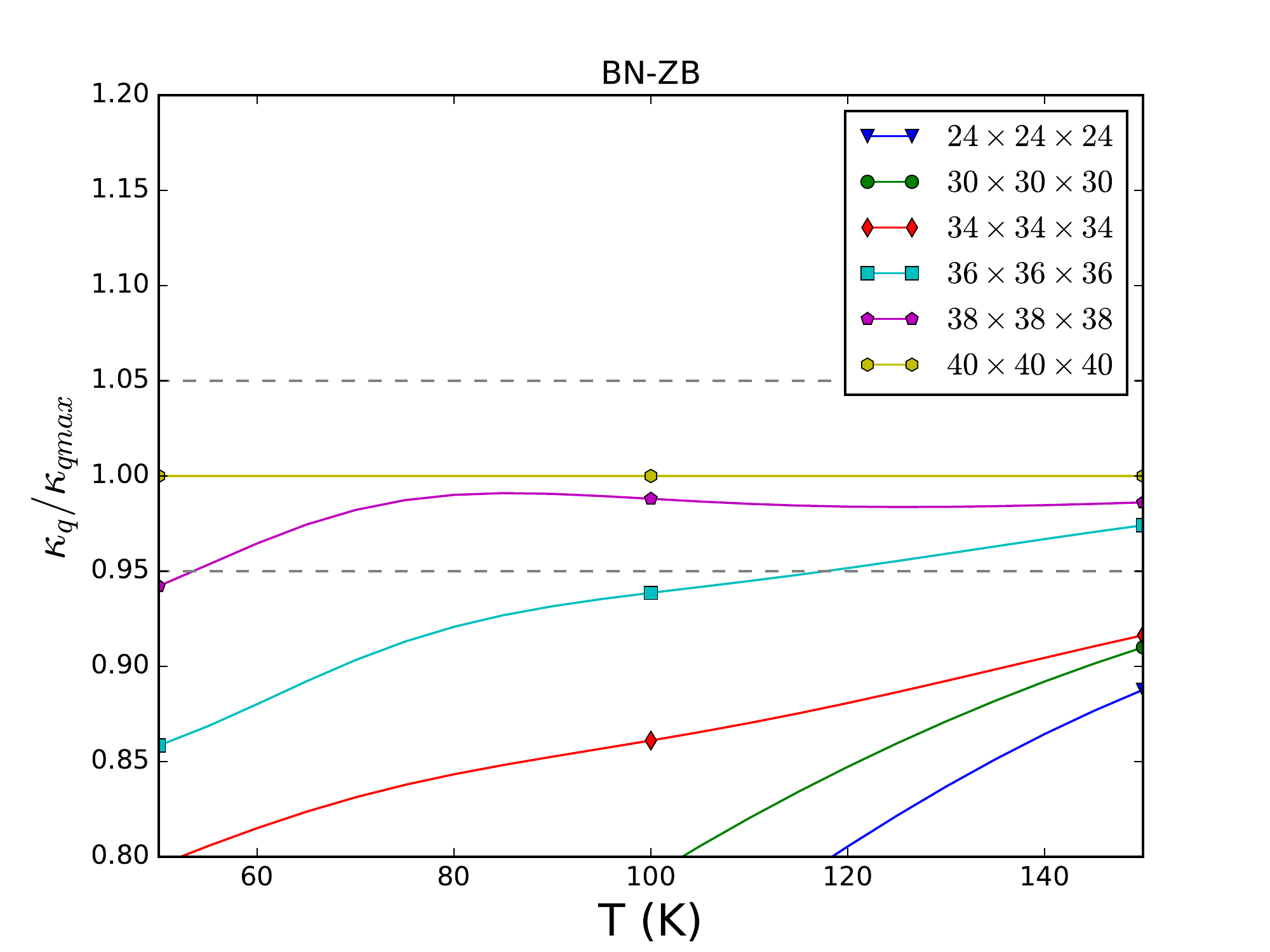}
	\end{subfigure}
	\begin{subfigure}{0.42\linewidth}
		\centering
		\includegraphics[width=\linewidth]{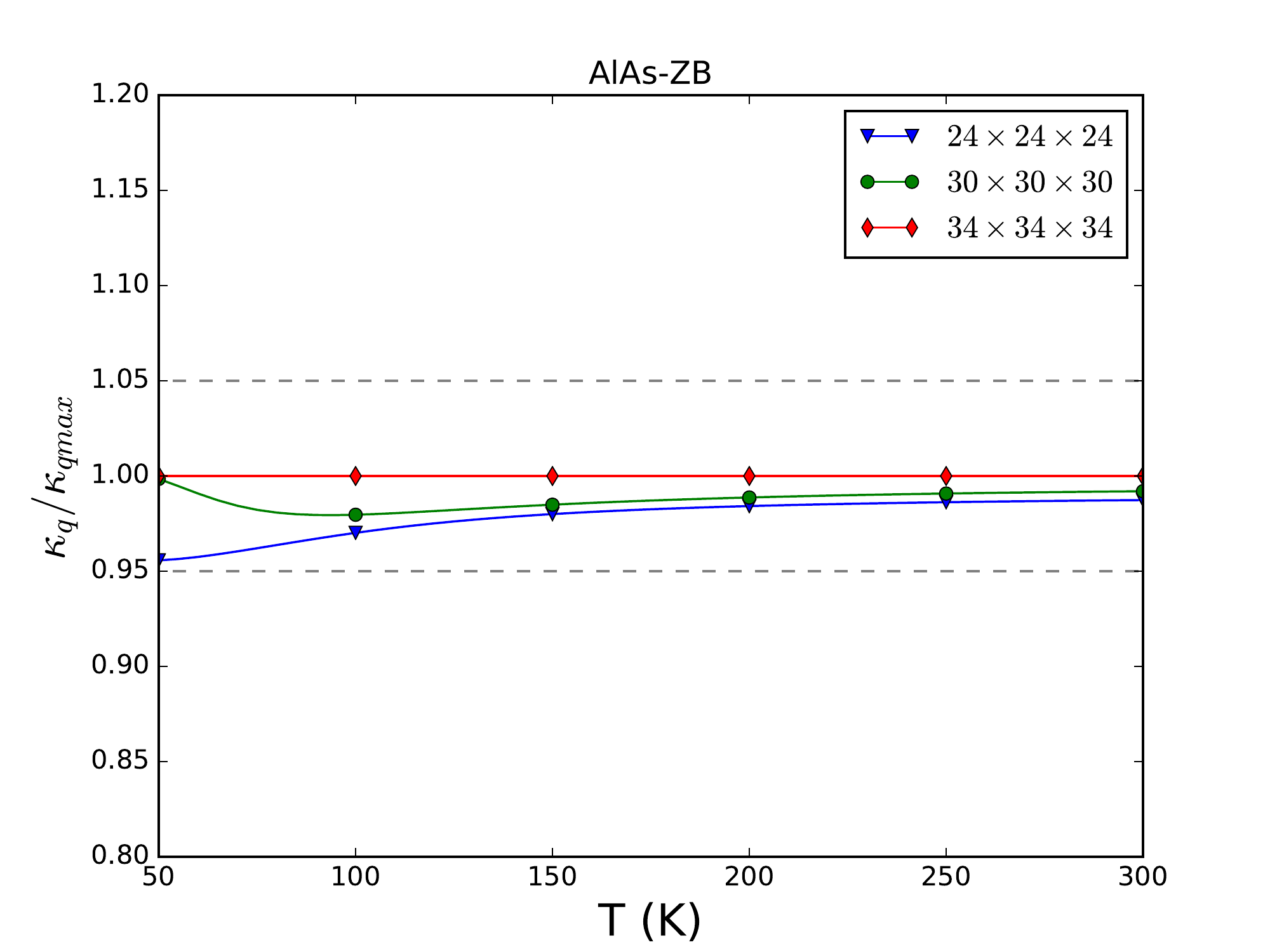}
	\end{subfigure}
	\begin{subfigure}[t]{0.42\textwidth}
		\centering
		\includegraphics[width=\linewidth]{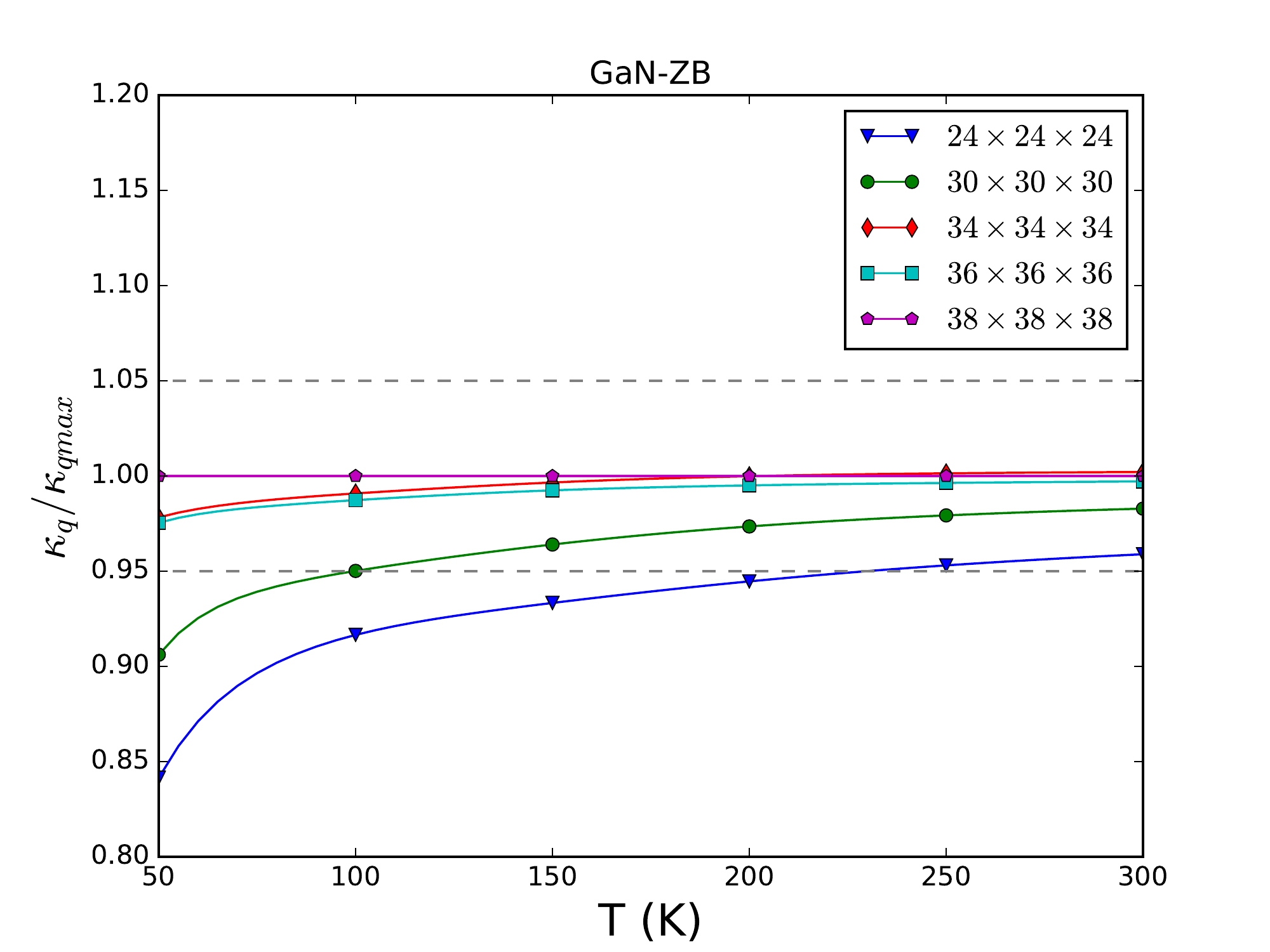}
	\end{subfigure}
	\begin{subfigure}[t]{0.42\textwidth}
		\centering
		\includegraphics[width=\linewidth]{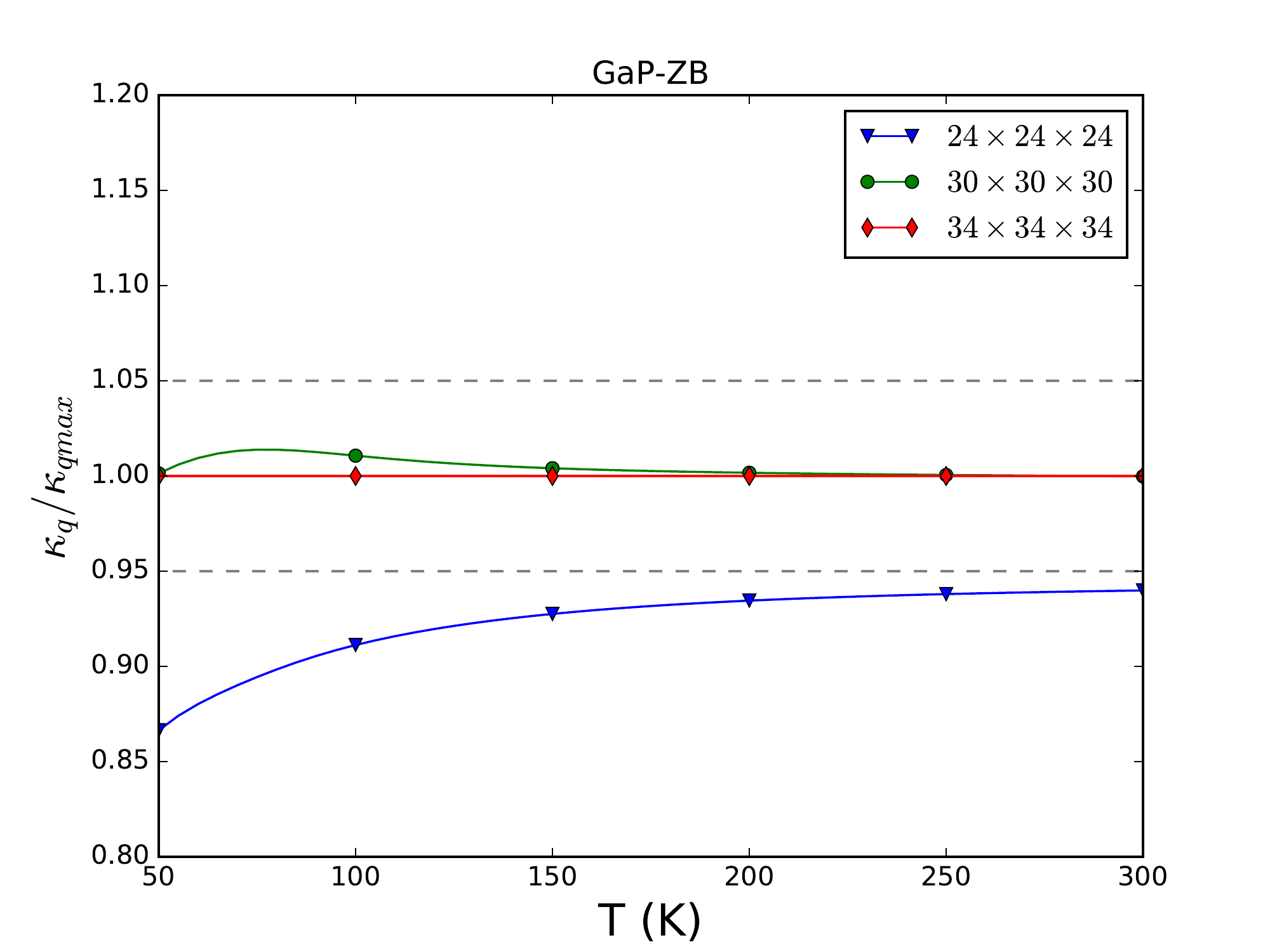}
	\end{subfigure}
	\begin{subfigure}[t]{0.42\textwidth}
		\centering
		\includegraphics[width=\linewidth]{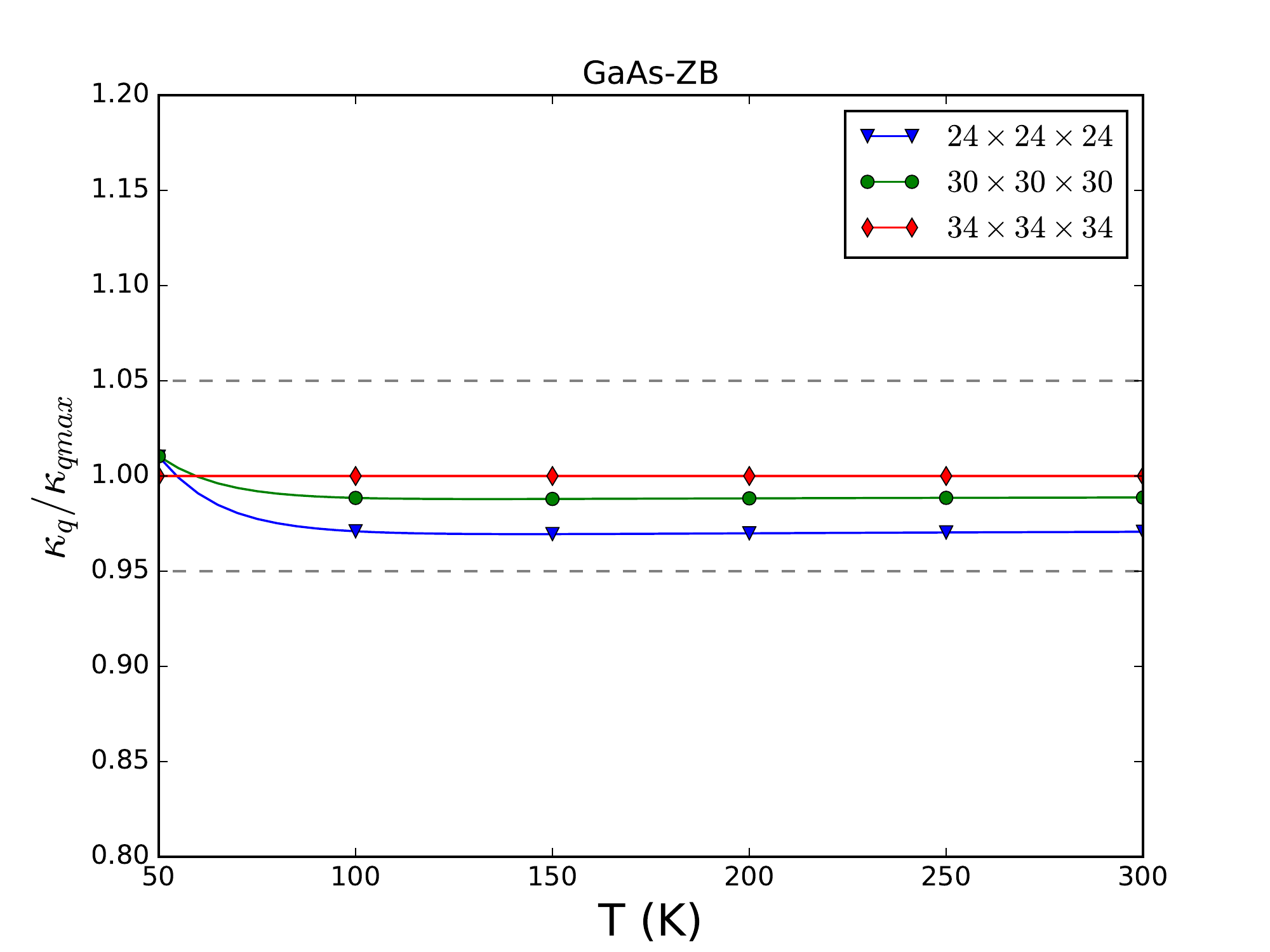}
	\end{subfigure}
	\begin{subfigure}[t]{0.42\textwidth}
		\centering
		\includegraphics[width=\linewidth]{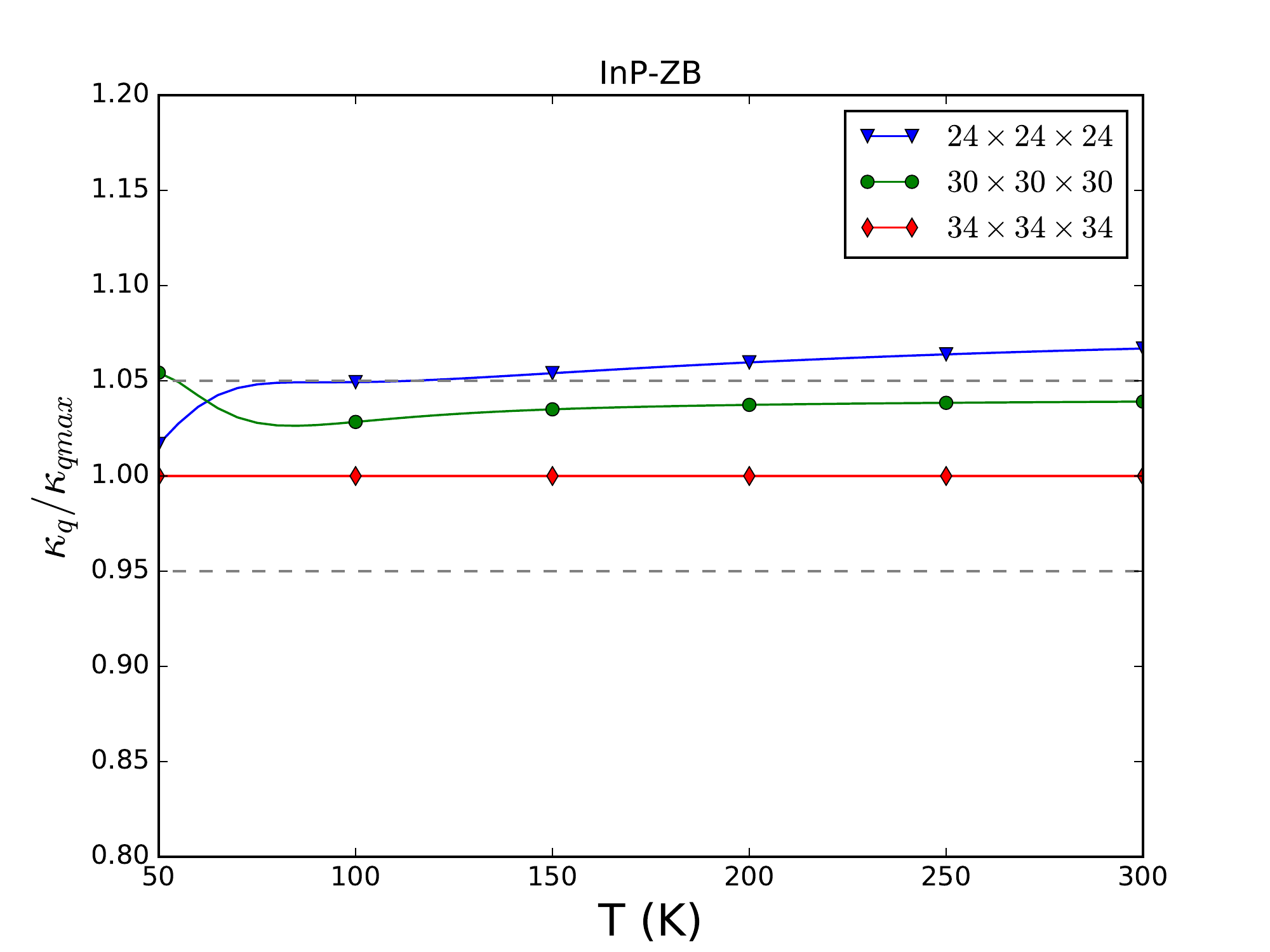}
	\end{subfigure}
	\begin{subfigure}[t]{0.42\textwidth}
		\centering
		\includegraphics[width=\linewidth]{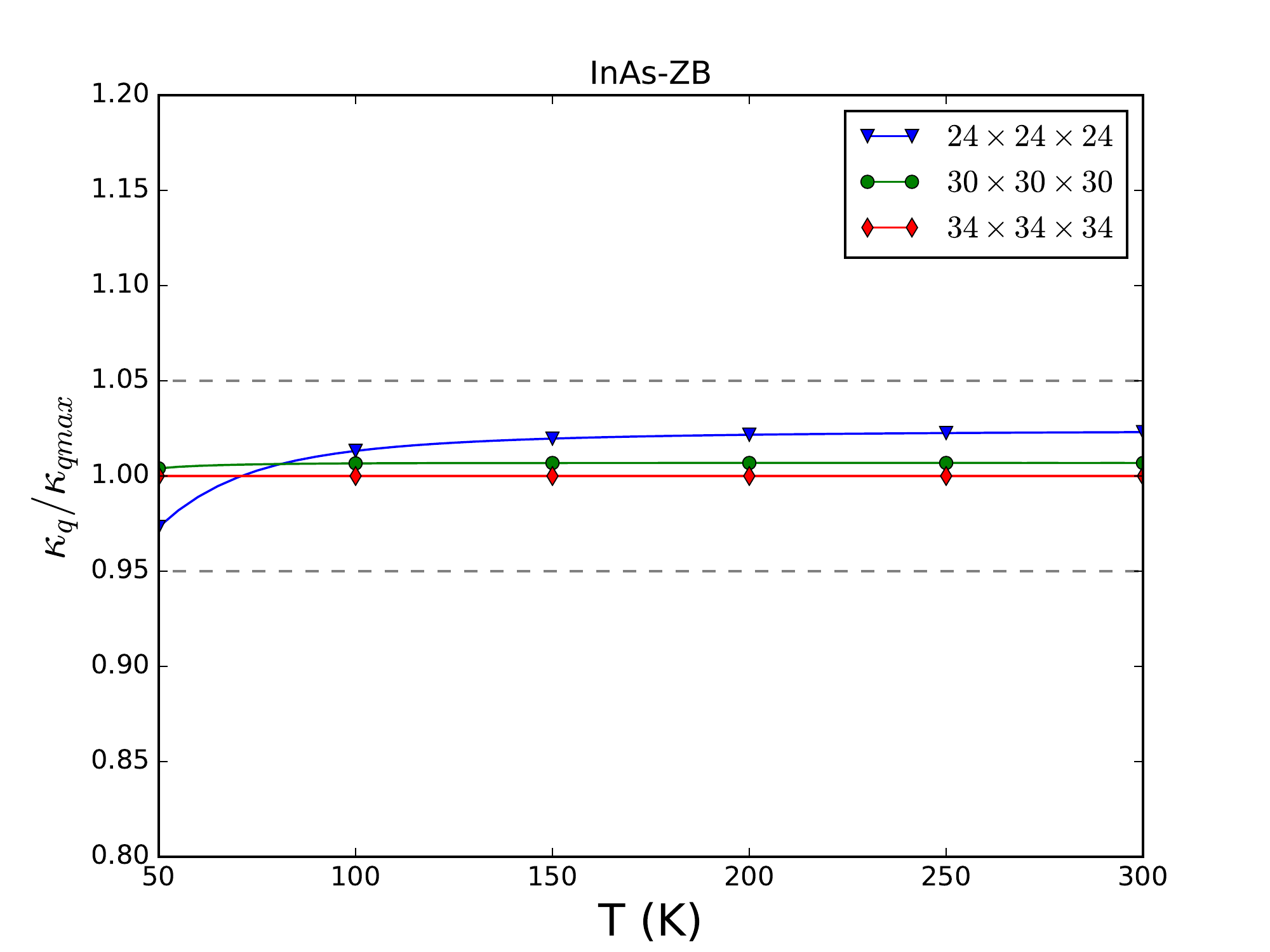}
	\end{subfigure}
	\begin{subfigure}[t]{0.42\textwidth}
		\centering
		\includegraphics[width=\linewidth]{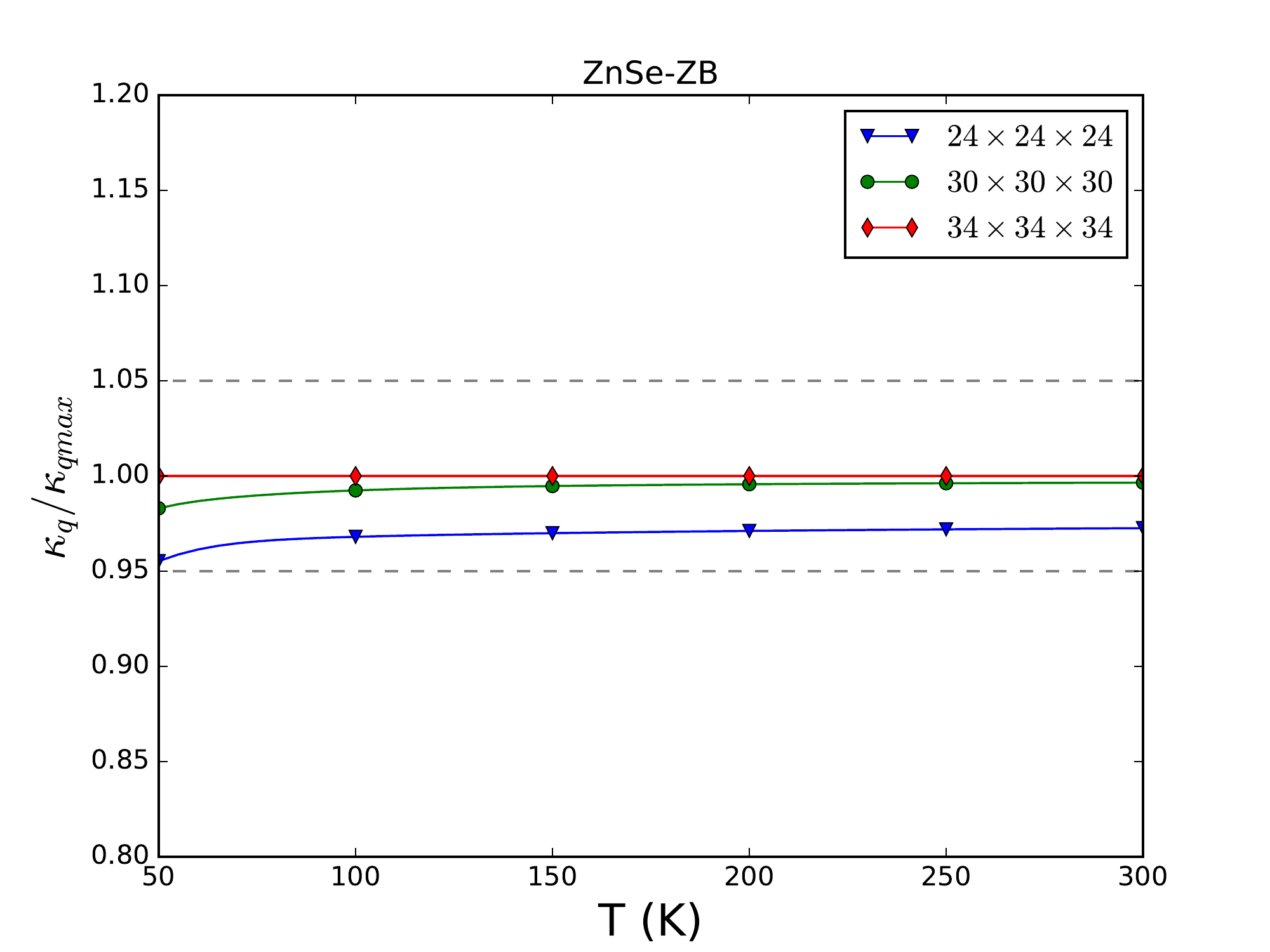}
	\end{subfigure}
	\caption{Like Fig.~\ref{conv-WZ}, but for the zinc-blende phase.}
	\label{conv-ZB}
\end{figure*}

We provide the convergence of thermal conductivity for both phases hexagonal (WZ) and cubic (ZB) with the reciprocal space mesh size in Figs.~\ref{conv-WZ} and \ref{conv-ZB}, respectively.

\section{Phonon dispersions}

\begin{figure*}
	\begin{subfigure}{\linewidth}
		\includegraphics[width=\textwidth]{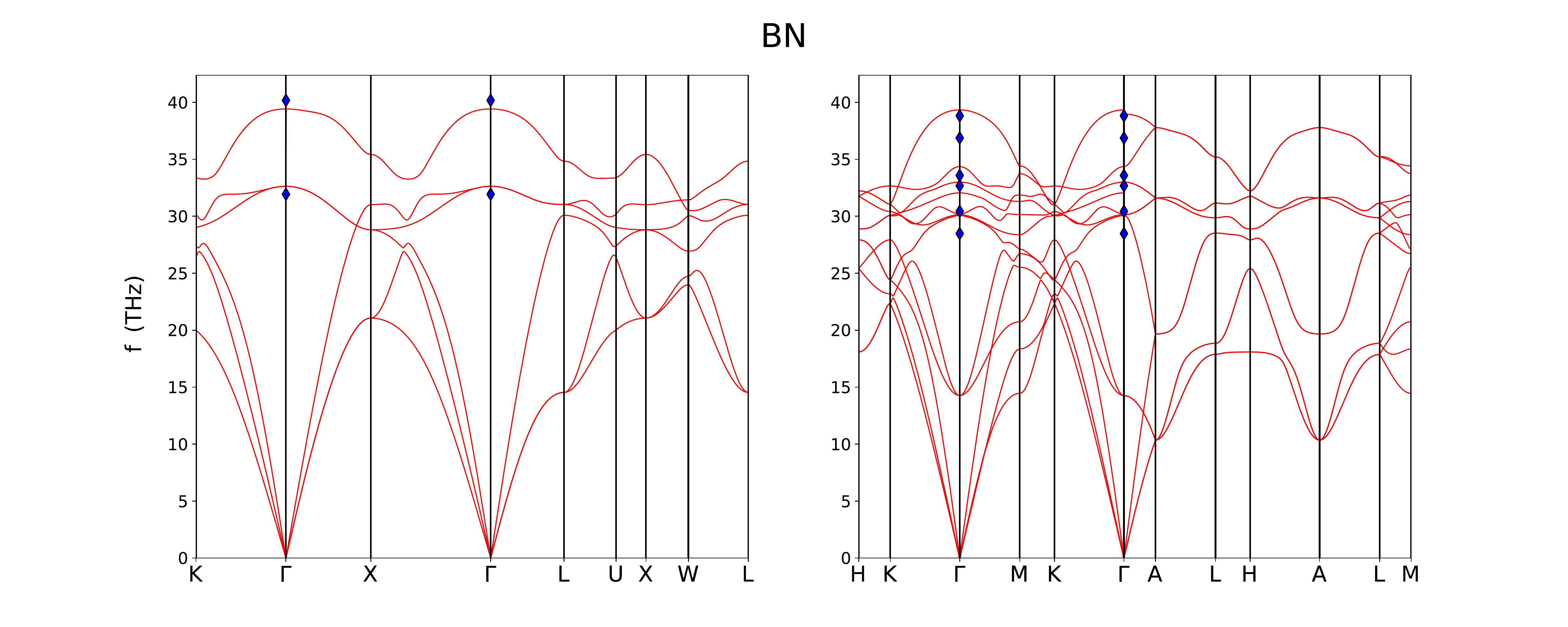}
	\end{subfigure}
	\begin{subfigure}{\linewidth}
		\includegraphics[width=\textwidth]{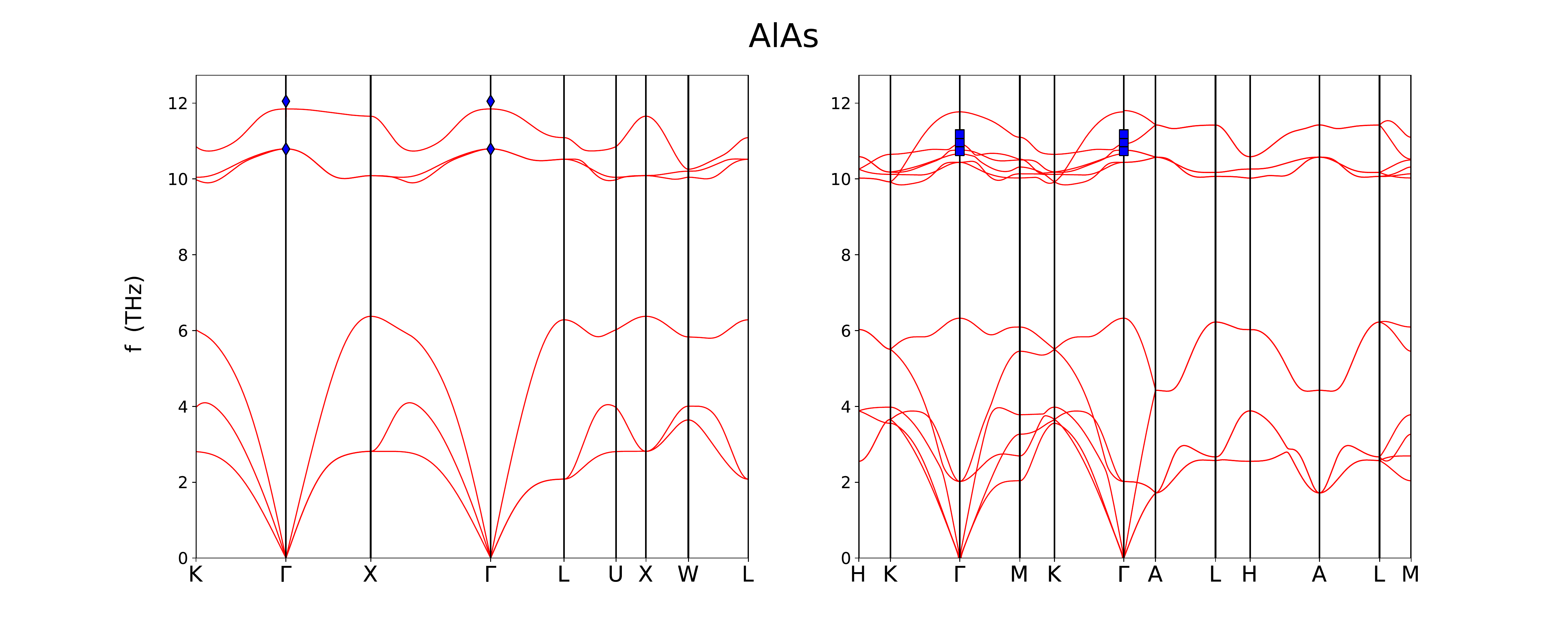}
	\end{subfigure}
	\caption{Dispersion relation for BN (top) and AlAs (bottom) in the cubic (left) and hexagonal (right) phases. Experimental results for BN are from Ref.~\citenum{EdgarP3N1994} (diamonds); and results for AlAs are from Ref.~\citenum{FunkACSNano2013} (squares) and Ref.~\citenum{AzuhataJPCM1949} (diamonds).}
	\label{disp1}
\end{figure*}

\begin{figure*}
	\begin{subfigure}{\linewidth}
		\includegraphics[width=\textwidth]{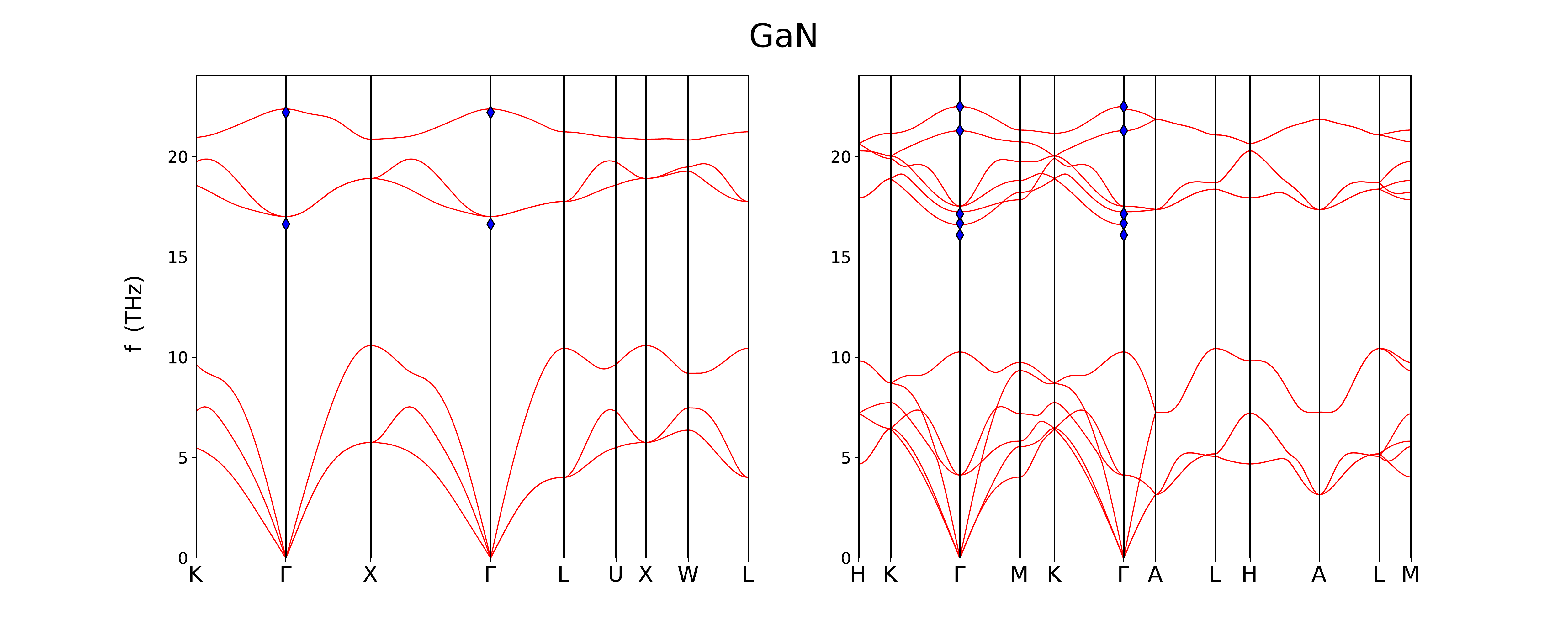}
	\end{subfigure}
	\begin{subfigure}{\linewidth}
		\includegraphics[width=\textwidth]{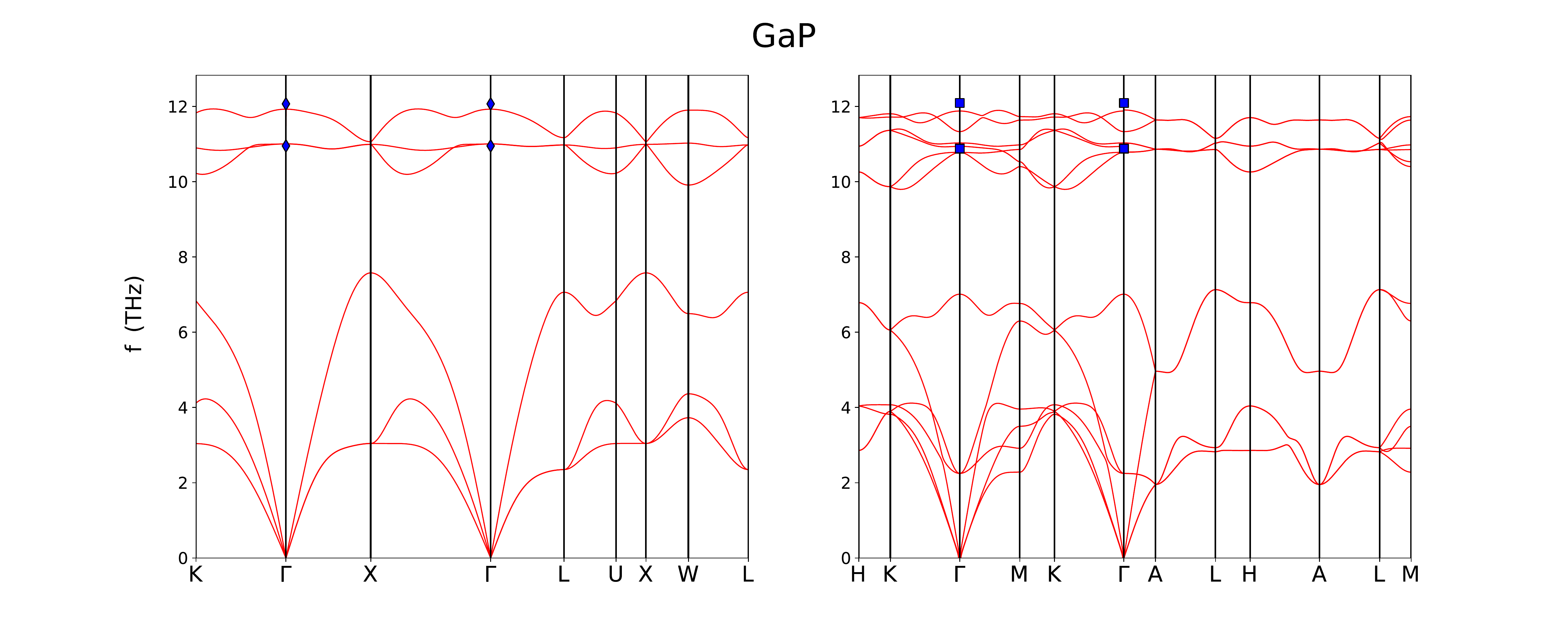}
	\end{subfigure}
	\begin{subfigure}{\linewidth}
		\includegraphics[width=\textwidth]{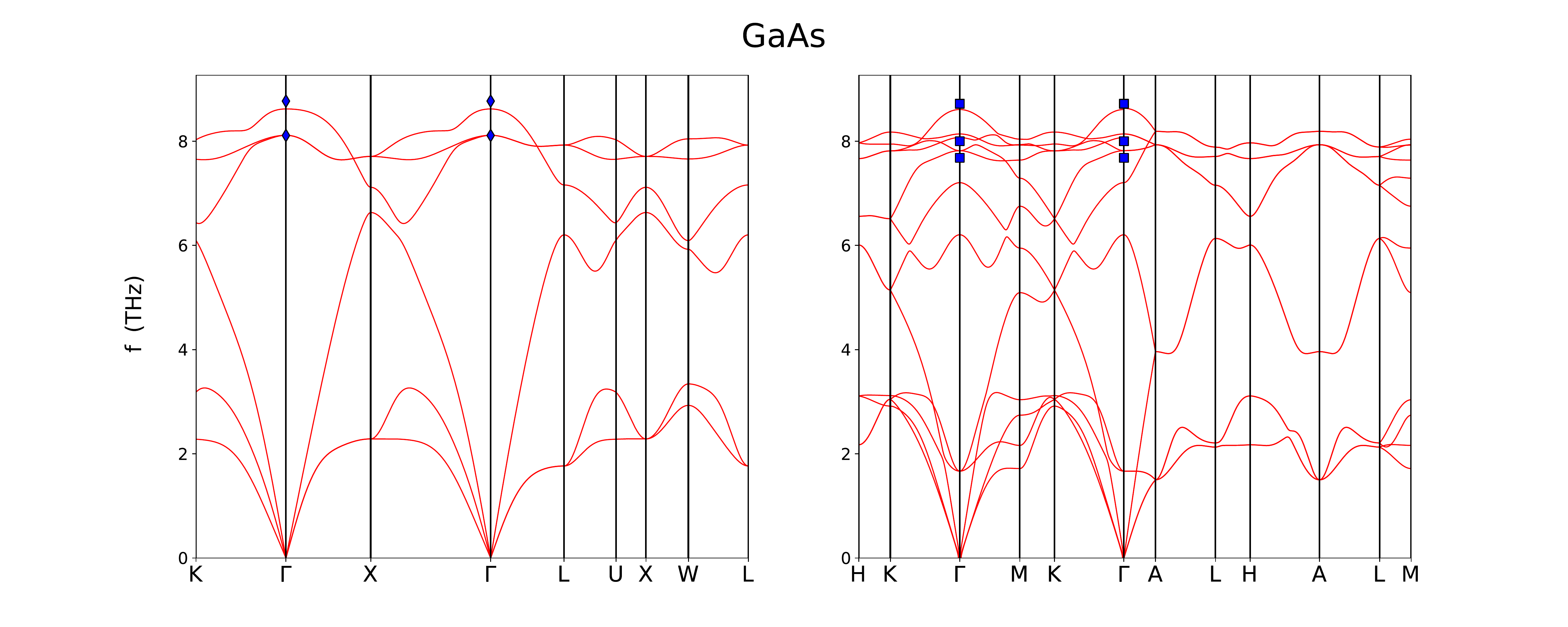}
	\end{subfigure}
	\caption{Dispersion relation for GaN (top), GaP (middle) and GaAs (bottom) in the cubic (left) and hexagonal (right) phases. Experimental results for GaN are from Ref.~\citenum{TabataJAP1996} (diamonds); results for GaP are from Ref.~\citenum{BergJCG2014} (squares) and Ref.~\citenum{gapzbref} (diamonds), and results for GaAs are from Ref.~\citenum{ZardoPRB2009} (squares) and Ref.~\citenum{StrauchJOPCM1990} (diamonds).}
	\label{disp2}
\end{figure*}

\begin{figure*}
	\begin{subfigure}{\linewidth}
		\includegraphics[width=\textwidth]{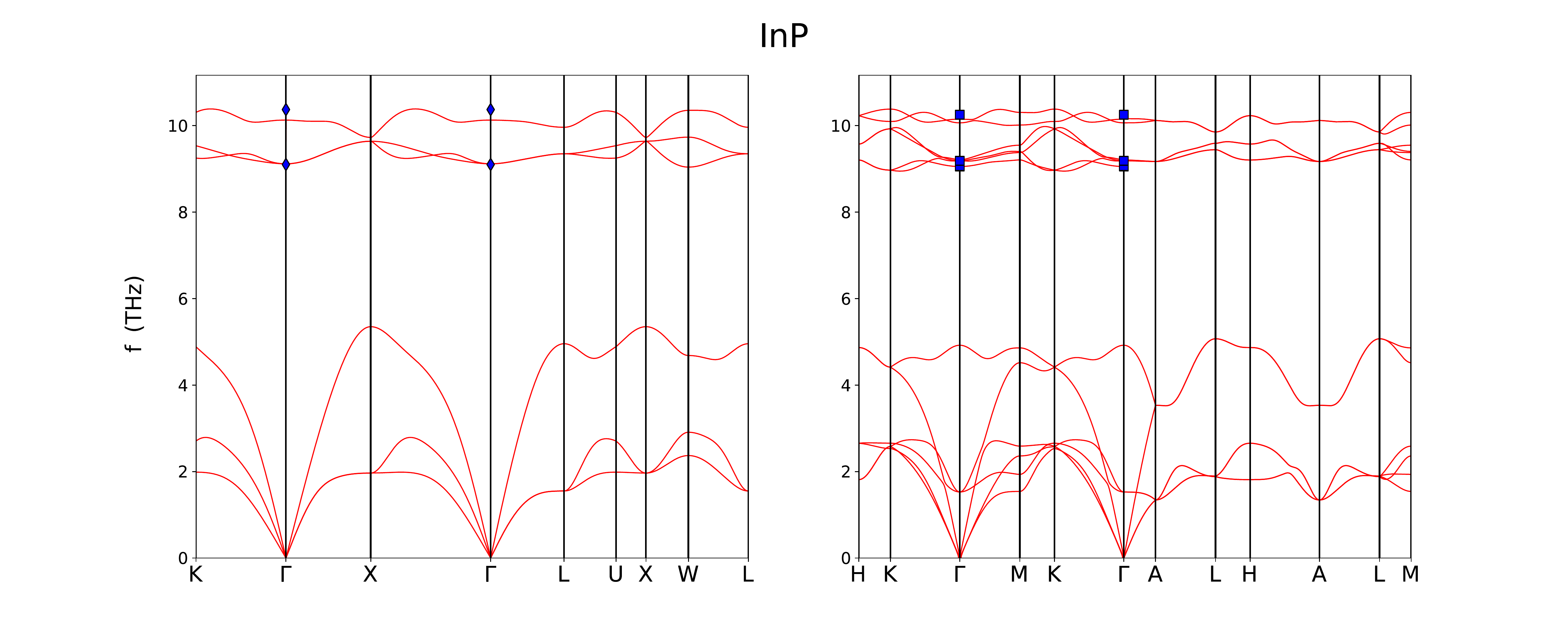}
	\end{subfigure}
	\begin{subfigure}{\linewidth}
		\includegraphics[width=\textwidth]{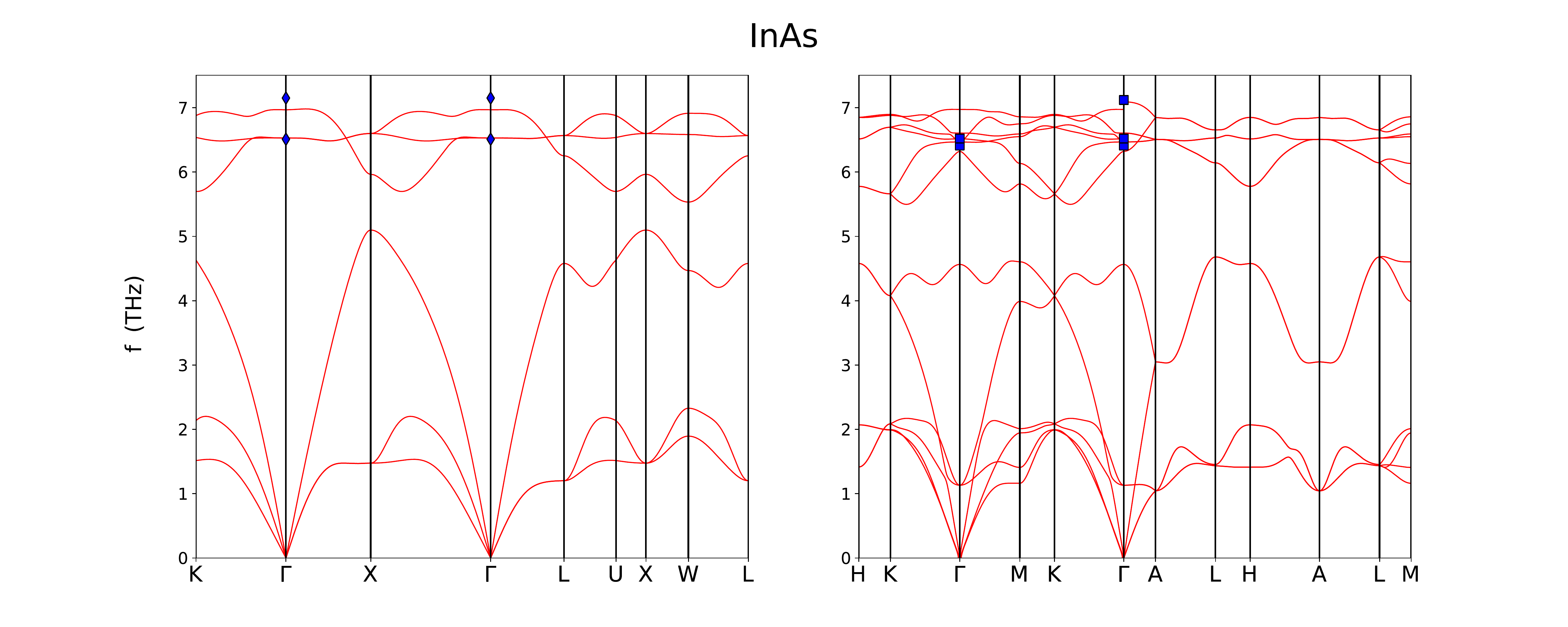}
	\end{subfigure}
	\begin{subfigure}{\linewidth}
		\includegraphics[width=\textwidth]{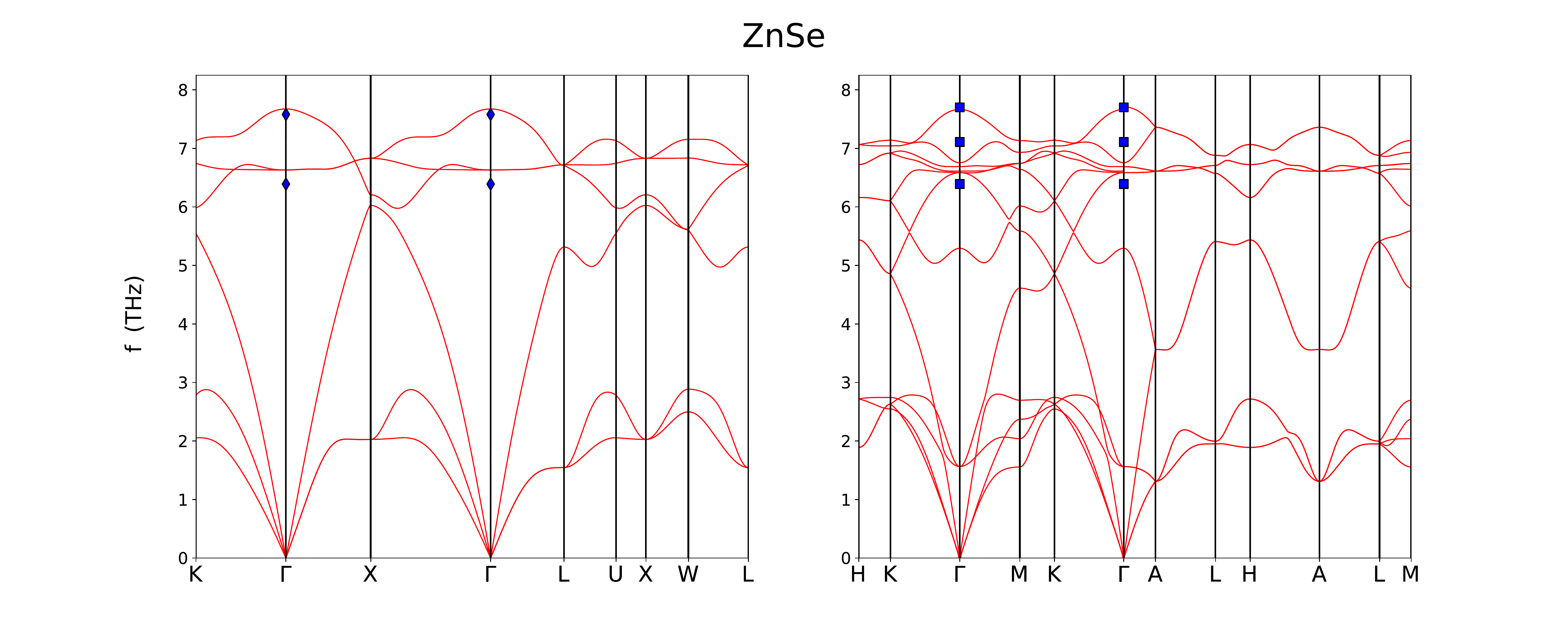}
	\end{subfigure}
	\caption{Dispersion relation for InP (top), InAs (middle) and ZnSe (bottom) in the cubic (left) and hexagonal (right) phases. Experimental results for InP are from Ref.~\citenum{GadretAPL2013} (squares) and Ref.~\citenum{BorcherdsJPC1975} (diamonds); results for InAs are from Ref.~\citenum{ZardoNLET2013} (squares) and Ref.~\citenum{CarlesPRB1980} (diamonds), and results for GaAs are from Ref.~\citenum{TeredesaiJNN2002} (squares) and Ref.~\citenum{HennionPLA1971} (diamonds).}
	\label{disp3}
\end{figure*}

We provide for completeness the computed phonon dispersion relations for the studied materials in Figs.~\ref{disp1}-\ref{disp3}, indicating the experimentally measured values of the zone-center phonon frequencies as well.

\section{Accessible phase space}
\begin{figure*}
	\begin{subfigure}[t]{1.00\textwidth}
		\centering
		\includegraphics[width=\linewidth]{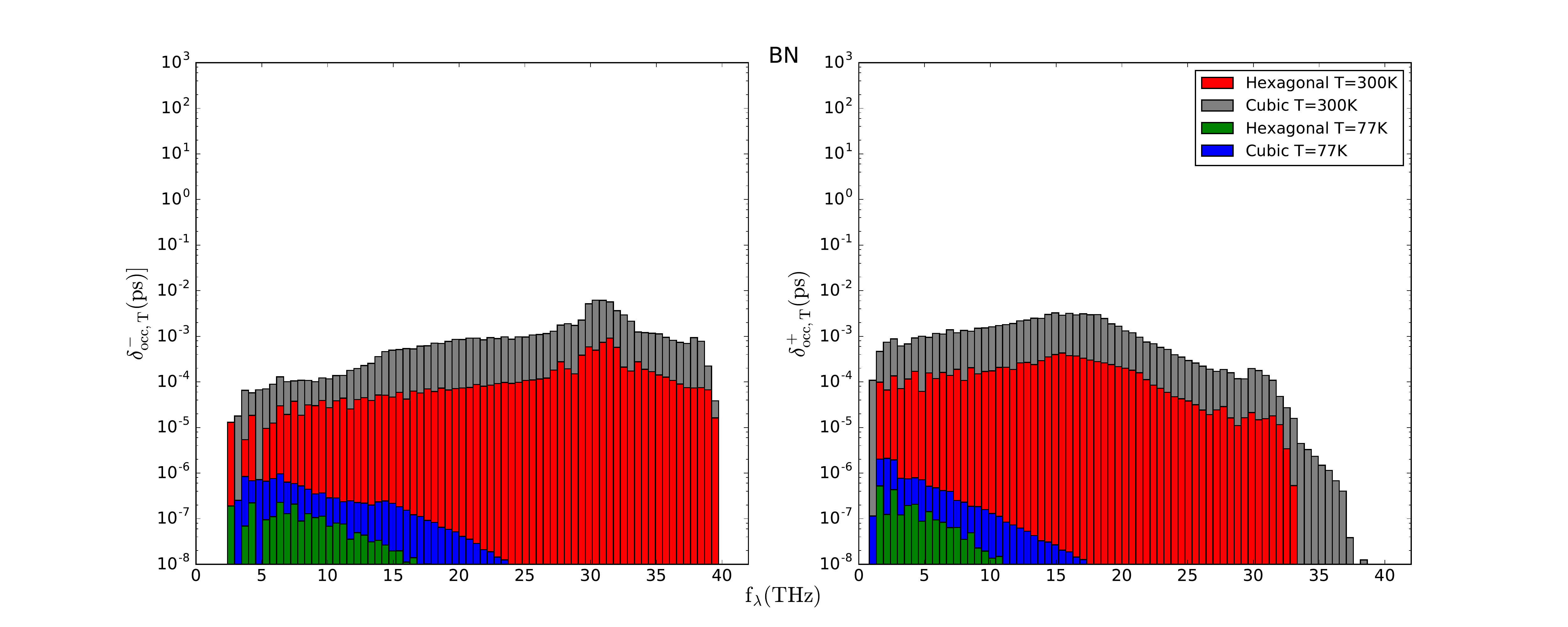}
	\end{subfigure}
	\begin{subfigure}[t]{1.00\textwidth}
		\centering
		\includegraphics[width=\linewidth]{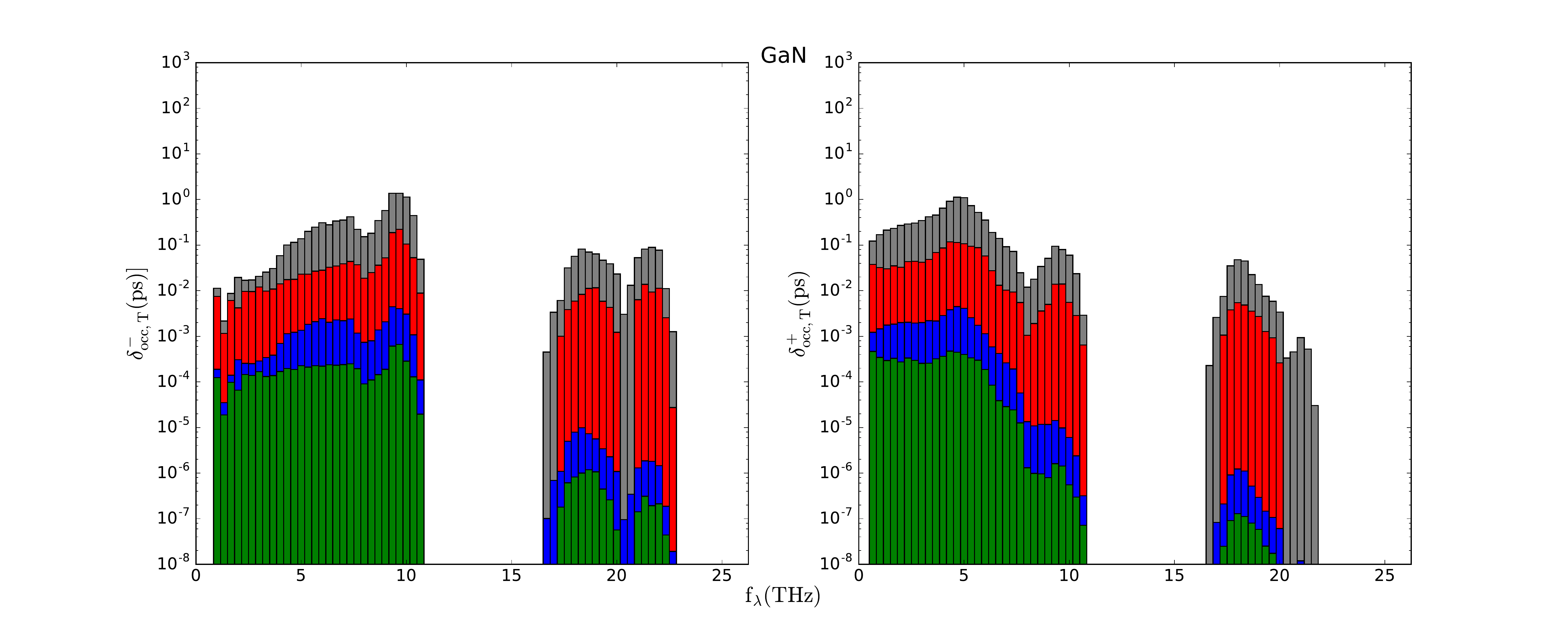}
	\end{subfigure}
	\caption{Contribution to the scattering space ($\delta_{\lambda\lambda'\lambda''}^{\pm\;occ,T}$) as a function of first phonon frequency for BN and GaN. Left images correspond to emission processes and right images to absorption processes.}
	\label{DELTAS1}
\end{figure*}

\begin{figure*}
	\begin{subfigure}[t]{1.00\textwidth}
		\centering
		\includegraphics[width=\linewidth]{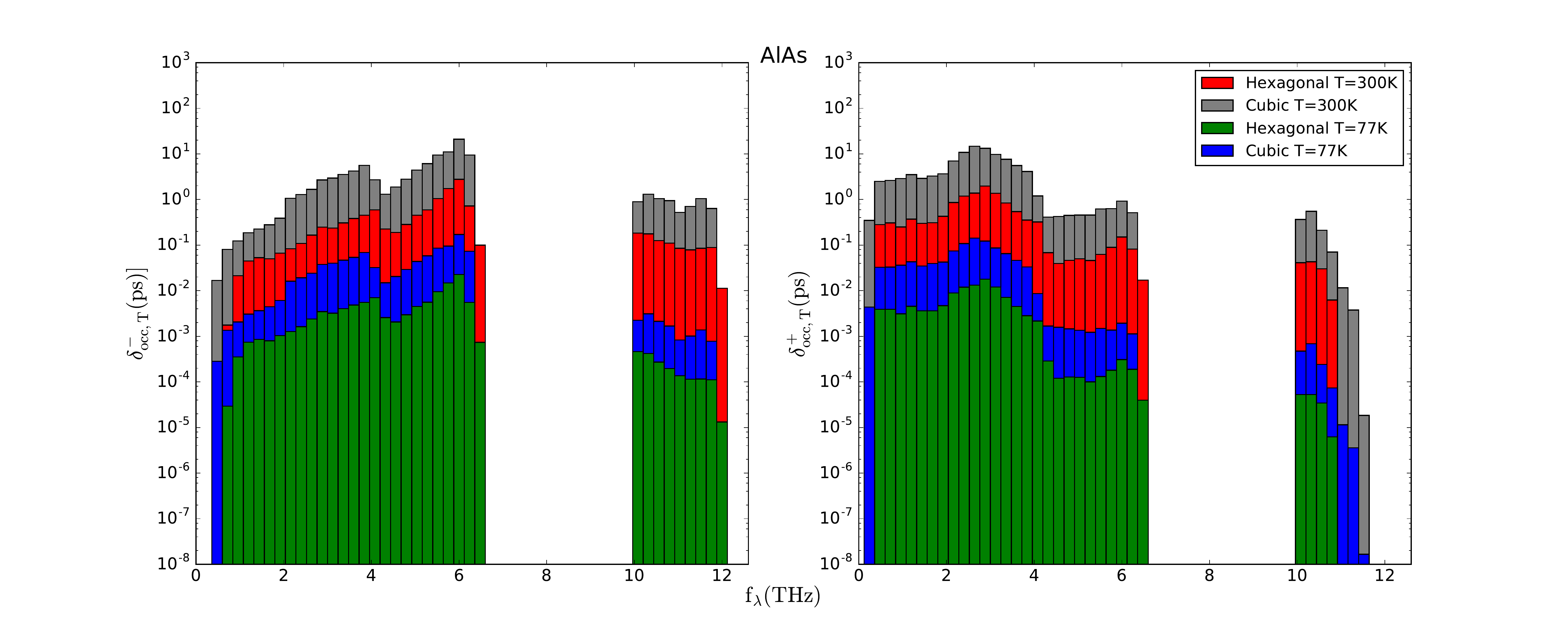}
		\label{alas-delta}
	\end{subfigure}
	\begin{subfigure}[t]{1.00\textwidth}
		\centering
		\includegraphics[width=\linewidth]{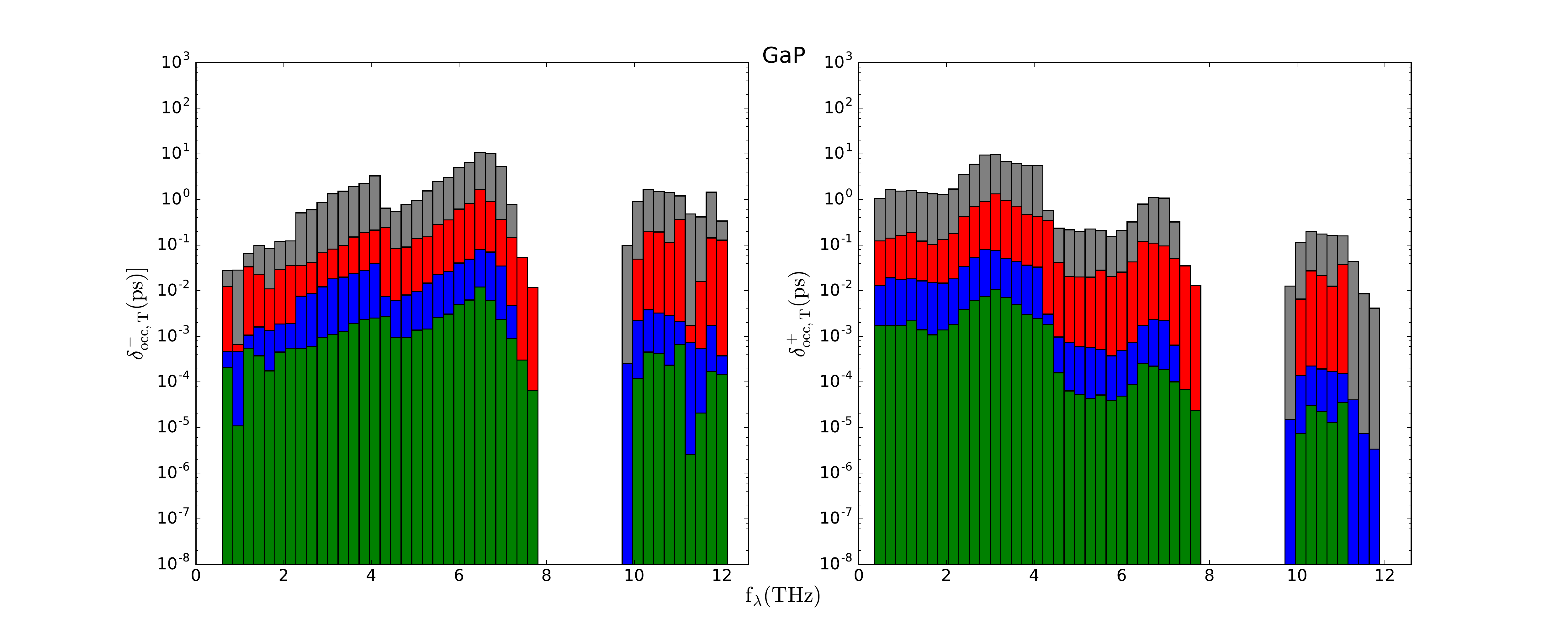}
		\label{gap-delta}
	\end{subfigure}
	\begin{subfigure}[t]{1.00\textwidth}
		\centering
		\includegraphics[width=\linewidth]{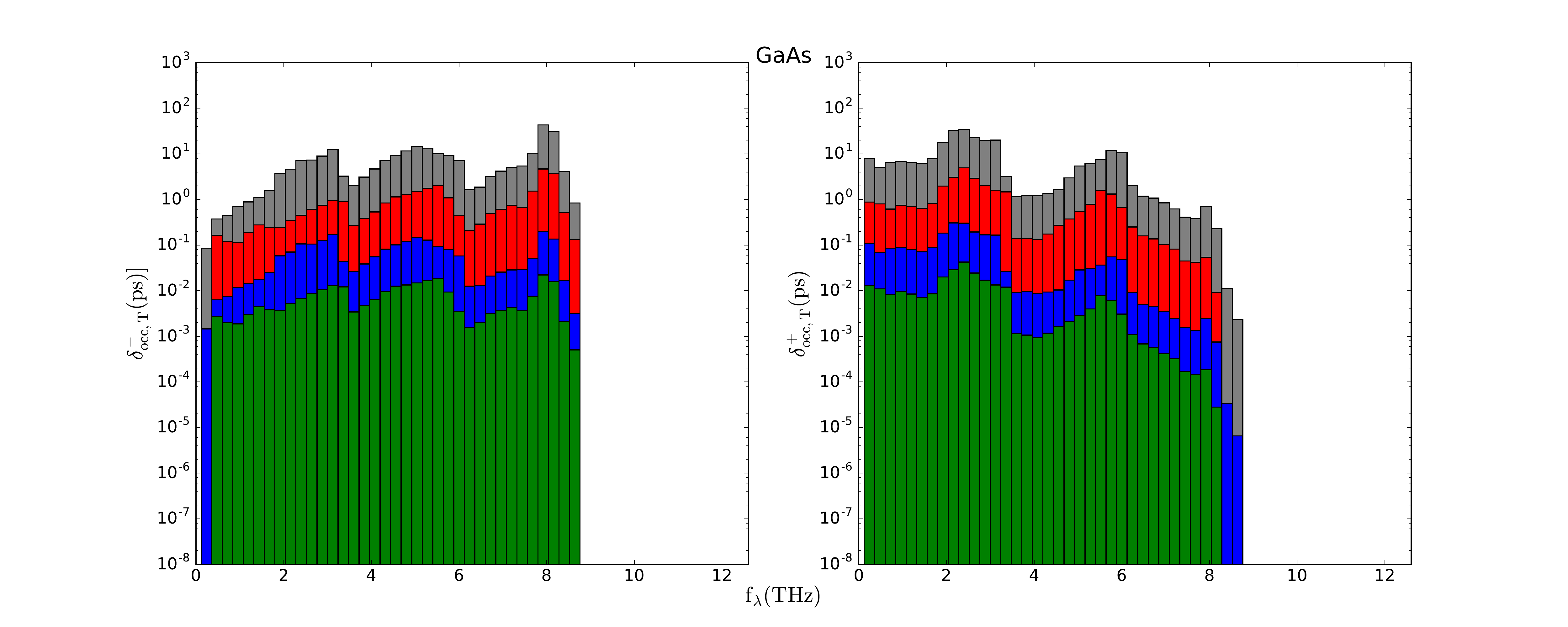}
		\label{gaas-delta}
	\end{subfigure}
	\caption{Contribution to the scattering space ($\delta_{\lambda\lambda'\lambda''}^{\pm\;occ,T}$) as a function of first phonon frequency for AlAs, GaP and GaAs. Left images correspond to emission processes and right images to absorption processes.}
	\label{DELTAS2}
\end{figure*}

\begin{figure*}
	\begin{subfigure}[t]{1.00\textwidth}
		\centering
		\includegraphics[width=\linewidth]{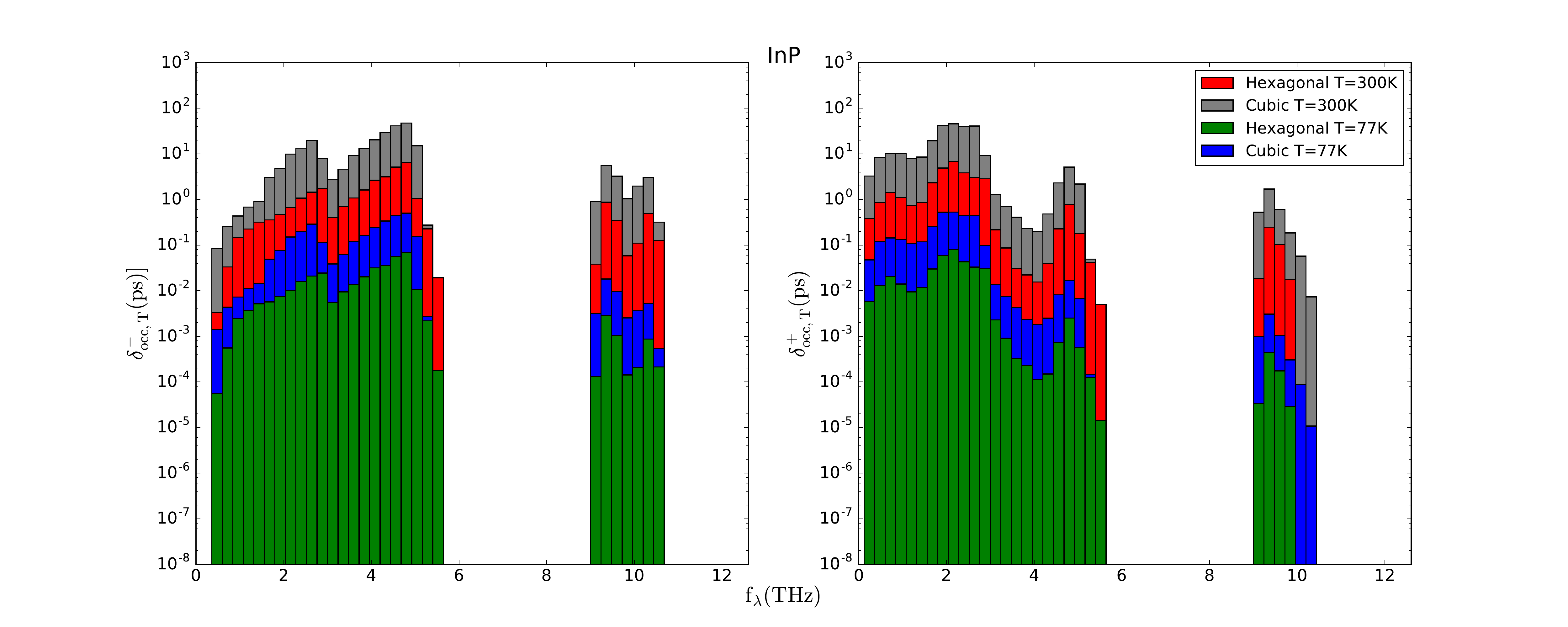}
		\label{inp-delta}
	\end{subfigure}
	\begin{subfigure}[t]{1.00\textwidth}
		\centering
		\includegraphics[width=\linewidth]{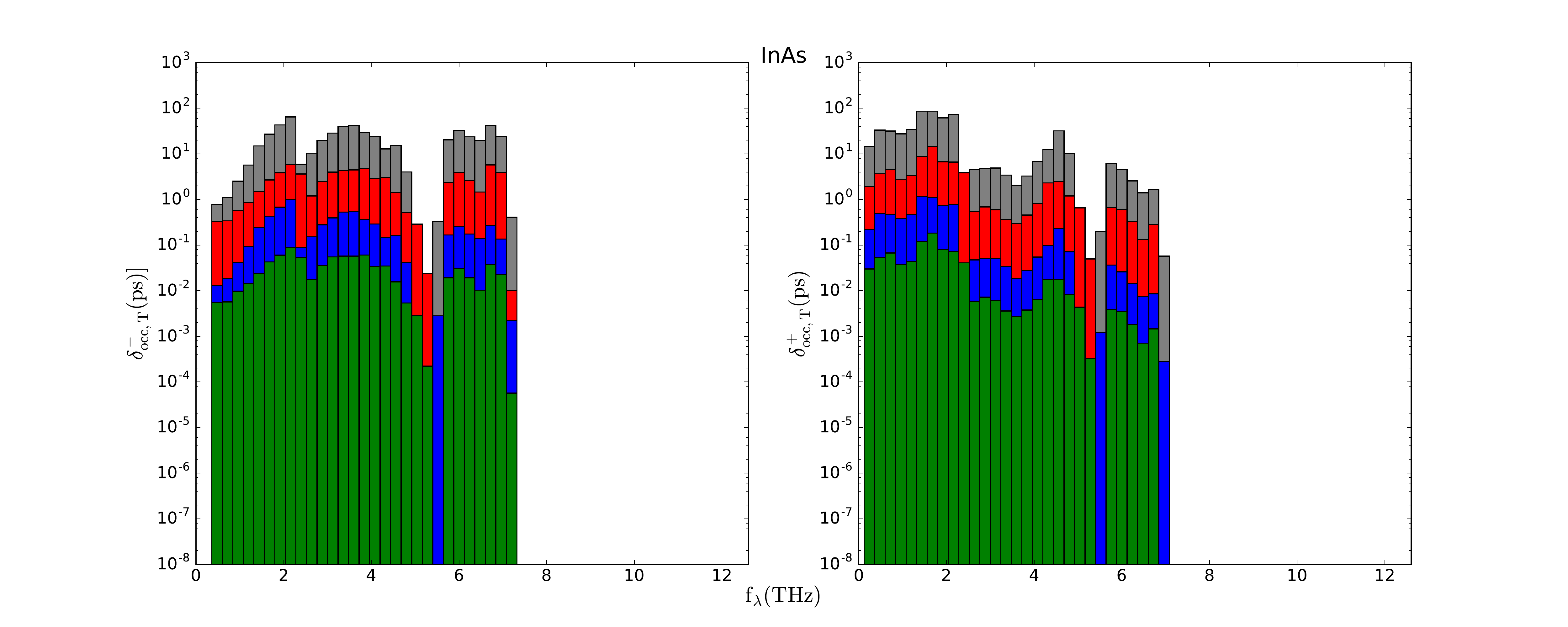}
		\label{inas-delta}
	\end{subfigure}
	\begin{subfigure}[t]{1.00\textwidth}
		\centering
		\includegraphics[width=\linewidth]{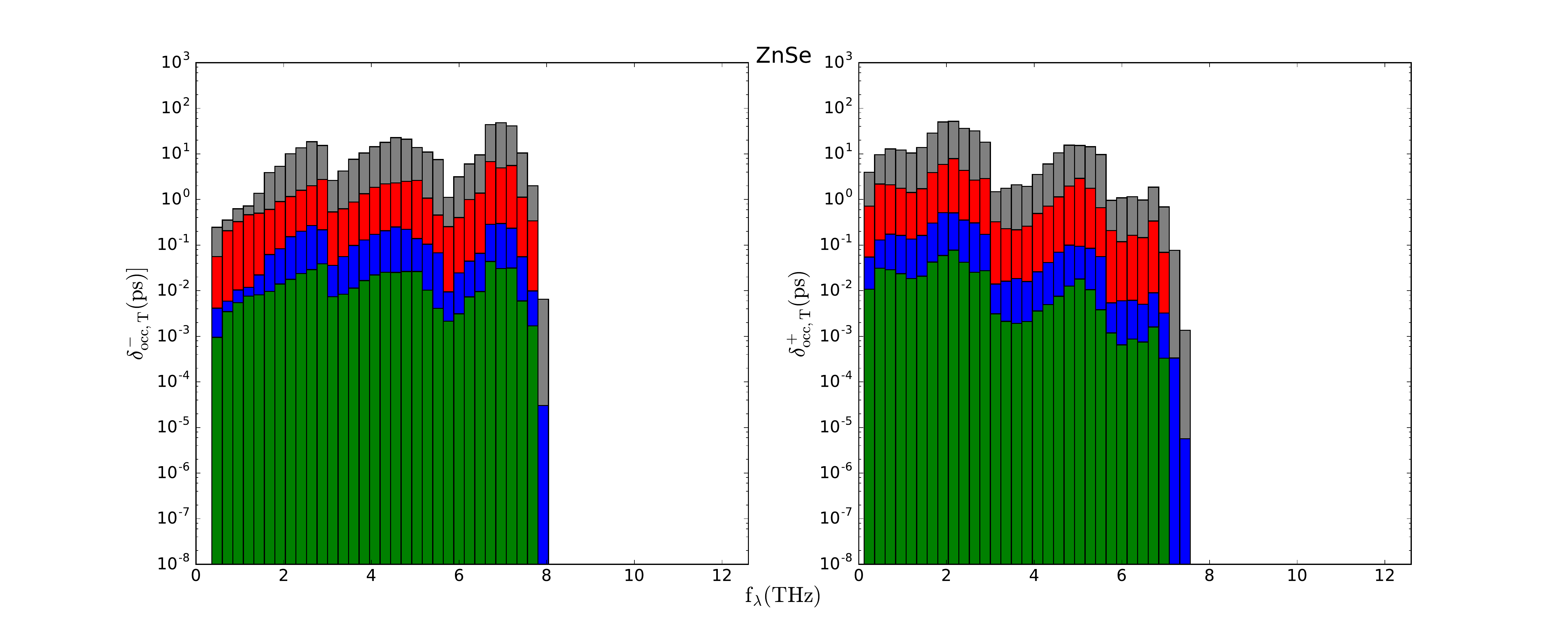}
		\label{znse-delta}
	\end{subfigure}
	\caption{Contribution to the scattering space ($\delta_{\lambda\lambda'\lambda''}^{\pm\;occ,T}$) as a function of first phonon frequency for InP, InAs and ZnSe. Left images correspond to emission processes and right images to absorption processes.}
	\label{DELTAS3}
\end{figure*}

The accessible phase space at 77 and 300 K as function of first phonon frequency of the involved triplet are provided Figs.~\ref{DELTAS1}-\ref{DELTAS3}.

\section{Mean effective anharmonicity}
\begin{figure*}
	\begin{subfigure}[t]{1.00\textwidth}
		\centering
		\includegraphics[width=\linewidth]{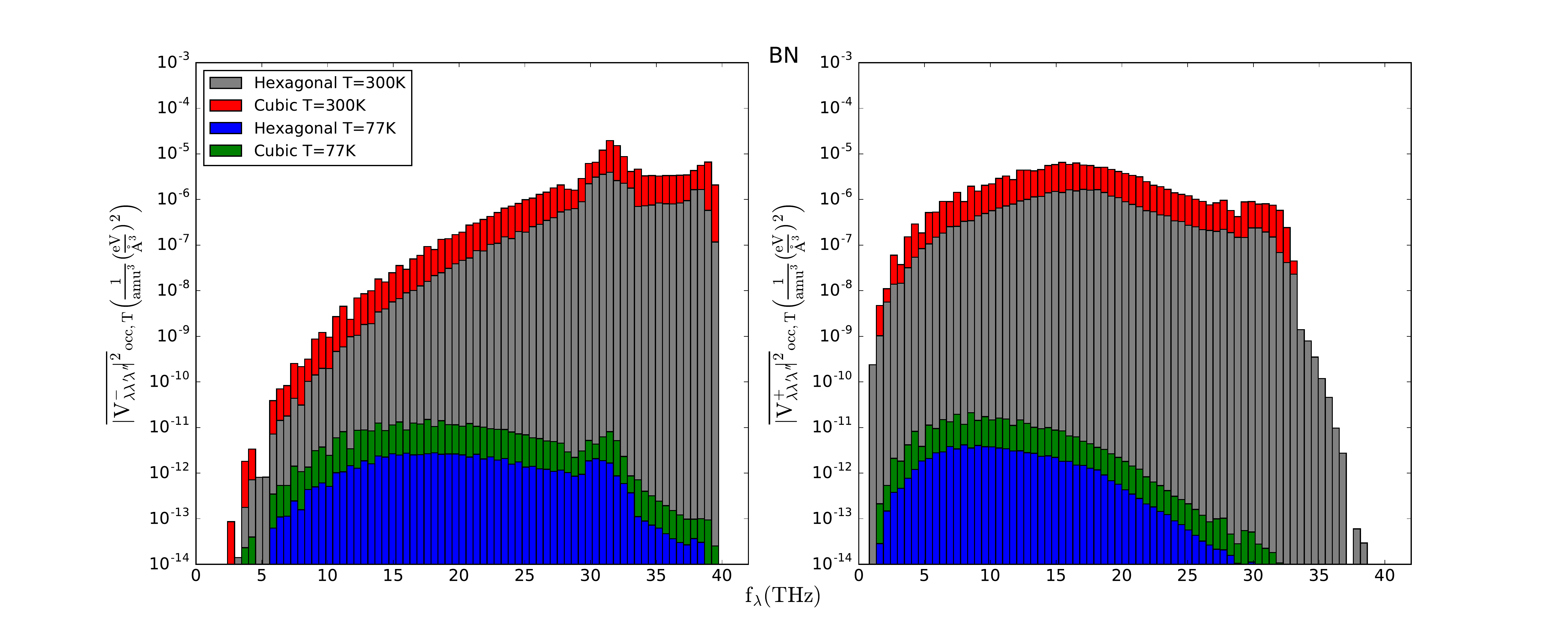}
		\label{bn-vpp}
	\end{subfigure}
	\begin{subfigure}[t]{1.00\textwidth}
		\centering
		\includegraphics[width=\linewidth]{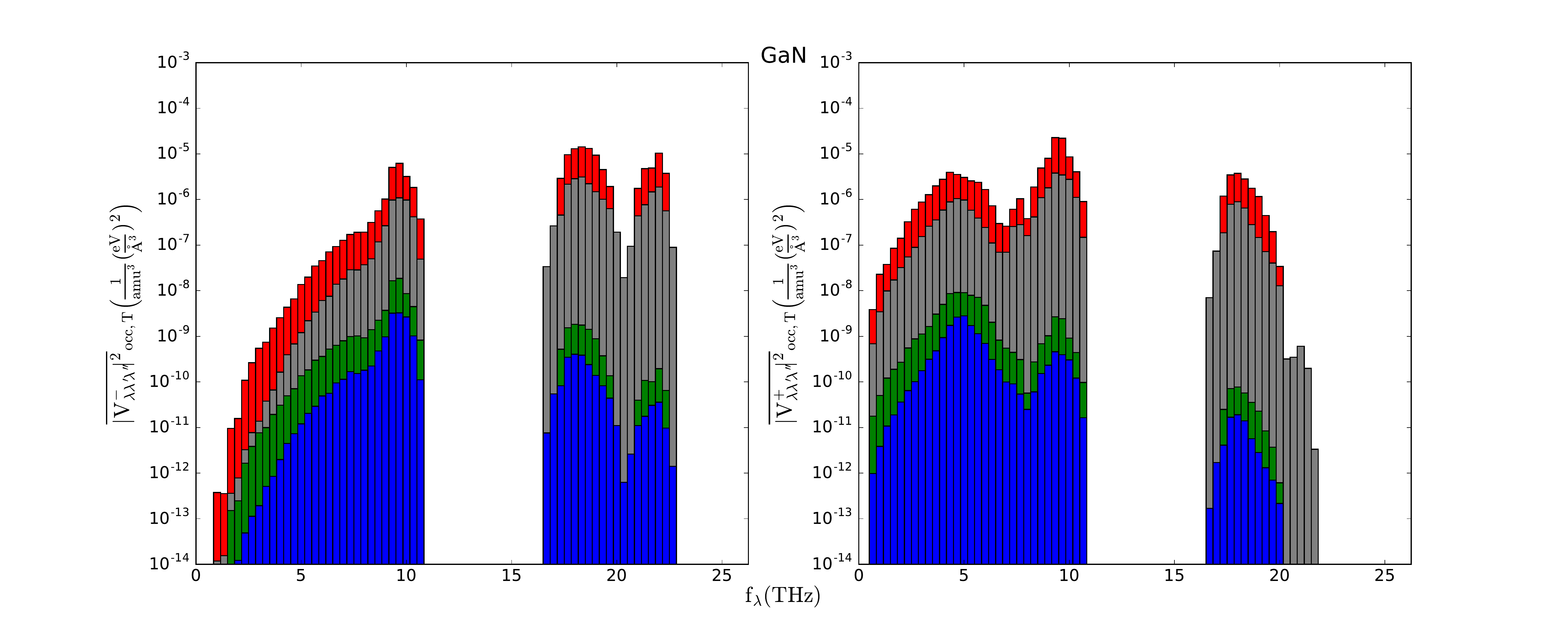}
		\label{gan-vpp}
	\end{subfigure}
	\caption{Mean of effective anharmonicity as a function of first phonon frequency for BN and GaN. Left images correspond to emission processes and right images to absorption processes.}
	\label{vpp_fulls1}	
\end{figure*}

\begin{figure*}
	\begin{subfigure}[t]{1.00\textwidth}
		\centering
		\includegraphics[width=\linewidth]{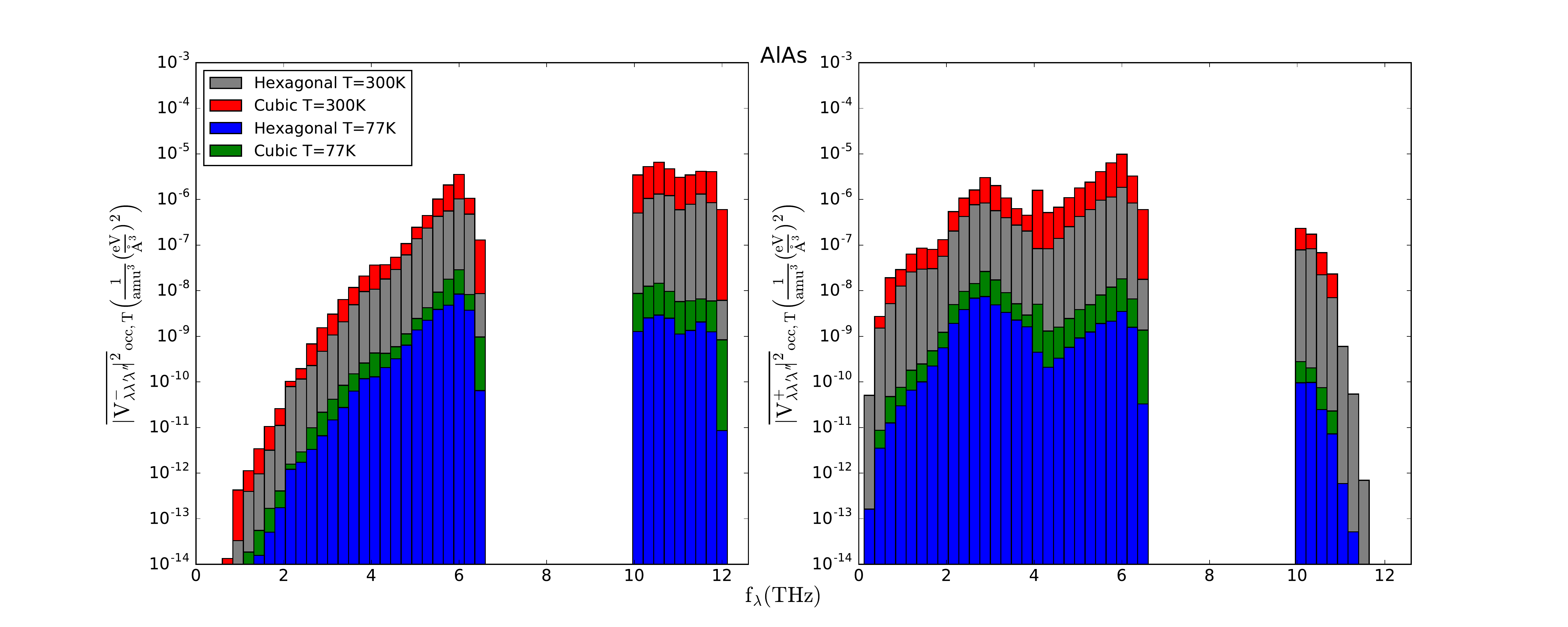}
		\label{alas-vpp}
	\end{subfigure}	
	\begin{subfigure}[t]{1.00\textwidth}
		\centering
		\includegraphics[width=\linewidth]{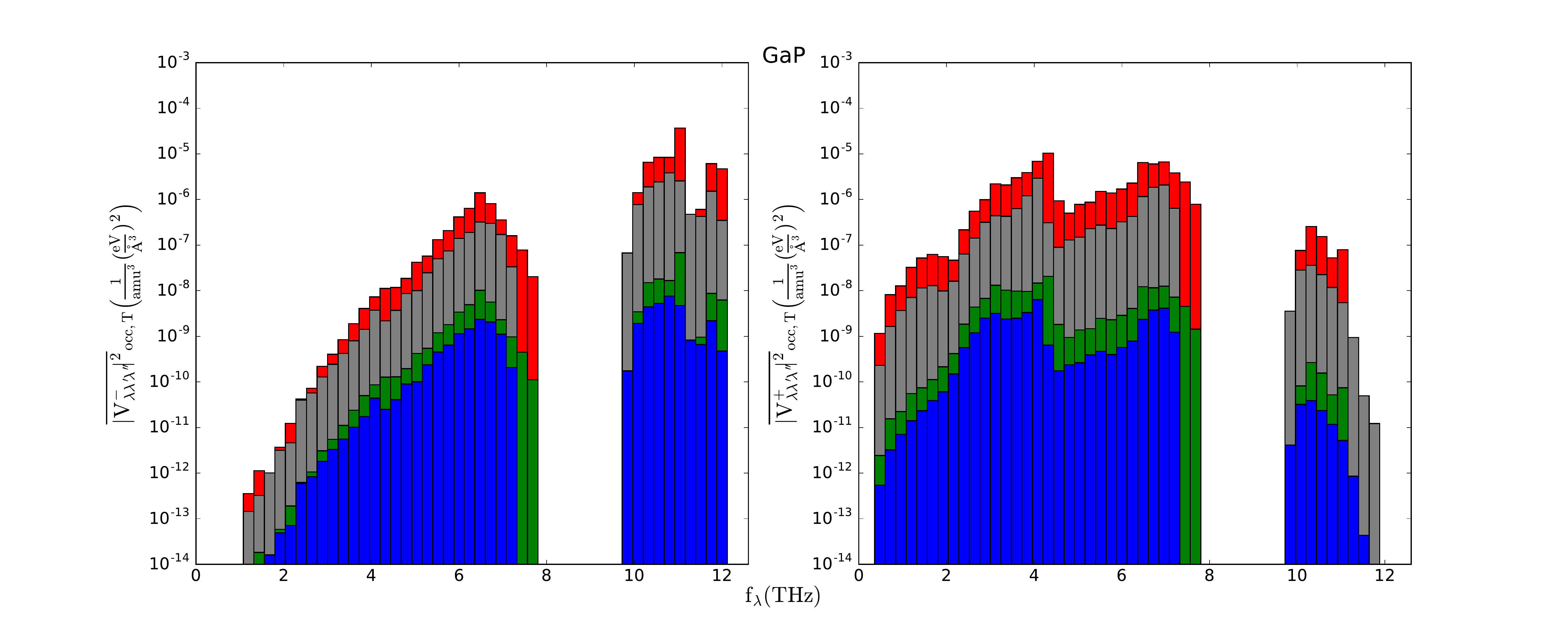}
		\label{gap-vpp}
	\end{subfigure}
	\begin{subfigure}[t]{1.00\textwidth}
		\centering
		\includegraphics[width=\linewidth]{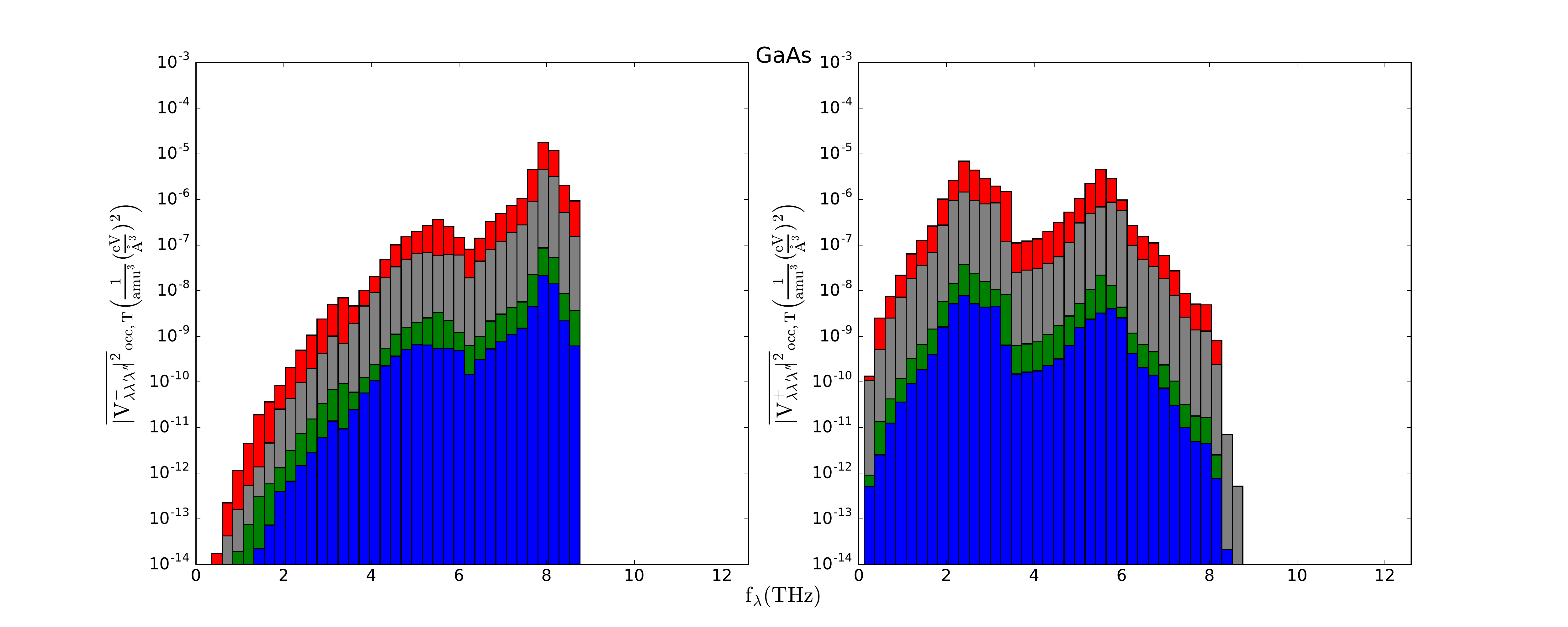}
		\label{gaas-vpp}
	\end{subfigure}
	\caption{Mean of effective anharmonicity as a function of first phonon frequency for AlAs, GaP and GaAs. Left images correspond to emission processes and right images to absorption processes.}
	\label{vpp_fulls2}	
	
\end{figure*}

\begin{figure*}
	\begin{subfigure}[t]{1.00\textwidth}
		\centering
		\includegraphics[width=\linewidth]{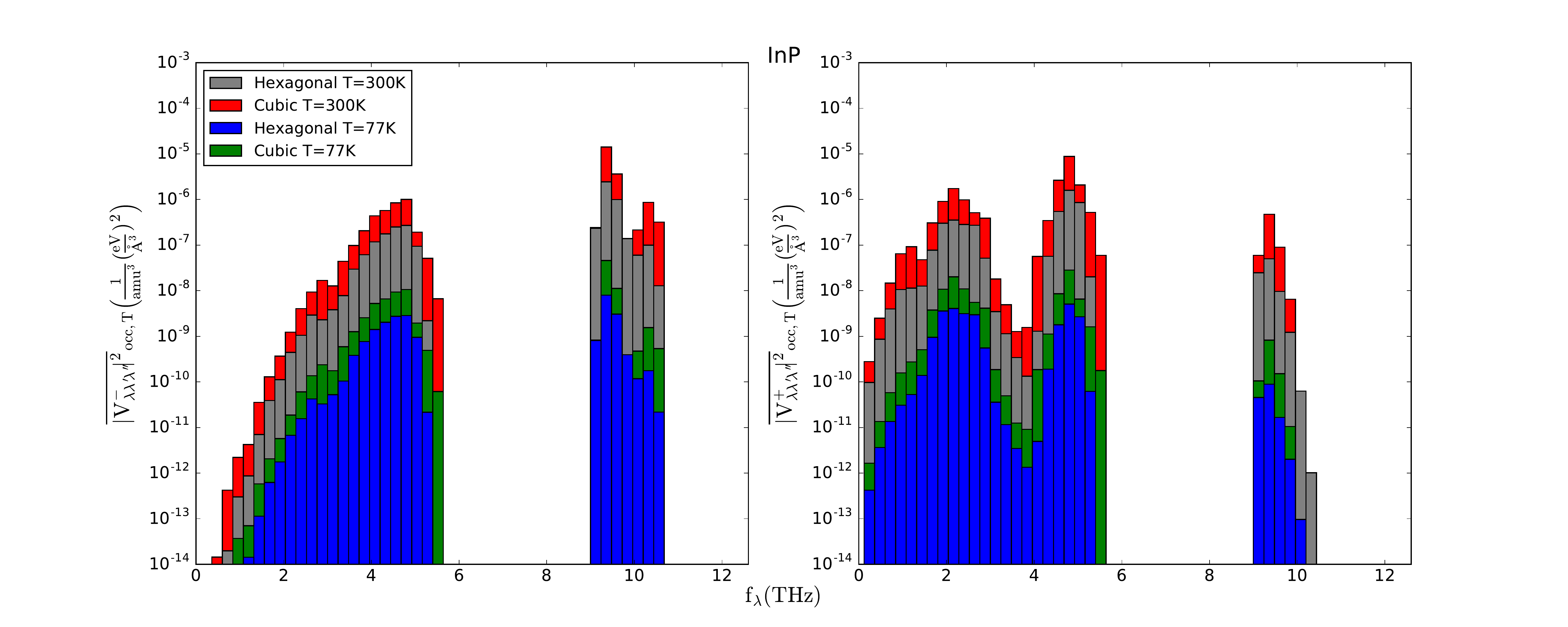}
		\label{inp-vpp}
	\end{subfigure}
	\begin{subfigure}[t]{1.00\textwidth}
		\centering
		\includegraphics[width=\linewidth]{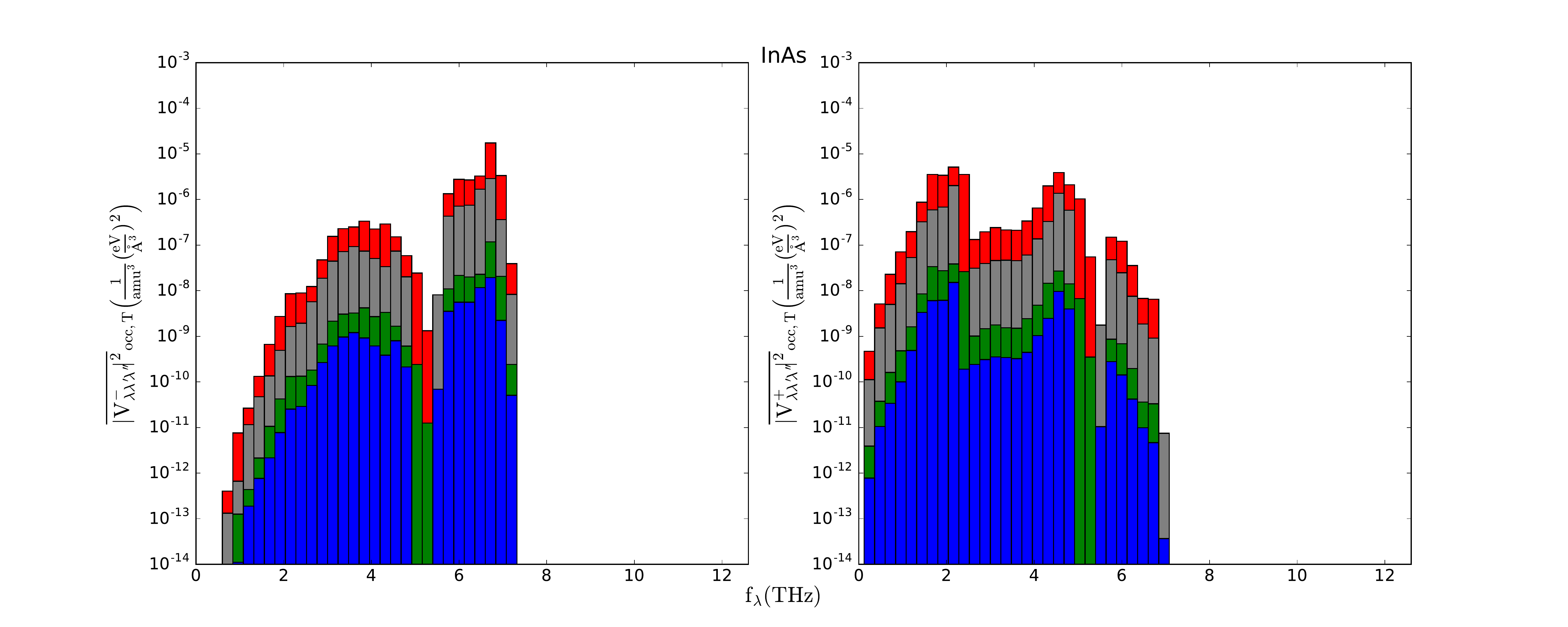}
		\label{inas-vpp}
	\end{subfigure}
	\begin{subfigure}[t]{1.00\textwidth}
		\centering
		\includegraphics[width=\linewidth]{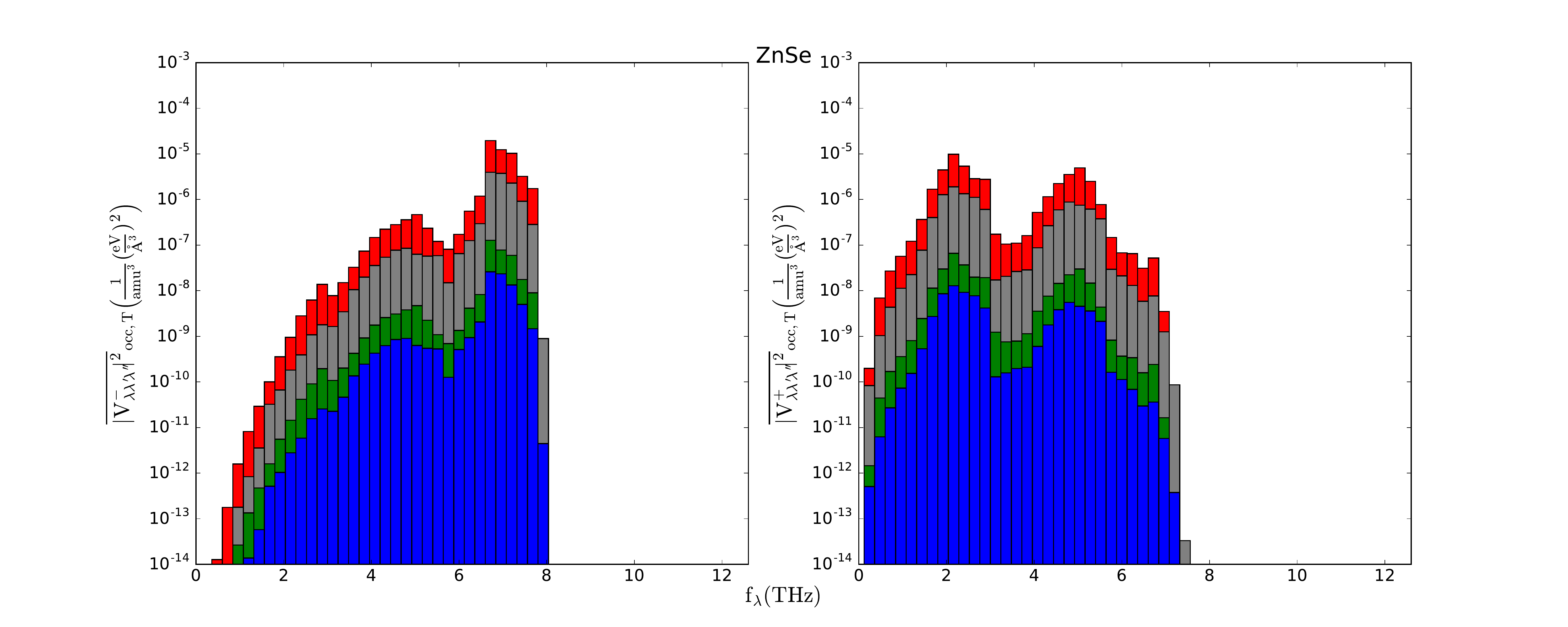}
		\label{znse-vpp}
	\end{subfigure}
	\caption{Mean of effective anharmonicity as a function of first phonon frequency for InP, InAs and ZnSe. Left images correspond to emission processes and right images to absorption processes.}
	\label{vpp_fulls3}
\end{figure*}

The mean effective anharmonicity as a function of first-phonon frequency is presented in Figs.~\ref{vpp_fulls1}-\ref{vpp_fulls3}. As noted in the main text, the mean is always higher in the zinc-blende phase.

\section{REAAPS at 700K}
\begin{figure*}
	\centering
	\includegraphics[width=1.00\textwidth]{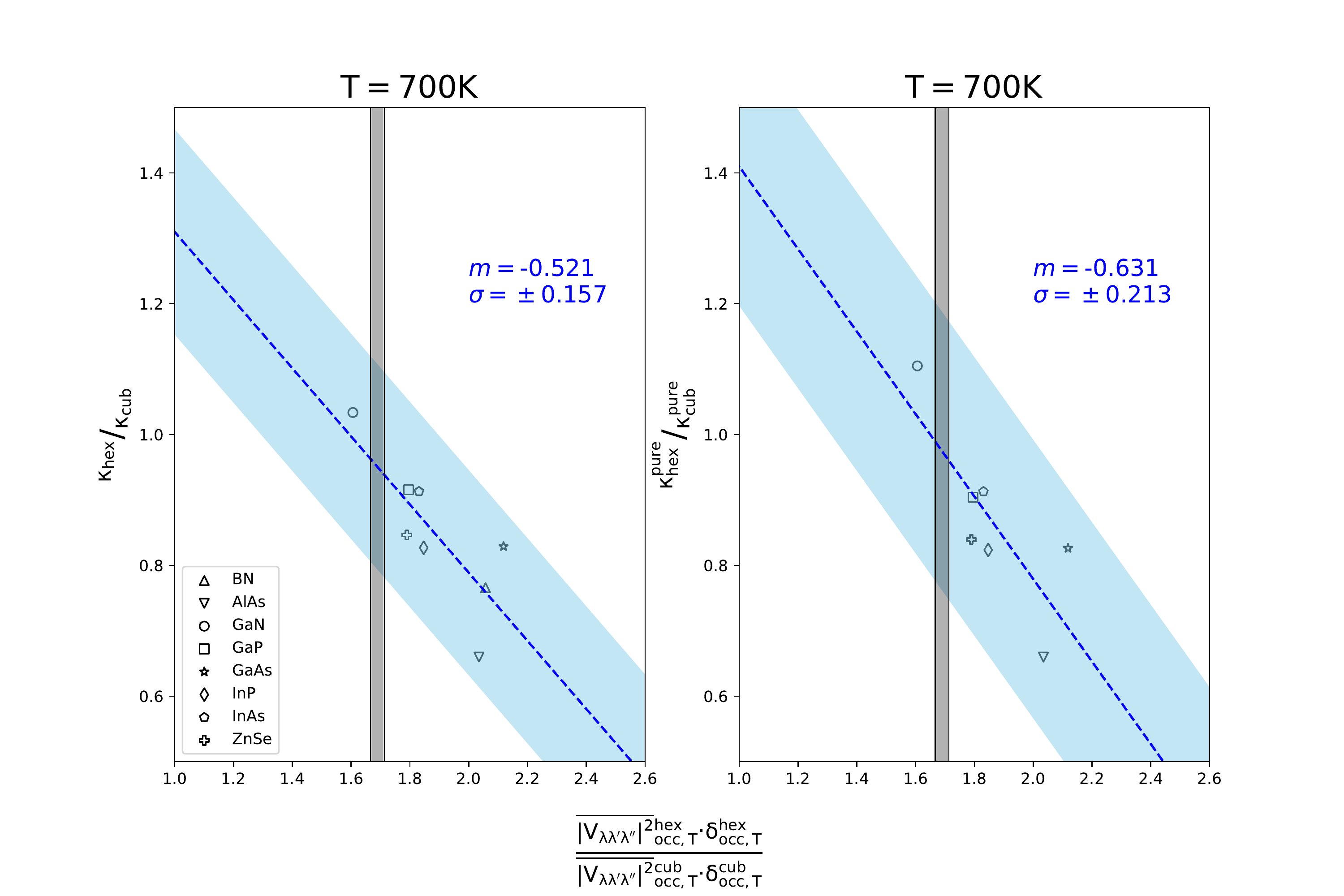}
	\caption{$\kappa_{hex}/\kappa_{cub}$ ratio as a function of the hexagonal-cubic ratio of the mean effective anharmonicity and the accessible phase space product at 700K for different materials with isotopic scattering (left) and without it (right). The factor that controls the ratio $\mathrm{\kappa_{hex}/\kappa_{cub}}$ (anharmonicity or phase space) for a given REAAPS is indicated together with which phase is the most conducting for that value. The grey filled region corresponds to the limiting region. The dashed blue line corresponds to a linear regression of the data, with the corresponding slope $m$ and standard deviation, $\sigma$ provided in the figure. The cyan area corresponds to an interval $\pm \sigma$.
		\label{reaaps700}
	}
\end{figure*}

The ratios with and without isotopes as function of REAAPS at 700K are shown in Fig.~\ref{reaaps700}.

\section{Nanowire and cumulative thermal conductivity ratios}
\begin{figure*}
	\begin{subfigure}{0.42\linewidth}
		\centering
		\includegraphics[width=\linewidth]{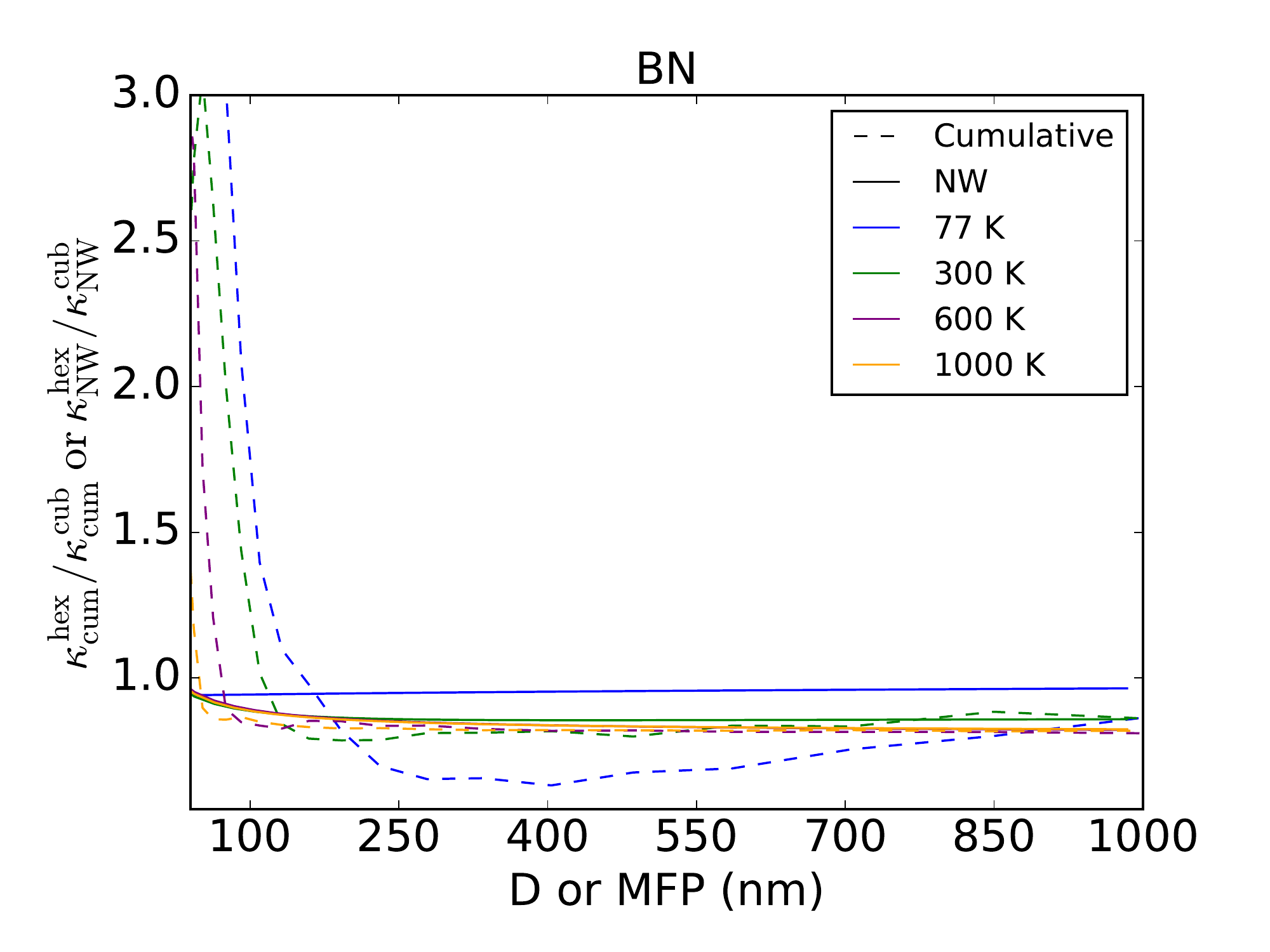}
	\end{subfigure}
	\begin{subfigure}{0.42\linewidth}
		\centering
		\includegraphics[width=\linewidth]{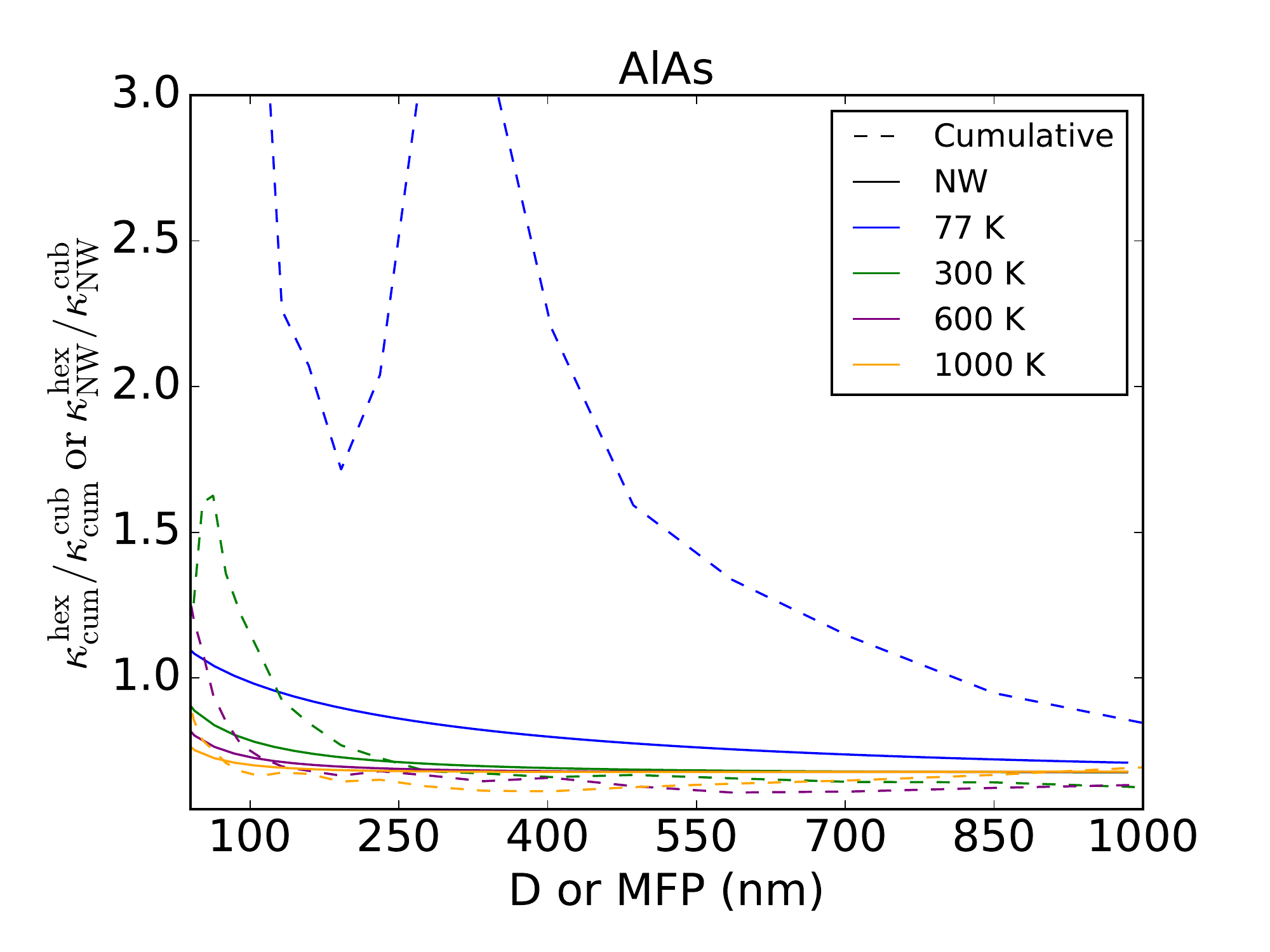}
	\end{subfigure}
	\begin{subfigure}[t]{0.42\textwidth}
		\centering
		\includegraphics[width=\linewidth]{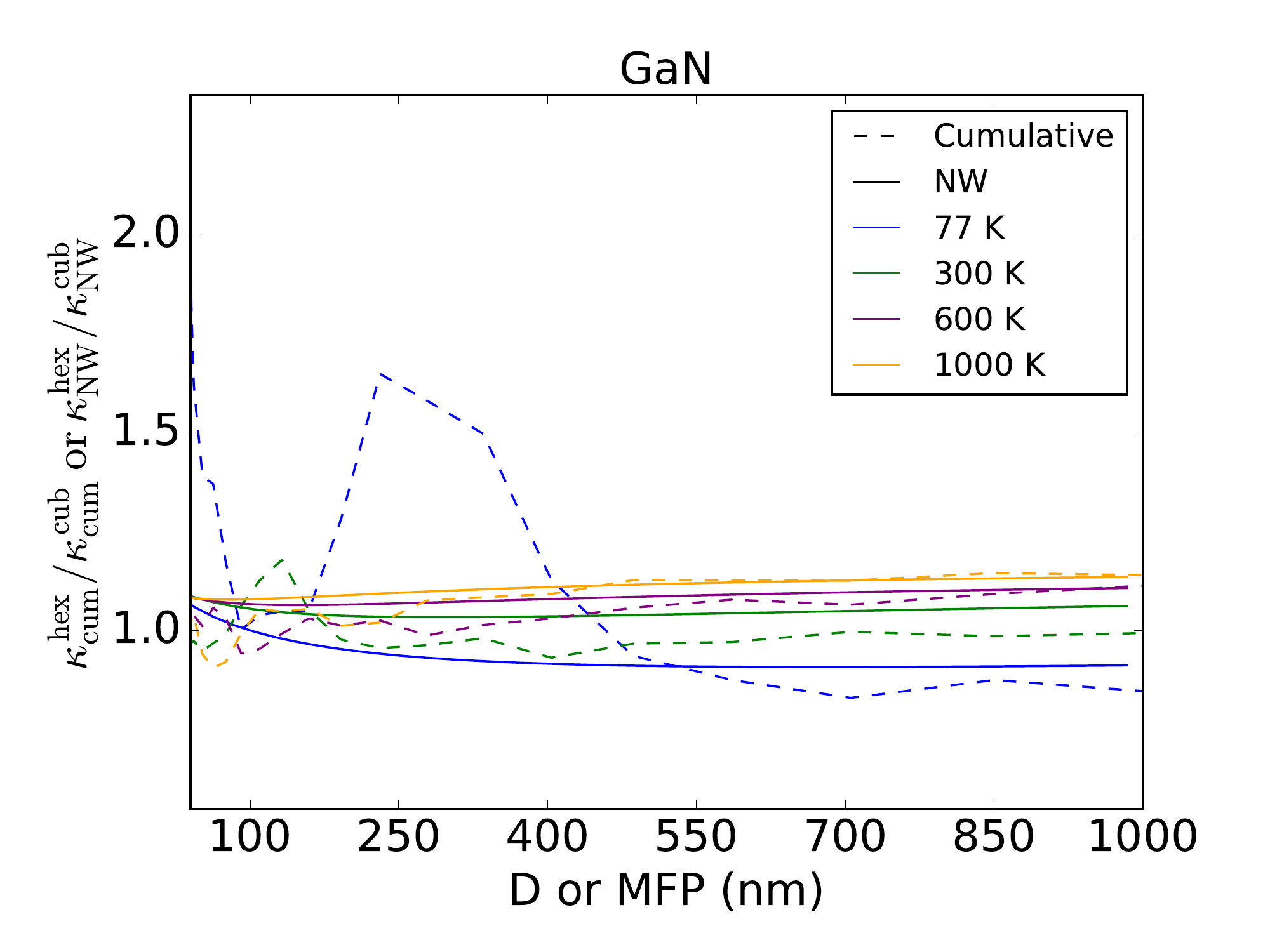}
	\end{subfigure}
	\begin{subfigure}[t]{0.42\textwidth}
		\centering
		\includegraphics[width=\linewidth]{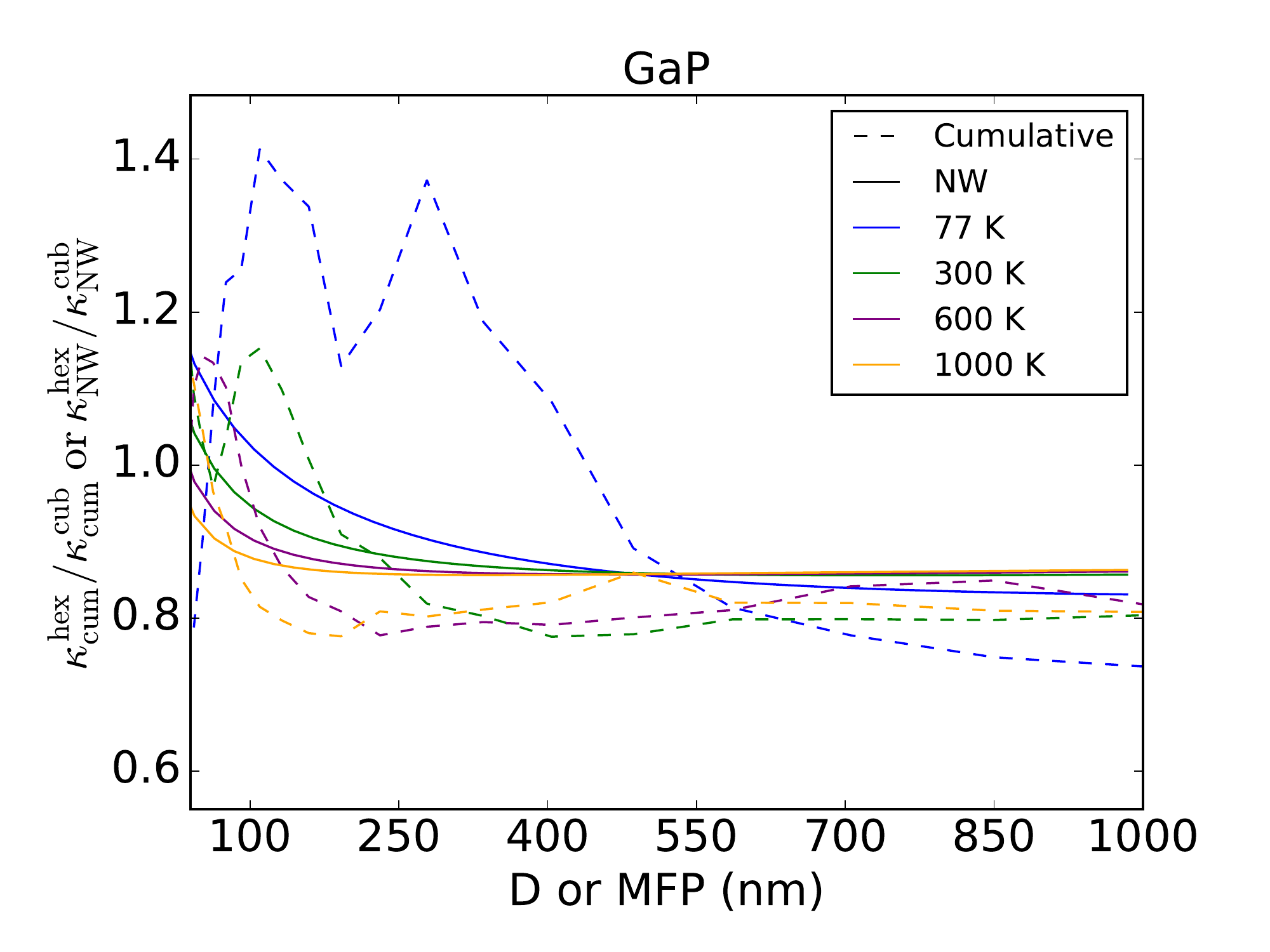}
	\end{subfigure}
	\begin{subfigure}[t]{0.42\textwidth}
		\centering
		\includegraphics[width=\linewidth]{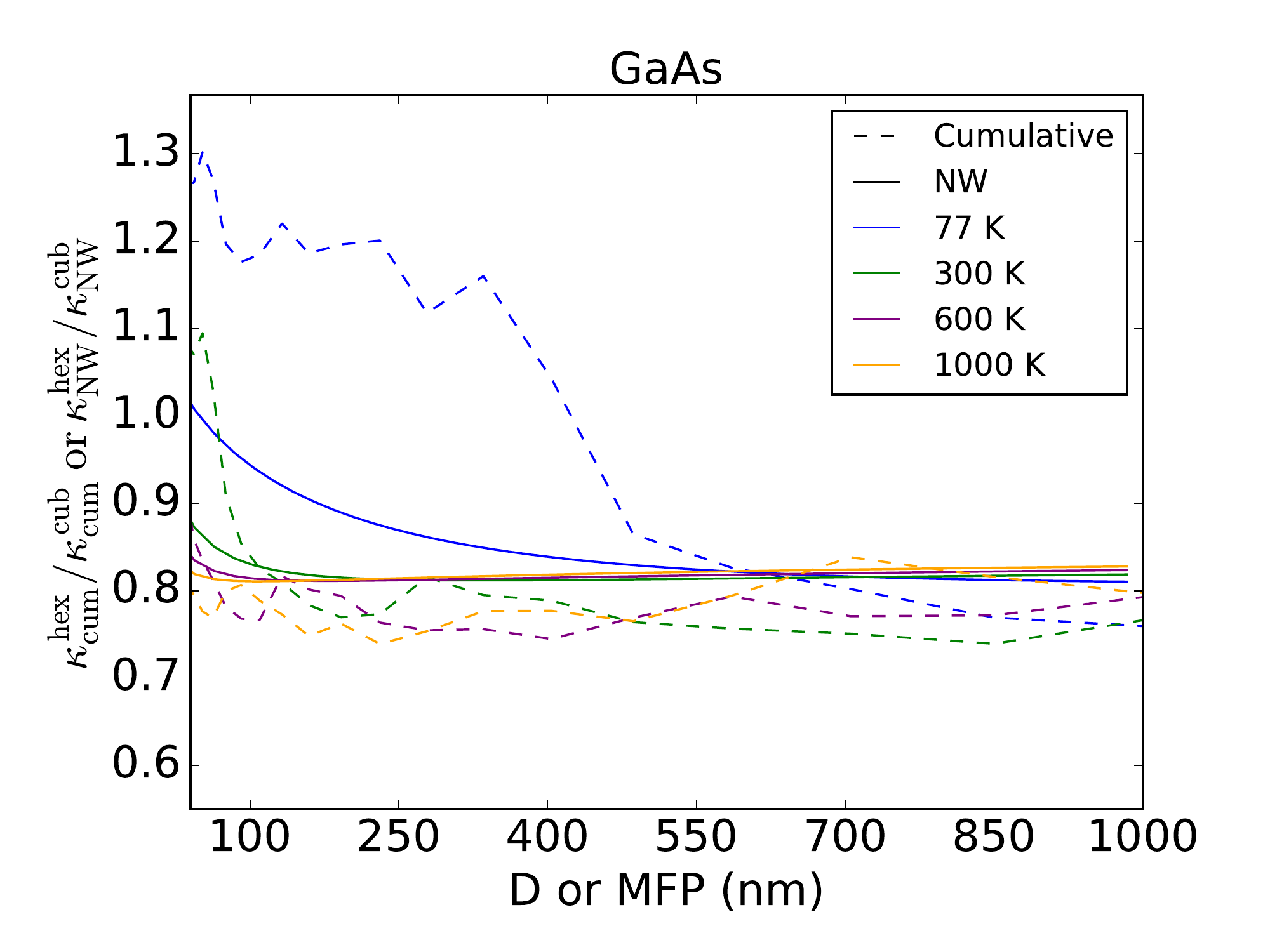}
	\end{subfigure}
	\begin{subfigure}[t]{0.42\textwidth}
		\centering
		\includegraphics[width=\linewidth]{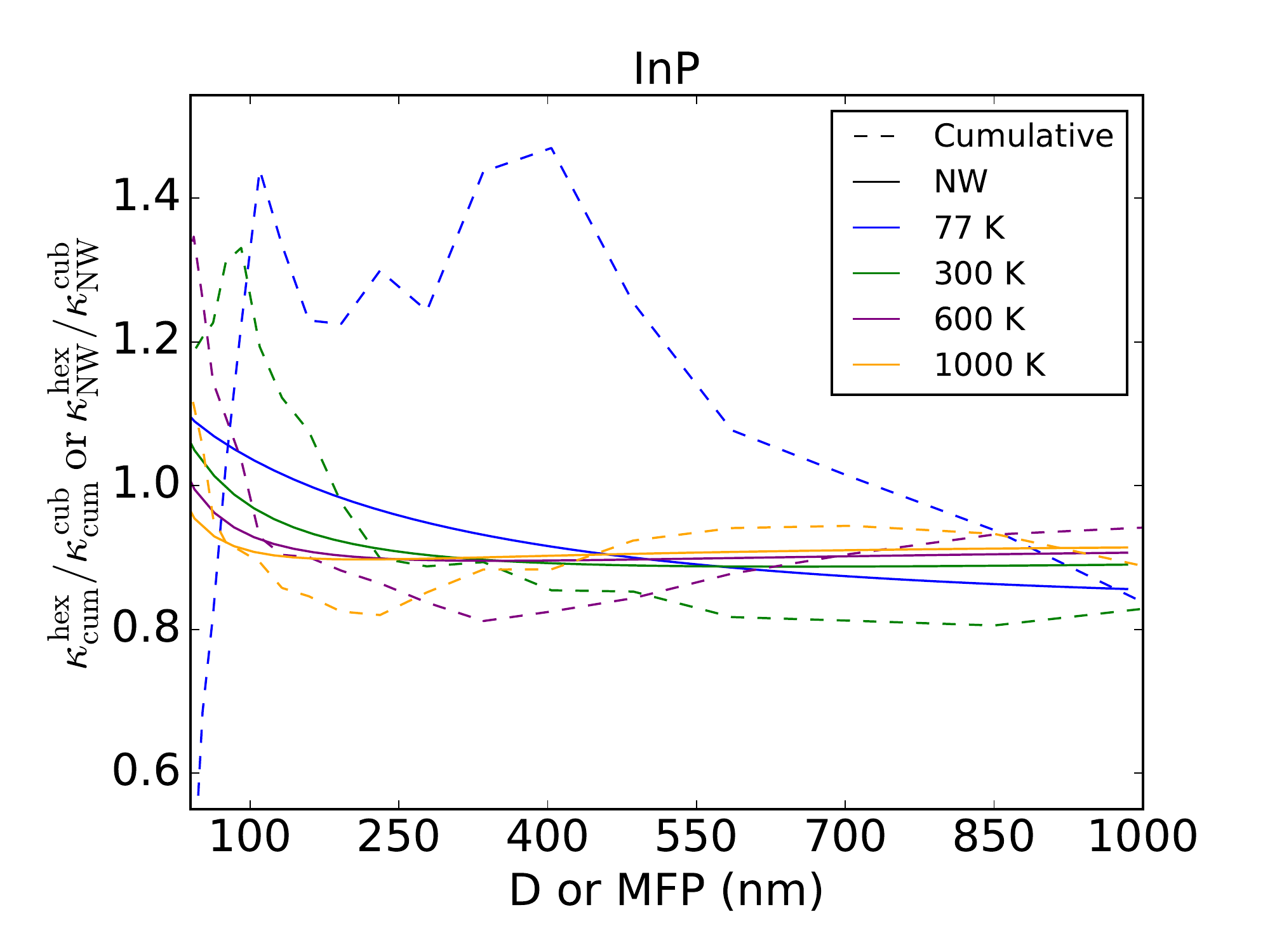}
	\end{subfigure}
	\begin{subfigure}[t]{0.42\textwidth}
		\centering
		\includegraphics[width=\linewidth]{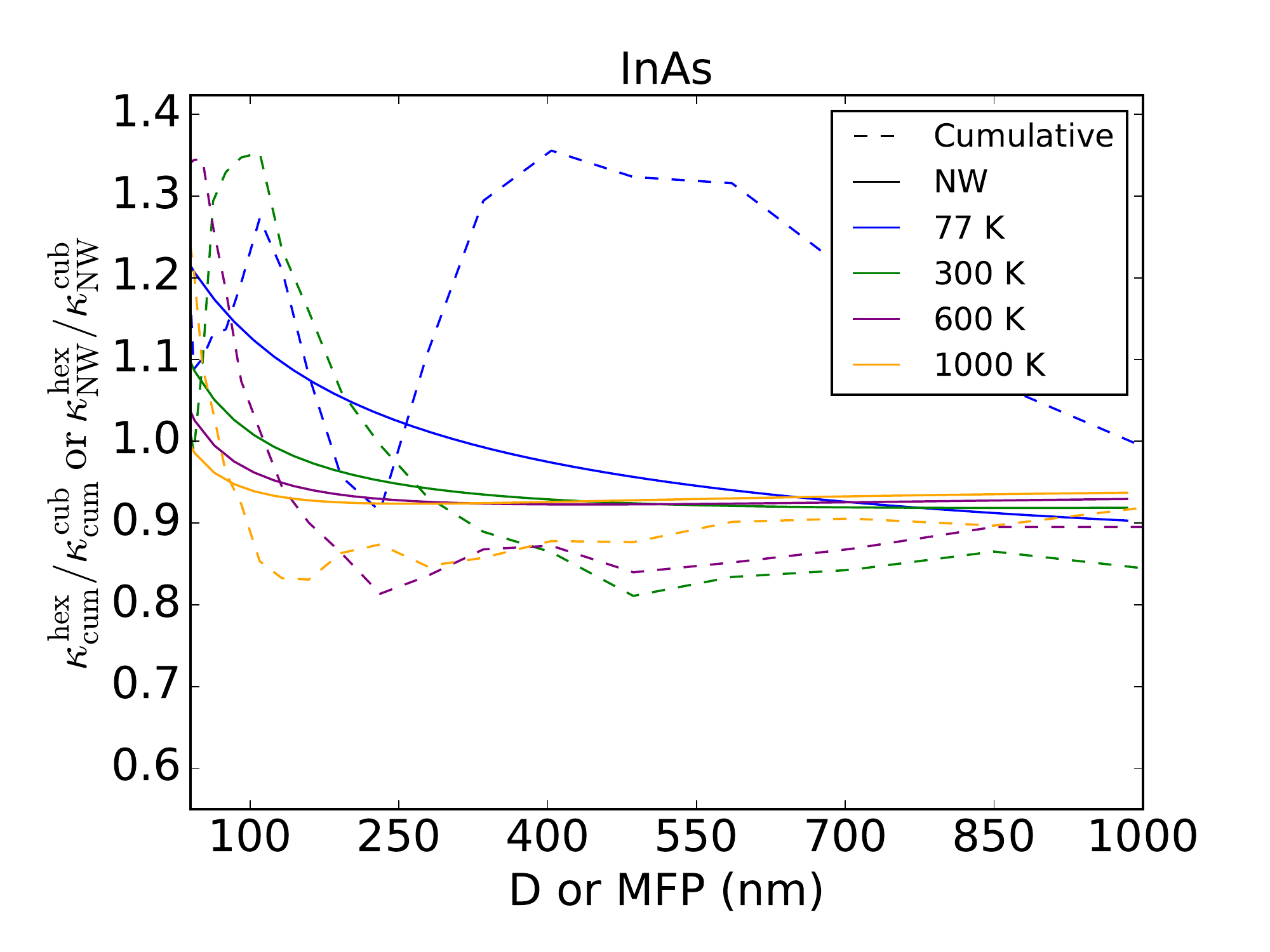}
	\end{subfigure}
	\begin{subfigure}[t]{0.42\textwidth}
		\centering
		\includegraphics[width=\linewidth]{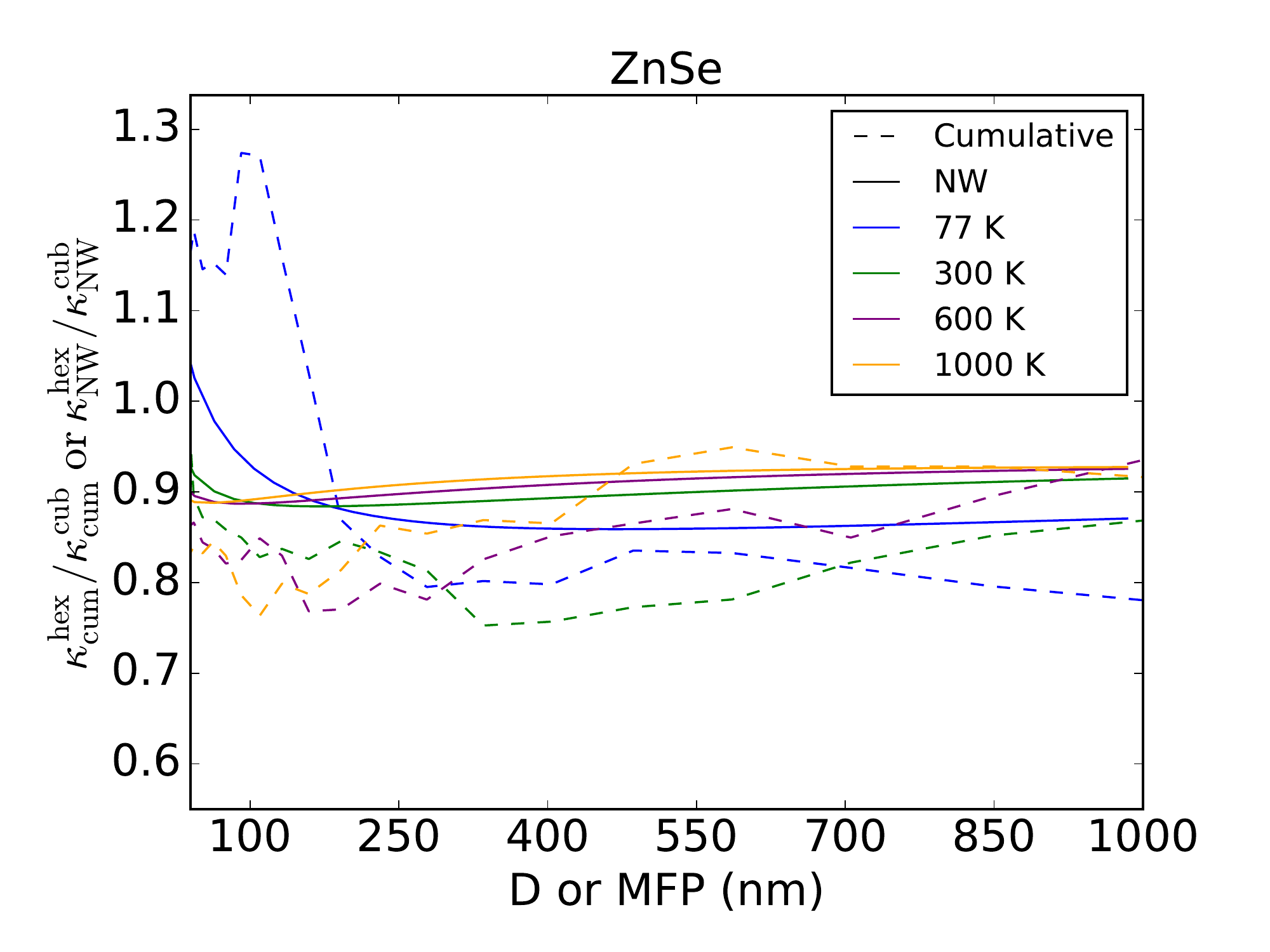}
	\end{subfigure}
	\caption{Solid lines: Hexagonal-cubic thermal conductivity ratio for nanowires at 77K (blue), 300K (green), 600K (purple) and 1000K (yellow) as function of diameter. Dashed lines:  Hexagonal-cubic cumulative thermal conductivity ratio at 77K (blue), 300K (green), 600K (purple) and 1000K (yellow) as function of phonon MFP.}
	\label{MFP-ratios}
\end{figure*}

The ratios of nanowire and cumulative thermal conductivity as function of diameter/MFP are presented in Fig.~\ref{MFP-ratios}

\bibliography{references}